%% file: demo.tex
\documentclass[12pt,english]{article}

\input{preamble}

\title{Optimal Seat Movement Strategies Among Modular Bus Units}

\def\shortauthors{Wang et al.}
\def\runningtitle{Optimal Seat Movement Strategies Among Modular Bus Units}

\author[a,c]{Shiyue Wang}
\author[b,d]{Yuhao Liu}
\author[b,$\ast$]{Zhibin Chen}
\author[e]{Mingyang Pei}
\author[a]{Changhong Lu}
\affil[a]{School of Mathematical Sciences, Key Laboratory of MEA (Ministry of Education), Shanghai Key Laboratory of PMMP, and Nantong Institute for Applied Mathematics and Artificial Intelligence, East China Normal University, Shanghai, 200241, China} 
\affil[b]{Center for Data Science, NYU Shanghai, NYU Shanghai, 567 West Yangsi Road, Pudong New District, Shanghai, 200126, China}
\affil[c]{Shanghai Innovation Institute, Shanghai, 200231, China}
\affil[d]{Department of Civil and Urban Engineering, New York University, 6 MetroTech Center, Brooklyn, NY 11201, United States}
\affil[e]{School of Civil Engineering \& Transportation, South China University of Technology, Guangzhou, 510640, China}

\def\corresemail{zc23@nyu.edu}

\def\abstract{
Movable seats enable passengers to transfer between coupled modular bus units while remaining seated, but require physical reconfiguration that existing modular transit models largely abstract away. We model seat reconfiguration as a colored generalized sliding-tile puzzle on the physical track graph, capturing simultaneous movements, seat indistinguishability within destination groups, and vertex, edge, and corner-following conflicts. A lexicographic objective minimizes makespan first and then the sum of costs (SoC), which penalizes movements and waiting at non-target positions. We develop an exact time-expanded mixed-integer linear programming formulation and a Boundary-Sequence Dynamic Programming (BSDP) algorithm. BSDP decomposes the problem at single-edge module interfaces and coordinates local schedules through compressed boundary-crossing sequences, provably preserving feasibility and optimal costs. On two- and three-module benchmarks, BSDP achieves mean per-instance speedups of 3.24× and 6.30× over Gurobi, respectively. In fixed-horizon tests with four to seven modules, both methods start from identical greedy schedules and receive a 300-s budget. BSDP obtains lower final SoC than Gurobi on 197 of 200 instances and ties on three, reducing mean final SoC by 17.93\%–69.07\% across the four size classes. Geometry sensitivity and ablation studies further clarify the contribution of boundary compression to computational performance. The framework links seat reconfiguration to the timing and operational cost of modular bus decoupling.
}

\def\keyword{Modular bus\\Seat movement\\Colored generalized sliding-tile puzzle\\Dynamic programming}

\begin{document}
\maketitle
\titlepageext

\section{Introduction}\label{sec:Intro}

With the emergence of ridesourcing, carpooling, and carsharing platforms offering door-to-door service at relatively low prices, an increasing number of travelers are opting out of traditional bus services. Reviving the bus service has thus become a significant and challenging question that remains unanswered. Recently, modular bus (MB) technology, developed by the Next Future Transportation Incorporation~\citep{NextFutureTransport}, has opened new possibilities for addressing the challenges faced by traditional bus services. An MB consists of separable modular units designed to seamlessly join or detach from one another on conventional city streets. When joined, these modules form a unified, open space resembling a traditional bus, allowing passengers to move freely across modules. When decoupled, the individual modules can travel to different destinations, offering door-to-door transit options for passengers. With this coupling-decoupling capability, MBs can easily adjust their size to match varying passenger demand, significantly reducing operational costs. 

Recognizing its potential, MB and autonomous modular vehicle (AMV) technologies have attracted increasing scholarly attention. Existing studies have investigated modular transit from several perspectives. Early studies focused on operational design and capacity optimization, jointly determining dispatch headways and the number of modular units in shuttle, corridor, and network settings~\citep{chen2019operational,chen2020operational,shi2020variable,chen2022continuous,dakic2021design,pei2021vehicle,liu2021improving}. Subsequent studies further examined vehicle formation and timetable design under time-varying or uncertain demand~\citep{ji2021scheduling,shi2021operations,tian2023joint,xia2023data,cong2025joint,wang2025optimal,yao2025collaborative}, as well as network-level, demand-responsive, and first-/last-mile service designs enabled by modularity~\citep{cheng2024autonomous,khan2024seamless,filippi2025exploiting,guo2023modular,guo2025operationalizing,zhang2020modular,fu2023dial,tang2024optimisation}. In addition, coupling–decoupling operations have been used to support transfer-based modular bus services and bus-bunching mitigation~\citep{gong2021transfer,khan2023application,khan2023bus,lin2024bunching,liu2024alleviating}, and recent studies have begun to explore stop-less or zero-dwell modular transit concepts relying on in-motion transfer capability~\citep{tian2022planning,khan2025no}.

These studies demonstrate the operational value of modularity, but they usually treat inter-module passenger movement as an abstract, instantaneous, or behaviorally simple process. In practice, however, passengers may not be able to move safely and efficiently across coupled modular units on their own, particularly when vehicles are in motion or crowded. To address this implementation gap, movable seats that slide along tracks connecting modular units provide a promising mechanism: passengers can remain seated while their seats move across modules. Despite this potential, existing studies have not optimized the detailed seat movement sequence required to reconfigure seats before modular units decouple. This paper therefore introduces and solves the optimal MB seat movement problem among coupled modular bus units.

This paper introduces and solves the optimal MB seat movement problem among coupled modular units. The objective is lexicographic: the primary objective is to minimize the makespan, which determines the earliest decoupling time of the modular bus units; among all makespan-optimal schedules, the secondary objective is to minimize the sum of costs (SoC), which captures the cumulative seat-operation cost before decoupling. The main contributions of this paper are summarized as follows.

\begin{itemize}
    \item We introduce the MB seat movement problem and formulate it as an MB-specific colored generalized sliding-tile puzzle (CGSP) on the physical track graph of coupled modular units. The formulation captures colored seat indistinguishability, empty aisle positions, vertex conflicts, edge conflicts, and corner-following conflicts.
    \item We develop an exact time-expanded mixed-integer linear programming model for the fixed-makespan seat movement problem. By searching over the makespan and minimizing SoC at the optimal makespan, the model provides a benchmark exact formulation for the lexicographic optimization problem.
    \item We propose an optimal \textbf{\underline{B}}oundary-\textbf{\underline{S}}equence \textbf{\underline{D}}ynamic \textbf{\underline{P}}rogramming (BSDP) algorithm that exploits the chain structure of coupled modular unites. By representing inter-module interactions through boundary crossing schedules, the global problem is decomposed into local module-level subproblems and composed through min-plus dynamic programming. This yields an exact algorithm tailored to the MB topology.
    \item We conduct computational experiments on 2-7 modules benchmark instances. The main results of 2 and 3 modules show that the proposed algorithm obtains the same lexicographically optimal solutions as the MILP formulation while achieving substantially lower runtimes (\textbf{$3.24\times$} speed up in 2-module instances and \textbf{$6.30\times$} speed up in 3-module instances), especially for the more difficult 3-module instances. For larger instances with 4-7 modules, BSDP reduces the mean final SoC relative to Gurobi by 17.93\%, 57.09\%, 64.16\%, and 69.07\%, respectively, under identical greedy initial schedules. It achieves a lower SoC on 197 of the 200 instances and matches Gurobi on the remaining three, demonstrating its effectiveness in improving solution quality at larger scales.
\end{itemize}

The remainder of this paper is organized as follows. Section~\ref{sec:literature-review} reviews related studies on modular bus, and pebble motion problem. Section~\ref{sec:problem_des_form} describes the MB seat movement problem and presents its CGSP-based formulation. Section~\ref{sec:milp} develops the time-expanded MILP model. Section~\ref{sec:bsdp} introduces the optimal boundary-sequence dynamic programming algorithm and proves its exactness. Section~\ref{sec:exp} reports computational results, and Section~\ref{sec:conclusion} concludes the paper and discusses future research directions.



\section{Literature Review}\label{sec:literature-review}

\subsection{Modular Bus}

MB systems have attracted increasing attention because they provide a flexible way to match transit capacity with passenger demand. The key feature of such systems is that multiple modular units can be coupled to form a larger vehicle or decoupled to serve different passenger groups. This feature allows transit operators to dynamically adjust vehicle capacity, reduce unnecessary empty-seat operation, and improve service responsiveness.

A core stream of MB research focuses on operational design, capacity allocation, and vehicle-formation scheduling. \cite{chen2019operational,chen2020operational} developed discrete and continuous models to jointly optimize dispatch headway and vehicle capacity for shuttle systems with modular vehicles. \cite{shi2020variable}, \cite{chen2021designing}, and \cite{chen2022continuous} extended modular-vehicle design to corridor-based systems, where modular units can serve multiple origin–destination patterns and perform station-wise docking operations. Further studies investigated scheduling and formation design under more general demand and operating conditions, including flexible scheduling on transit routes~\citep{ji2021scheduling}, oversaturated corridor operations~\citep{shi2021operations}, and joint scheduling and formation design under time-dependent demand~\citep{tian2023joint}. Other studies examined distributionally robust timetabling~\citep{xia2023data}, joint timetabling and formation design~\citep{cong2025joint}, timetabling with coupling and decoupling tactics~\citep{wang2025optimal}, and continuous approximation methods under demand uncertainty~\citep{yao2025collaborative}. Recent work has also considered real-time modular-vehicle operations through rolling-horizon and learning-based approaches~\citep{tian2025toward}.

Another stream has extended MB design from single route and corridor to networks and flexible service settings. At the network level, \cite{dakic2021design} and \cite{pei2021vehicle} studied dispatching and composition decisions for modular units across multiple fixed bus lines. \cite{cheng2024autonomous} investigated the design and system costs of modular transit on grid networks, accounting for pod joining and disjoining and en-route passenger transfers. \cite{khan2024seamless} developed a network-wide service that uses in-motion passenger transfers and pod exchanges between bus lines to eliminate the need for passengers to disembark when changing lines. \cite{filippi2025exploiting} optimized module allocation across bus lines through inter-line sharing and empty-module rebalancing to minimize the number of modules required to meet demand. \cite{wang2025optimizing} studied routing, scheduling, and repositioning in a hub-and-spoke system integrating first-/last-mile services with scheduled mainline operations. MBs have also been used to support flexible-route, customized, and demand-responsive services~\citep{gong2021transfer,liu2021improving,guo2023modular,guo2025operationalizing}. In addition, several studies have explored the potential of MBs for first-/last-mile integration and hybrid transit services~\citep{zhang2020modular,fu2023dial,tang2024optimisation}. These studies show that modular vehicles are being investigated not only as variable-capacity buses, but also as a broader platform for flexible, demand-responsive, and integrated public transportation services.

A closely related line of work investigates passenger transfer and service control enabled by coupling--decoupling operations. Specifically, \cite{gong2021transfer} proposed a transfer-based customized MB system with passenger-route assignment optimization. \cite{tian2022planning} optimized the locations and capacities of stations supporting module assembly and disassembly, together with vehicle formations under multi-period demand. \cite{zhang2024optimising} integrated coupling--decoupling operations with skip-stop strategies to accommodate imbalanced demand along a route. \cite{khan2023application}, \cite{khan2023bus}, \cite{lin2024bunching}, and \cite{liu2024alleviating} investigated modular splitting and merging strategies for bus-bunching mitigation. In-motion passenger transfer supports several of these operational innovations. For example, \cite{liu2024alleviating} incorporated passenger transfers between modules into catch-up and holding split strategies, while \cite{khan2025no} used this capability to develop a stop-less service in which supplementary modules serve boarding and alighting passengers. \cite{zhu2026joint} jointly optimized departure schedules, stop-wise vehicle formations, and reserved seat allocations to coordinate seating capacity with reservation demand. The present study addresses a complementary problem of physical seat reconfiguration: given seat-to-module assignments, it optimizes simultaneous seat movements along the physical track graph within and between coupled modules, subject to occupancy and movement-conflict constraints. The objective is to minimize the reconfiguration makespan and, among makespan-optimal schedules, the cumulative seat-operation cost.

Beyond service planning and operations, another group of studies focuses on the physical maneuvering and control of MBs. \cite{li2022trajectory} studied trajectory planning for MB docking and platooning operations, and \cite{li2023trajectory} further examined trajectory optimization for split operations. \cite{han2024planning} planned two-dimensional trajectories for MB en-route docking, while \cite{ma2025safety} studied safety-assurance adaptive control for modular autonomous vehicles. 
However, existing modular vehicle studies mainly focus on operational decisions at the system, route, network, vehicle-formation, or vehicle-control level. Even when passenger transfer, seat allocation, or station-wise docking is considered, the physical process of moving passengers or seats inside coupled modules is usually abstracted away. In other words, previous studies typically determine where passengers should go, which modular units should be coupled or decoupled, or how vehicle capacity should be adjusted, but they do not pay attention to the detailed movement sequence required to reconfigure seats within the coupled vehicle. This omission is important because safe and efficient in-vehicle reconfiguration is a prerequisite for realizing many operational benefits promised by MB systems, especially when in-motion transfer is involved. The present paper addresses this missing layer by studying the optimal physical movement of movable seats among modular units.

\subsection{Pebble Motion Problem}

The seat movement problem studied in this paper is rooted in the pebble motion problem (PMP), a classical graph-based rearrangement problem. In PMP, a set of pebbles occupies distinct vertices of a graph, and each move transfers one pebble to an adjacent unoccupied vertex. The objective is to transform an initial arrangement into a target arrangement while preserving the constraint that each vertex can contain at most one pebble~\citep{kornhauser1984coordinating}. This abstraction has been widely used to study multi-object rearrangement, multi-robot motion planning, and related routing problems. Later studies extended PMP by considering rotations, feasibility tests, and planning algorithms for distinguishable pebbles on graphs~\citep{yu2015pebble}. These studies provide the fundamental graph-theoretic basis for modeling seat movement as a constrained rearrangement process.

Two research streams that generalize the PMP perspective are particularly relevant to the MB seat movement problem: multi-agent path finding (MAPF) and generalized sliding-tile puzzles (GSTP). MAPF views each moving object as an agent that must travel from a start vertex to a target vertex while avoiding conflicts with other agents. Compared with classical PMP, MAPF emphasizes simultaneous time-indexed path planning and optimization objectives such as makespan and sum of costs (SoC). This makes MAPF a natural reference point for the proposed problem, because the earliest decoupling time of coupled modular units is determined by the makespan, while seat-operation effort can be represented by an SoC-type objective. For pebble motion on trees, \cite{ardizzoni2024algorithm} developed a constructive algorithm with a refined upper bound on the number of moves in the resulting feasible plans, explicitly accounting for the number of pebbles and the tree's corridor structure. A closely related branch of MAPF studies target interchangeability and target assignment. In team-based or colored MAPF, agents are partitioned into teams or colors, and agents within the same team may be assigned to interchangeable targets. \citet{ma2016optimal} studied the combined target-assignment and path-finding problem (TAPF), which generalizes both anonymous and non-anonymous MAPF by assigning agents in each team to team-specific targets while planning collision-free paths. \citet{bartak2021colored} and \citet{bartak2021ozomorph} investigated colored MAPF and SAT-based solving approaches, where agents and targets are specified by colors rather than fixed identities. More recent studies further improved anonymous MAPF algorithms and target-assignment MAPF solvers~\citep{ali2024improved,tang2023solving}. These studies are relevant to MB seat movement because seats assigned to the same destination module are indistinguishable, and the desired final configuration is color-based rather than identity-based. 

However, MAPF-based formulations are not fully consistent with the physical movement mechanism of MB seats. Standard MAPF usually assumes point-like agents and represents collisions through vertex conflicts, edge conflicts, and sometimes following or swapping conflicts. Such rules are appropriate for mobile robots or abstract agents, but they do not explicitly model the physical volume of sliding seats on tracks. In the MB seat movement process, a seat cannot simply move into an occupied vertex unless the seat currently occupying that vertex simultaneously moves forward in the same direction. This requirement follows from the adopted sliding-tile abstraction, in which each seat is represented by a non-rotating, cell-sized square footprint and all moving seats translate uniformly by one grid spacing over the same time interval. If the leading seat moves perpendicular to the trailing seat's motion, their footprints overlap during the transition, even though their destination cells are distinct. Moving both seats by one cell in the same direction preserves their relative displacement throughout the step and avoids this overlap. The corner-following constraint therefore prevents an intermediate collision that is not captured by vertex and edge conflicts alone~\citep{gozon2024computing}. If this rule is ignored, an MAPF formulation may admit movement patterns that are collision-free for point agents but infeasible for physical seats. Therefore, although colored MAPF, team MAPF, and TAPF provide useful concepts for target interchangeability and collision-free coordination, they are not the most faithful model for MB seat movement. By contrast, the GSTP provides a more suitable modeling framework for this problem. GSTP extends the classical sliding-tile puzzle by allowing multiple escorts and synchronous tile movements, while enforcing geometric collision constraints among neighboring tiles~\citep{gozon2024computing}. These features are directly aligned with physical seat movement: empty aisle cells function as escorts or swap spaces, and multiple seats may slide simultaneously along the track graph. The CGSP further allows tiles to have different colors, so tiles of the same color are interchangeable~\citep{gozon2024optimally}. This colored structure matches the MB setting, where seats assigned to the same modular unit are equivalent in the target state. 

\begin{figure}[t]
    \centering
    \begin{subfigure}[t]{0.47\textwidth}
        \centering
        \includegraphics[width=\linewidth]{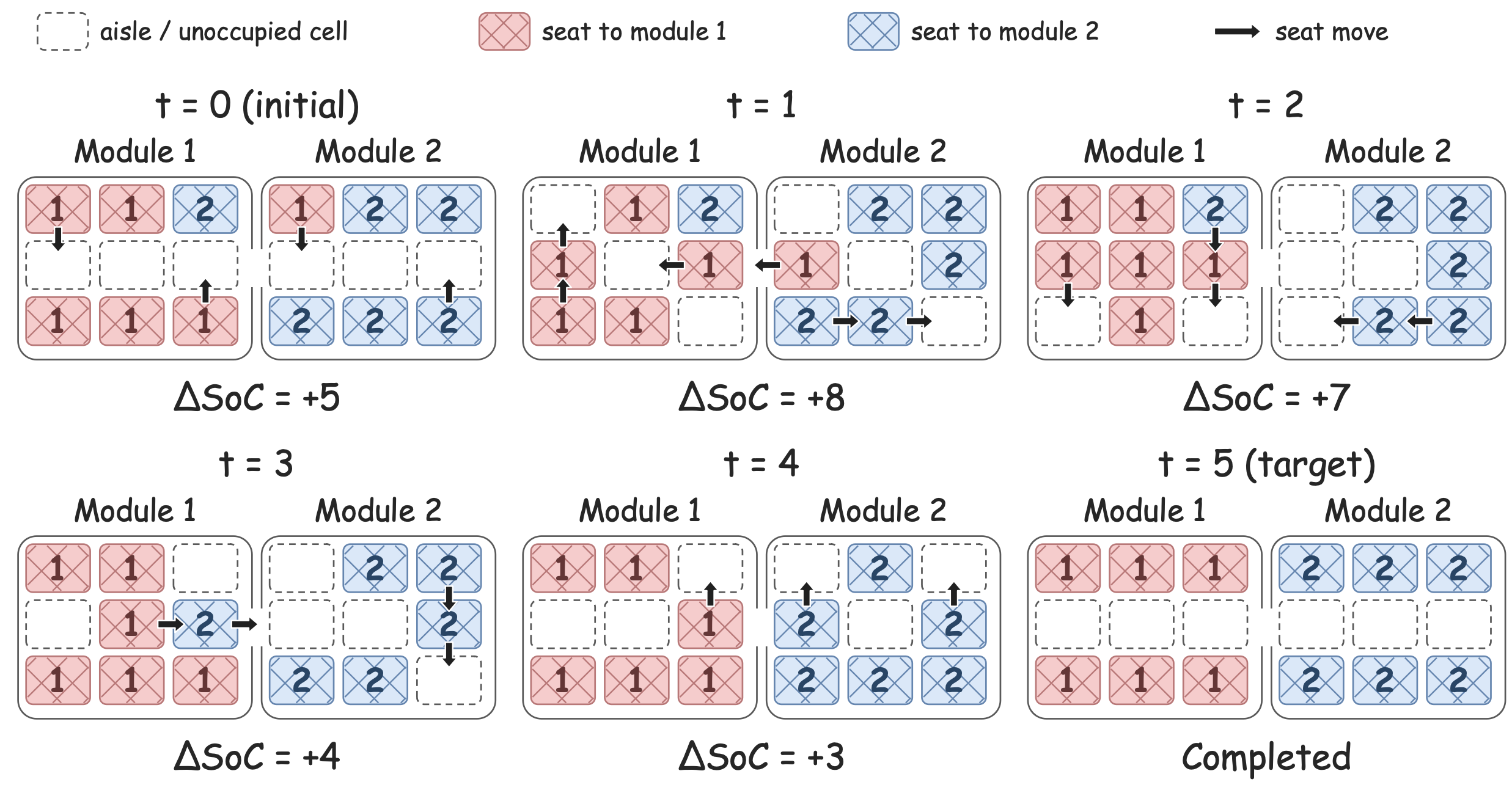}
        \caption{A makespan-optimal schedule with $\mathrm{SoC}=27$.}
        \label{fig:sameK-soc-exp1}
    \end{subfigure}
    \hfill
    \begin{subfigure}[t]{0.49\textwidth}
        \centering
        \includegraphics[width=\linewidth]{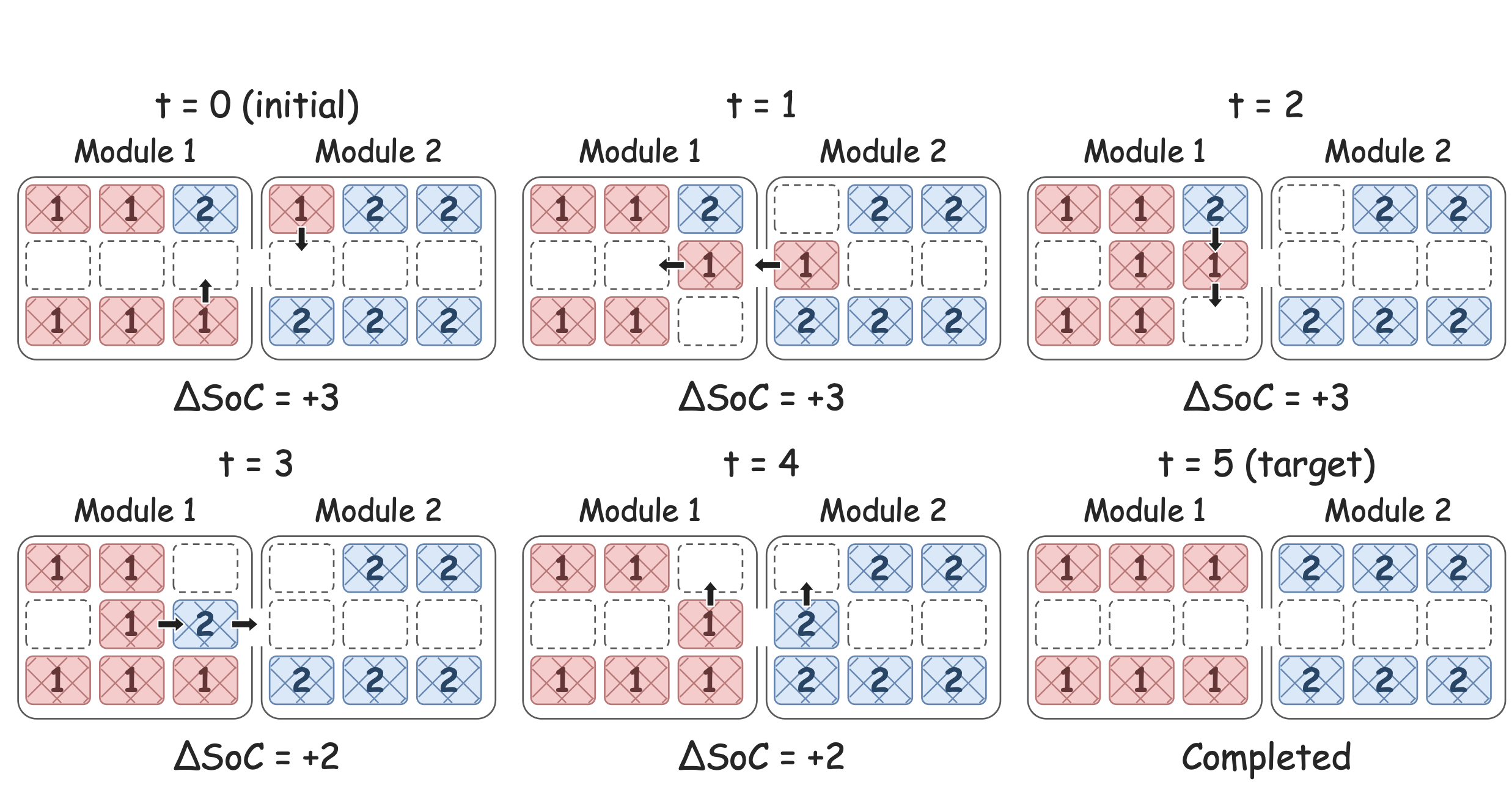}
        \caption{A lexicographically optimal schedule with $\mathrm{SoC}=13$.}
        \label{fig:sameK-soc-exp2}
    \end{subfigure}
    \caption{Comparison of seat movement schedules with the same minimum makespan $T^*=5$.}
    \label{fig:schedule-comparison}
\end{figure}

Existing GSTP and CGSP studies, however, have primarily focused on makespan-related questions. \citet{gozon2024computing} studied makespan-optimal solutions for GSTP and established complexity results and approximation bounds. \citet{gozon2024optimally} extended the framework to CGSP and characterized the lower and upper bounds of its complexity and makespan under different distinguishability settings. These works provide the closest theoretical foundation for MB seat movement, but they do not address fixed-makespan SoC minimization or lexicographic makespan--SoC optimization. 
This distinction is important for MB operations. Makespan determines
the earliest decoupling time permitted by the seat-reconfiguration
constraints. SoC measures the cumulative cost of physical seat
movements and waiting at non-target positions, while waiting at a
correct target position incurs no cost. It therefore serves as an
operational proxy for passenger disturbance and seat-operation effort.
Minimizing makespan alone does not distinguish between schedules
that achieve the same decoupling time but incur different amounts
of movement and non-target waiting.

Figure~\ref{fig:schedule-comparison} illustrates this distinction
using a two-module instance. Both schedules start from the same
initial configuration, reach the same target configuration, and
attain the minimum makespan $T^*=5$. Seat colors and numbers indicate
destination modules, and arrows show physical movements in the next
transition. The label $\Delta\mathrm{SoC}$ at time $t$ denotes the
cost incurred during the transition from $t$ to $t+1$, including
physical movements and non-target waiting. The legend in
Figure~\ref{fig:sameK-soc-exp1} applies to both panels.
The makespan-optimal schedule in Figure~\ref{fig:sameK-soc-exp1}
requires 22 physical seat movements and 5 seat-time units of
non-target waiting, giving a total SoC of 27. In contrast, the
lexicographically optimal schedule in
Figure~\ref{fig:sameK-soc-exp2} requires only 10 movements and
3 seat-time units of non-target waiting, giving a total SoC of 13.
Thus, optimizing the secondary objective reduces SoC by 51.85\%
in this example without delaying decoupling. This comparison
demonstrates that makespan optimality alone does not guarantee
a low-cost seat-reconfiguration schedule and motivates SoC
minimization among makespan-optimal schedules. This paper therefore positions the MB seat movement problem as an application-driven CGSP with an additional SoC optimization layer. Compared with standard MAPF, the CGSP model better captures the physical volume and sliding interactions of seats through the corner-following rule. Compared with existing CGSP studies, the proposed model introduces a lexicographic objective that first minimizes makespan and then minimizes SoC among makespan-optimal schedules. The proposed BSDP algorithm further exploits the chain topology of coupled modular units, thereby providing an exact solution approach for a problem that lies at the intersection of MB operations, pebble-motion-based rearrangement, and CGSP optimization.


\section{Colored Generalized Sliding-tile Puzzle (CGSP)}\label{sec:cgsp}
Before formulating the MB seat movement problem, we
introduce the colored generalized sliding-tile puzzle or CGSP in
operational terms. The purpose of this section is to explain the
underlying movement model to readers who may not be familiar with
sliding-tile puzzles or multi-agent motion planning.

The classical sliding-tile puzzle consists of a set of tiles and one
empty position on a rectangular board. A tile can move only by sliding into the adjacent empty position. The generalized sliding-tile puzzle or GSTP extends this setting in two important ways: it permits multiple empty positions, referred to as \emph{escorts}, and permits multiple tiles to move synchronously within the same time step~\citep{gozon2024computing}. The CGSP further partitions the tiles into colors. Tiles of the same color are interchangeable and may occupy any target position assigned to that color~\citep{gozon2024optimally}.

\paragraph{Board, tiles, and configurations.} A CGSP is defined on a board graph $\mathcal G=(\mathcal V,\mathcal E)$. Each vertex $v\in\mathcal V$ represents one discrete physical position, and each edge $\{u,v\}\in\mathcal E$ indicates that a tile can move directly between the two corresponding positions. A vertex may be occupied by
at most one tile. Vertices that are not occupied are called \emph{escorts} or \emph{empty spaces}; they provide the free space needed for the tiles to move.

Let $\mathcal C$ denote the finite set of tile colors. A configuration is a map
\begin{equation}
X:\mathcal V\longrightarrow\mathcal C\cup\{0\},
\end{equation}
where $X[v]=c$ means that vertex $v$ is occupied by a tile of color $c\in\mathcal C$, and $X[v]=0$ means that the vertex is empty. The number of tiles of each color remains unchanged throughout the movement process.

Each color $c\in\mathcal C$ is associated with a set $\mathcal D_c\subseteq\mathcal V$ of admissible target vertices. The target sets are pairwise disjoint, and the cardinality of $\mathcal D_c$ equals the number of color-$c$ tiles. A configuration is a target configuration if every vertex in $\mathcal D_c$ is occupied by a color-$c$ tile for every $c\in\mathcal C$. Because tiles of the same color are indistinguishable, it is unnecessary to specify which individual color-$c$ tile occupies which vertex in $\mathcal D_c$.

\begin{figure}[t]
    \centering
    \setlength{\cgspImageWidth}{0.29\textwidth}
    \captionsetup[subfigure]{font=small,labelfont=bf,skip=3pt}

    \begin{subfigure}[t]{0.25\textwidth}
        \centering
        \includegraphics[width=\cgspImageWidth]{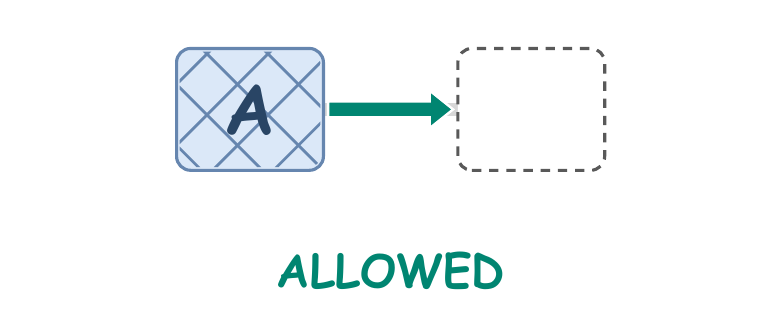}
        \caption{Move into an empty cell.}
        \label{fig:cgsp-empty-cell}
    \end{subfigure}
    \hspace{0.06\textwidth}
    \begin{subfigure}[t]{0.25\textwidth}
        \centering
        \includegraphics[width=\cgspImageWidth]{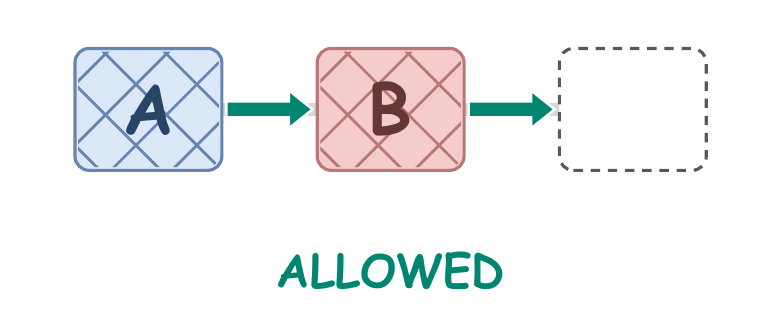}
        \caption{Straight simultaneous following.}
        \label{fig:cgsp-straight-following}
    \end{subfigure}

    \begin{subfigure}[t]{0.2\textwidth}
        \centering
        \includegraphics[width=\cgspImageWidth]{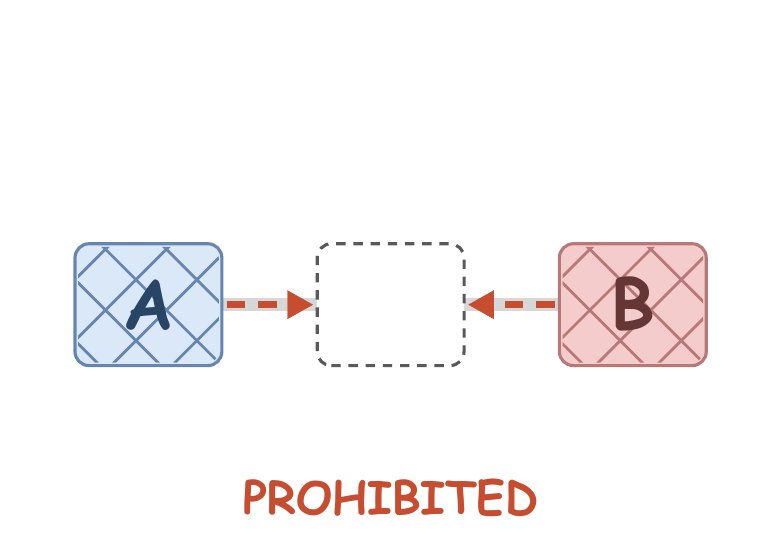}
        \caption{Shared destination.}
        \label{fig:cgsp-shared-destination}
    \end{subfigure}
    \hfill
    \begin{subfigure}[t]{0.2\textwidth}
        \centering
        \includegraphics[width=\cgspImageWidth]{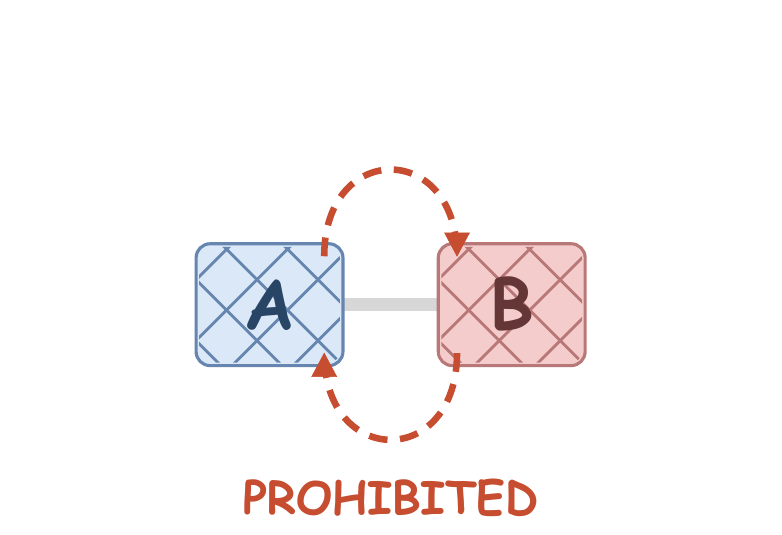}
        \caption{Opposite use of one edge.}
        \label{fig:cgsp-opposite-edge}
    \end{subfigure}
    \hfill
    \begin{subfigure}[t]{0.2\textwidth}
        \centering
        \includegraphics[width=\cgspImageWidth]{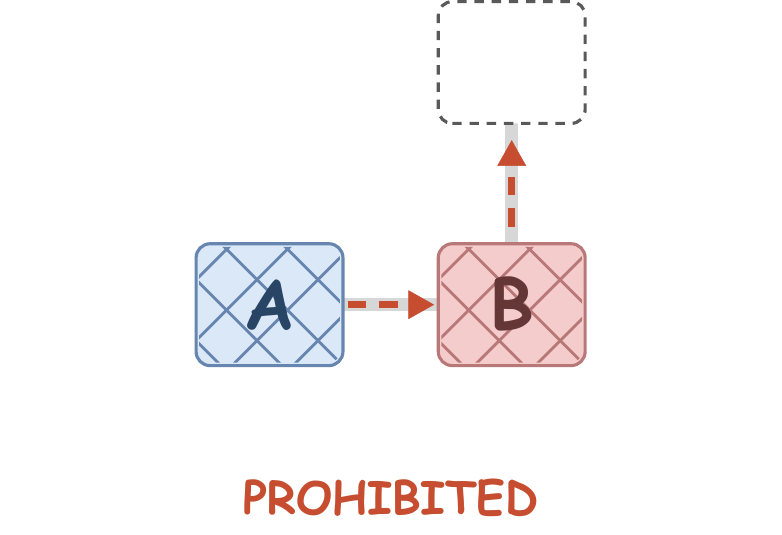}
        \caption{Turning while following.}
        \label{fig:cgsp-turning-following}
    \end{subfigure}

    \caption{Allowed and prohibited simultaneous CGSP movements.
    Solid teal arrows indicate feasible movements; dashed vermillion
    arrows indicate prohibited combinations. All arrows within a panel
    occur simultaneously in one time step.}
    \label{fig:cgsp_moves}
\end{figure}

\paragraph{A simultaneous time step.}
Figure~\ref{fig:cgsp_moves} provides a visual overview of allowed and prohibited tile movements. Tiles are labeled A and B, and empty vertices are represented by unfilled cells with dashed outlines. Arrows indicate proposed movements during a single time step. Solid teal arrows represent feasible movements, whereas dashed vermillion arrows represent prohibited combinations.

Time is discretized into equal-length steps. During one time step, each tile either remains at its current vertex or moves to an adjacent vertex. All selected movements are evaluated with respect to the configuration at the beginning of the time step and are then applied simultaneously. Accordingly, the arrows within each panel of Figure~\ref{fig:cgsp_moves} describe actions proposed for the same time interval, rather than a sequence of successive movements.

Under this simultaneous interpretation, a tile may move into a vertex that is empty at the beginning of the time step. It may also move into an initially occupied vertex, provided that the occupying tile simultaneously moves one position forward in the same direction. This permits a straight row of tiles to advance together, similar to a coordinated platoon advancing along a constrained guideway. The selected actions must satisfy the following rules.

\begin{enumerate}
    \item \textbf{Adjacency and single-action rule.}
    A moving tile may traverse only one edge of the board graph during
    a time step. Each tile selects exactly one action: either waiting
    at its current vertex or moving to one adjacent vertex.

    \item \textbf{Vertex-capacity rule.}
    No two tiles may have the same destination vertex during the same
    time step. Equivalently, every vertex has unit occupancy capacity
    at the end of the step. This prohibits two tiles from moving
    simultaneously into the same empty position.

    \item \textbf{Edge-capacity rule.}
    Two tiles may not traverse the same undirected edge in opposite
    directions during the same time step. In transportation terms, a
    single track segment cannot simultaneously support two opposing
    movements.

    \item \textbf{Straight-following rule.}
    Suppose that a trailing tile moves from $u$ to $v$, where $v$ is
    occupied at the beginning of the time step. The tile initially at
    $v$ must simultaneously move to the next adjacent vertex directly
    ahead of $v$ in the direction from $u$ to $v$. The three positions
    must therefore be aligned in the same direction. The leading tile
    may not turn left or right while the trailing tile enters its
    current position.

    The rule can be applied repeatedly to a line of tiles. Hence,
    several consecutive tiles may move forward simultaneously,
    provided that they all move in the same direction and the leading
    end of the moving chain has sufficient empty space. If the leading tile turns while being followed, the swept physical spaces of the two tiles may overlap even though their final vertices are different. Such a prohibited movement is called a
    \emph{corner-following conflict}.
\end{enumerate}

Figure~\ref{fig:cgsp_moves}(a) shows a tile moving into an initially empty vertex, while panel (b) shows straight simultaneous following. Panels (c), (d), and (e) illustrate violations of the vertex-capacity, edge-capacity, and straight-following rules, respectively. In particular, panel (e) shows a corner-following conflict: the leading tile turns while the trailing tile enters its original vertex, which is prohibited even though their destination vertices
are distinct.

The MB seat movement problem introduced next is an application-specific CGSP. Its physical track graph determines adjacency, the destination modular unit determines a seat's color, and the middle-aisle vertices provide the empty escorts.

\section{Problem Description and Formulation}\label{sec:problem_des_form}
\subsection{Problem Description}

\begin{figure}[h!]
    \centering
    \includegraphics[width=1\linewidth]{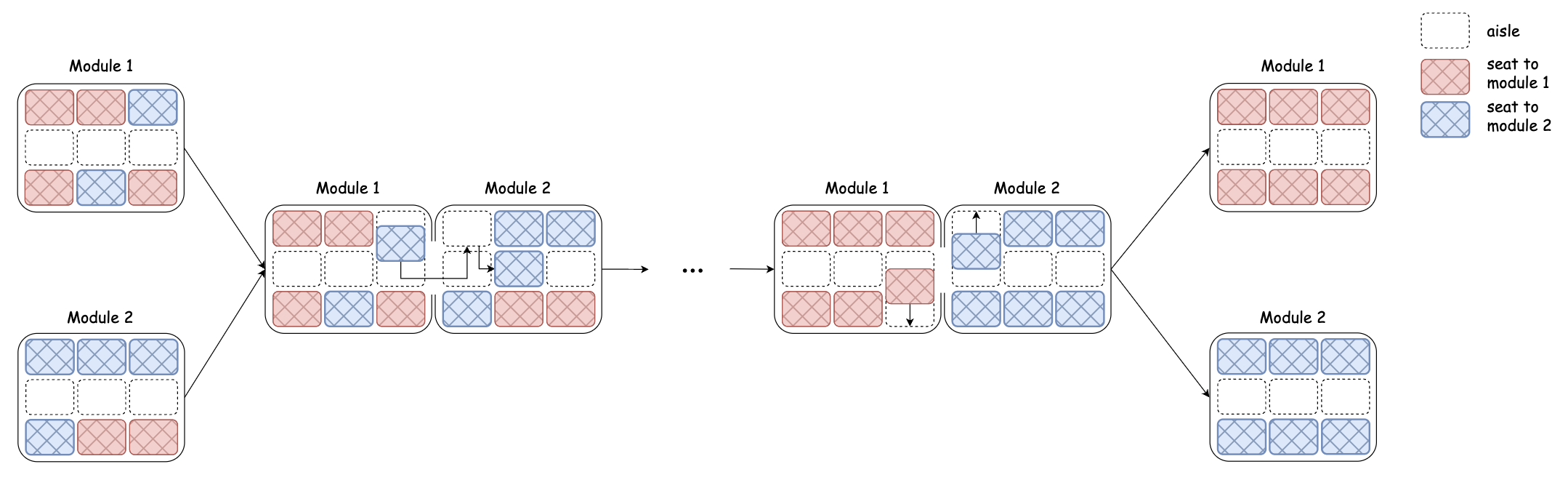}
    \caption{Seat movement process in an MB with two modular units and $m=3$}
    \label{fig:SeatsMovement}
\end{figure}

Figure~\ref{fig:SeatsMovement} illustrates the seat movement process for two coupled modular units in the representative case $m=3$. More generally, each modular unit has $m\ge1$ columns, with $2m$ movable seats located in the top and bottom rows and $m$ track positions in the middle row forming the central aisle. Seats are colored according to their assigned modular units, with the assignments respecting passengers' destinations or routes. The seat--unit assignment is provided as an input to the seat movement problem. When modular units are coupled, seats are treated as sliders that move along the physical tracks. By the end of the process, every seat must be transferred to its assigned modular unit, so that each unit contains only seats of its designated color. The primary objective is to minimize the overall movement duration. The formulation developed below applies to any $m\in\mathbb Z_{\ge1}$ and any number $N\ge2$ of coupled modular units; Figure~\ref{fig:SeatsMovement} is only an illustration of the case $m=3$.

\subsection{Problem Formulation}
The MB seat movement process can be modeled as CGSP~\citep{gozon2024optimally} on the physical track graph of the coupled modular bus units. In the CGSP terminology, feasible seat positions are vertices of the puzzle board, movable seats are tiles, empty track positions are escorts or swap spaces, and the destination modular unit assigned to a seat determines its color. Therefore, seats assigned to the same modular unit are treated as indistinguishable tiles of the same color. Compared with the standard CGSP, the present problem has an MB-specific board graph and an MB-specific target pattern: each color must finally occupy the non-aisle seat positions of its assigned modular unit.

Consider $N\ge2$ coupled modular units, each with $m\ge1$ columns, where $m$ is a vehicle-design parameter. Each modular unit therefore has a $3\times m$ track layout, with $2m$ seat positions in the top and bottom rows and $m$ central-aisle positions in the middle row. Let $\mathcal G_{N,m}=(\mathcal V_{N,m},\mathcal E_{N,m})$ denote the physical track graph induced by the coupled MB. Its vertex set is
\begin{equation}
\mathcal V_{N,m}=\left\{(r,j):r\in\{1,2,3\},\ j\in\{1,2,\ldots,mN\}\right\},
\end{equation}
where $r$ and $j$ are the row and column indices, respectively. The $i$-th modular unit $\mathcal M_i$ occupies
\begin{equation}
\mathcal W_i =\left\{(r,j)\in\mathcal V_{N,m}:r\in\{1,2,3\},\ (i-1)m+1\le j\le im\right\},\ i=1,2,\cdots,N.
\end{equation}
Within each modular unit, all horizontally and vertically adjacent vertices are connected. Two consecutive modular units are connected only through the central aisle. Hence,
\begin{equation}
\mathcal E_{N,m} =\bigcup_{i=1}^{N} \left\{\{u,v\}:u,v\in\mathcal W_i,\ \lVert u-v\rVert_1=1\right\} \cup \left\{\{(2,mi),(2,mi+1)\}:i=1,2,\cdots,N-1\right\}.
\end{equation}
For each $i=1,2,\ldots,N-1$, the edge $\{(2,mi),(2,mi+1)\}$ is the unique physical track connection between $\mathcal M_i$ and $\mathcal M_{i+1}$. Because an edge is an unordered vertex pair, $\mathcal G_{N,m}$ is undirected and this connection can be traversed in either direction. In particular, for $i=2,3,\ldots,N$, movement from $\mathcal M_i$ to $\mathcal M_{i-1}$ uses the same boundary edge $\{(2,m(i-1)),(2,m(i-1)+1)\}$.

\begin{figure}[h!]
	\centering
	\begin{minipage}[c]{0.4\textwidth}
		\centering
        
        \includegraphics[width=\textwidth]{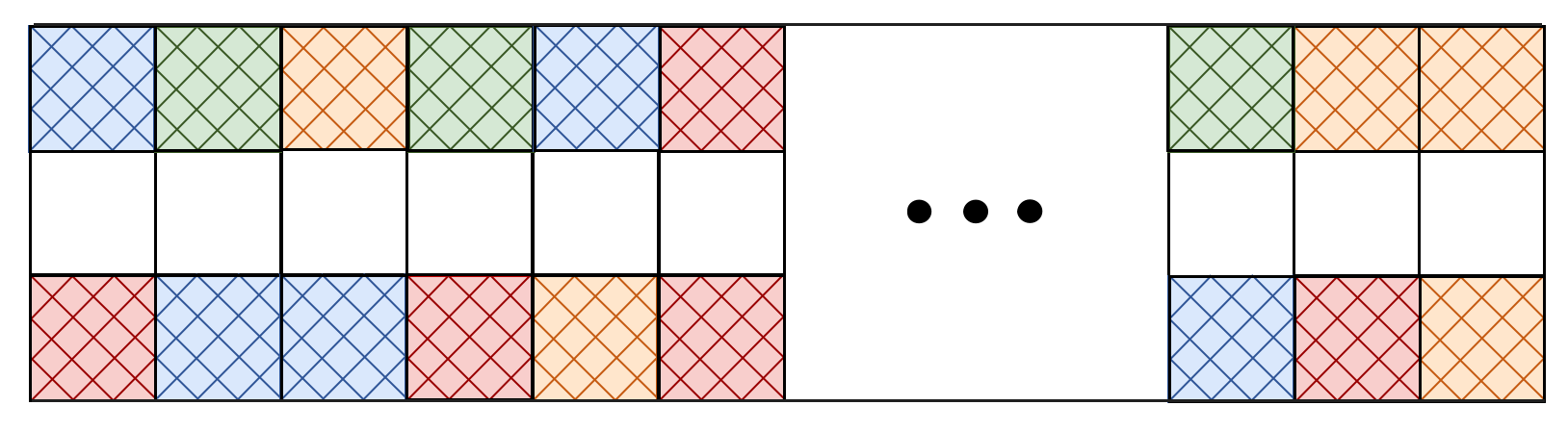}
		\subcaption{Initial state}
		\label{fig:org}
	\end{minipage}
	\begin{minipage}[c]{0.4\textwidth}
		\centering
		\includegraphics[width=\textwidth]{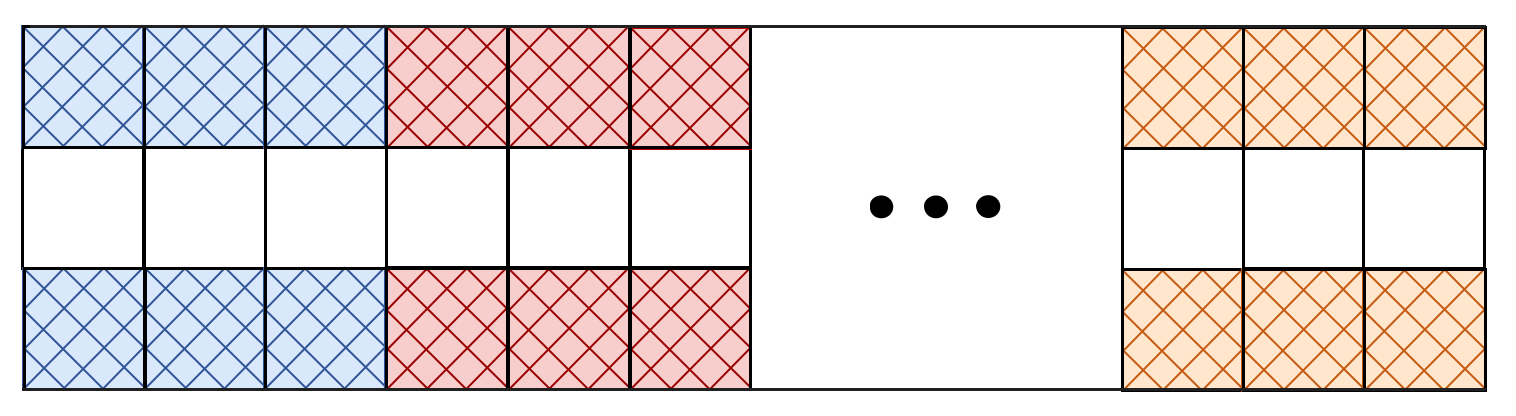}
		\subcaption{Target state}
		\label{fig:target}
	\end{minipage} 
    \caption{CGSP for MB seat movement}
    \label{fig:orgAndTarget}
\end{figure}

Let $\mathcal C=\{1,2,\ldots,N\}$ be the color set of seats. A seat of color $c\in\mathcal C$ is assigned to modular unit $\mathcal M_c$. Each modular unit contains $2m$ movable seats; hence, each color occurs exactly $2m$ times and the coupled system contains $2mN$ seats. The remaining $mN$ vertices are empty track positions and serve as escorts in the CGSP representation. Figure~\ref{fig:orgAndTarget} illustrates the special case $m=3$, whereas the formulation applies to every $m\in\mathbb Z_{\ge1}$.

To facilitate the formulation, we model each state of the CGSP as a state matrix, which is defined below:

\begin{definition}[State matrix]
A state matrix $X$ is a $3\times mN$ matrix with entries in $\mathcal C\cup\{0\}$, where $X[r,j]=c$ indicates that vertex $(r,j)$ is occupied by a color-$c$ seat and $X[r,j]=0$ indicates an empty track position. The feasible state space is
\begin{equation}
\mathcal X_{N,m}=\left\{X\in(\mathcal C\cup\{0\})^{3\times mN}: \sum_{v\in\mathcal V_{N,m}}\mathbf 1\{X[v]=c\}=2m,\ \forall c\in\mathcal C,\ \sum_{v\in\mathcal V_{N,m}}\mathbf 1\{X[v]=0\}=mN \right\}.
\end{equation}

When $N$ and $m$ are fixed, we write $\mathcal X=\mathcal X_{N,m}$ for brevity.
\end{definition}

Seats of the same color are indistinguishable, so the state matrix records only the color occupying each vertex rather than the identity of an individual seat. The state-matrix definition assumes that the seat--unit assignment is predetermined and provided as an input to the optimal seat movement problem. Joint optimization of seat--unit assignment and seat movement is beyond the scope of this study.

For the illustrative case $m=3$, the state of the CGSP in Figure~\ref{fig:org} can be represented by the following state matrix:
\begin{equation*}\label{mat:original_state_matrix}
    \begin{pmatrix}
    1 & 3 & N & 3 & 1 & 2 & \ & 3 & N & N \\ 
    0 & 0 & 0 & 0 & 0 & 0 & \cdots & 0 & 0 & 0 \\ 
    2 & 1 & 1 & 2 & N & 2 & \ & 1 & 2 & N  
    \end{pmatrix}
\end{equation*}


\begin{definition}[Target state matrix]\label{def:finalMatrix}
The target state matrix $X^{\mathcal G}\in\mathcal X_{N,m}$ is defined by
\begin{equation}
X^{\mathcal G}[r,j]=
\begin{cases}
i, & r\in\{1,3\},\ (i-1)m+1\le j\le im,\ i=1,2,\cdots,N,\\
0, & r=2,\ j=1,2,\cdots,mN.
\end{cases}
\end{equation}
\end{definition}

Thus, all non-aisle positions of $\mathcal M_i$ are occupied by color-$i$ seats and all central-aisle positions are empty. Because each color occurs exactly $2m$ times, the target state matrix is unique for every $m\ge1$. For each color $c\in\mathcal C$, define its target-cell set as $\mathcal G_c=\{v\in\mathcal V_{N,m}:X^{\mathcal G}[v]=c\}$. Waiting at a vertex in $\mathcal G_c$ means that a color-$c$ seat is already at a correct target position.

We next specialize the simultaneous CGSP movement rules in Section~\ref{sec:cgsp} to the MB physical track graph. Let
\begin{equation}
\mathcal A_{N,m}^{\mathrm{move}}
=\left\{(u,v)\in\mathcal V_{N,m}\times\mathcal V_{N,m}:\{u,v\}\in\mathcal E_{N,m}\right\}
\end{equation}
be the set of directed physical movement arcs, and let
\begin{equation}
\mathcal A_{N,m}^{\mathrm{wait}}=\{(v,v):v\in\mathcal V_{N,m}\},
\ \mathcal A_{N,m}=\mathcal A_{N,m}^{\mathrm{move}}\cup\mathcal A_{N,m}^{\mathrm{wait}}
\end{equation}
be the waiting-arc set and the complete action-arc set, respectively. Each undirected physical edge induces two directed movement arcs, so a boundary edge can be traversed in either direction.

An \emph{action arc} is a pair $(u,v)\in\mathcal A_{N,m}$. A \emph{color-labeled seat action} is a triple $(c,u,v)$ indicating that a color-$c$ seat at vertex $u$ takes action arc $(u,v)$. If $u\neq v$, the seat moves physically from $u$ to $v$; if $u=v$, it waits at $u$.

For $(u,v)\in\mathcal A_{N,m}^{\mathrm{move}}$, where $u=(r_u,j_u)$ and $v=(r_v,j_v)$, define
\begin{equation}
F(u,v)=2v-u=(2r_v-r_u,\,2j_v-j_u).
\end{equation}
The vertex $F(u,v)$ is immediately ahead of $v$ in the movement direction from $u$ to $v$.

\begin{definition}[One-step transition]\label{def:one-step}
For $X,X'\in\mathcal X$, $X'$ is reachable from $X$ in one time step if there exists a simultaneous action set $\mathcal Q\subseteq\mathcal C\times\mathcal A_{N,m}$ satisfying all of the following conditions:
\begin{enumerate}
    \item \textbf{Tail consistency and completeness.} For every $(c,u,v)\in\mathcal Q$, $X[u]=c$. Moreover, every occupied vertex $u$ is the tail of exactly one action:
    \begin{equation}
    \left|\{(c,u,v)\in\mathcal Q:v\in\mathcal V_{N,m}\}\right|=1,
    \ \forall u\in\mathcal V_{N,m}\text{ with }X[u]\neq0.
    \end{equation}
    \item \textbf{No vertex conflict.} No two actions have the same head vertex:
    \begin{equation}
    \left|\{(c,u,v)\in\mathcal Q:c\in\mathcal C,\ u\in\mathcal V_{N,m}\}\right|\le1,\ \forall v\in\mathcal V_{N,m}.
    \end{equation}
    \item \textbf{No edge conflict.} If $(c,u,v)\in\mathcal Q$ with $u\neq v$, then $(c',v,u)\notin\mathcal Q$ for every $c'\in\mathcal C$.
    \item \textbf{Straight simultaneous following.} If $(c,u,v)\in\mathcal Q$, $u\neq v$, and $X[v]=c'\neq0$, then the leading seat must simultaneously take action $(c',v,F(u,v))\in\mathcal Q$.
    \item \textbf{Simultaneous state update.} All actions in $\mathcal Q$ are applied simultaneously, and
    \begin{equation}
    X'[v]=c\ \Longleftrightarrow\ (c,u,v)\in\mathcal Q\text{ for some }u
    \end{equation}
\end{enumerate}
\end{definition}

The fourth condition already implies that $F(u,v)$ is a valid adjacent vertex because every triple in $\mathcal Q$ must use an arc in $\mathcal A_{N,m}$. Therefore, the additional conditions $F(u,v)\in\mathcal V_{N,m}$ and $\{v,F(u,v)\}\in\mathcal E_{N,m}$ would be redundant. If $v$ is empty in state $X$, a seat may move into it without requiring a forward action. If $v$ is occupied, however, the leading action must occur in the same time step and in the same direction. This rule is the key difference between it and Colored Multi-Agent Pathfinding (also known as combined target-assignment and pathfinding)~\citep{bartak2021ozomorph, ma2016optimal, wang2025paths}.

Given an initial state matrix $X^0\in\mathcal{X}$, we evaluate a seat movement schedule using two criteria. The primary criterion is the makespan, i.e., the number of time steps required to reach the target state matrix. Operationally, the makespan determines the earliest time at which the modular bus units can be decoupled. The secondary criterion is the sum of costs (SoC), which measures the cumulative seat-operation cost before decoupling. The SoC is minimized only among schedules that attain the minimum makespan, so improving passenger experience does not delay the decoupling operation.

For a color-$c$ seat taking action $(u,v)\in\mathcal A_{N,m}$, define
\begin{equation}
\omega_{uv}^c=
\begin{cases}
0, & u=v,\ v\in\mathcal G_c,\\
1, & \text{otherwise}.
\end{cases}
\end{equation}
Thus, a physical movement has unit cost, waiting at a non-target vertex has unit cost, and waiting at a correct target vertex has zero cost.

For a nonnegative integer $T$, let $\mathcal P_T(X^0,X^{\mathcal G})$ be the set of schedules that transform $X^0$ into $X^{\mathcal G}$ in exactly $T$ time steps. A schedule is
\begin{equation}
\Pi=(X^0,\mathcal Q^0,X^1,\mathcal Q^1,\cdots,\mathcal Q^{T-1},X^T)
\in\mathcal P_T(X^0,X^{\mathcal G}),
\end{equation}
where $X^T=X^{\mathcal G}$ and each $\mathcal Q^t$ is a simultaneous action set that induces the one-step transition from $X^t$ to $X^{t+1}$.

Let $y_{uvt}^c(\Pi)=1$ if $(c,u,v)\in\mathcal Q^t$ and $y_{uvt}^c(\Pi)=0$ otherwise. The SoC is
\begin{equation}
\operatorname{SoC}(\Pi)=\sum_{t=0}^{T-1}\sum_{c\in\mathcal C}\sum_{(u,v)\in\mathcal A_{N,m}}\omega_{uv}^c y_{uvt}^c(\Pi).
\end{equation}

The minimum decoupling time is
\begin{equation}
T^*=\min\left\{T\in\mathbb Z_{\ge0}:\mathcal P_T(X^0,X^{\mathcal G})\neq\emptyset\right\}.
\end{equation}
At time $T^*$, all seats are in their assigned modules and all central-aisle vertices are empty; therefore, no further inter-module seat movement is required and the modular bus units can decouple.

For every feasible fixed makespan $T$, define
\begin{equation}
\operatorname{SoC}^*(T)=\min_{\Pi\in\mathcal P_T(X^0,X^{\mathcal G})}
\operatorname{SoC}(\Pi).
\end{equation}
The optimization criterion in this study is lexicographic:
\begin{equation}
\Pi^*\in \arg\min_{\Pi\in\bigcup_{T\in\mathbb Z_{\ge0}}\mathcal P_T(X^0,X^{\mathcal G})}^{\mathrm{lex}} \bigl(T(\Pi),\operatorname{SoC}(\Pi)\bigr).
\end{equation}
Equivalently, the primary objective determines $T^*$ and the secondary objective is
\begin{equation}
\operatorname{SoC}^*=\operatorname{SoC}^*(T^*)=\min_{\Pi\in\mathcal P_{T^*}(X^0,X^{\mathcal G})}\sum_{t=0}^{T^*-1}\sum_{c\in\mathcal C}
\sum_{(u,v)\in\mathcal A_{N,m}}\omega_{uv}^c y_{uvt}^c(\Pi).
\end{equation}
Thus, improving the secondary passenger-comfort criterion never delays the operationally critical decoupling time.

\paragraph{Why the two criteria cannot generally be minimized simultaneously.}
Consider the following three-seat CGSP toy instance on a $2\times3$ orthogonal grid, where $0$ denotes an empty vertex. This auxiliary example is not an MB capacity instance; its sole purpose is to demonstrate that the two objectives can conflict.
\begin{equation}
X^0=
\begin{bmatrix}
B&0&A\\
C&0&0
\end{bmatrix},
\ 
X^{\mathcal G}=
\begin{bmatrix}
A&B&0\\
0&C&0
\end{bmatrix}.
\end{equation}
Under the simultaneous-movement rules in Section~\ref{sec:cgsp}, exhaustive enumeration of the $120$ possible labeled-seat states shows that the minimum makespan is $T^*=4$. One makespan-optimal schedule is
\begin{equation}
\begin{bmatrix}B&0&A\\ C&0&0\end{bmatrix}
\rightarrow
\begin{bmatrix}0&B&0\\ C&0&A\end{bmatrix}
\rightarrow
\begin{bmatrix}0&0&B\\ C&A&0\end{bmatrix}
\rightarrow
\begin{bmatrix}0&A&B\\ C&0&0\end{bmatrix}
\rightarrow
\begin{bmatrix}A&B&0\\ 0&C&0\end{bmatrix}.
\end{equation}
All three seats incur unit cost in each of the four time steps, so this schedule has SoC $12$. Exhaustive enumeration further shows that every feasible length-$4$ schedule has SoC at least $12$; hence, $\operatorname{SoC}^*(4)=12$.

If one additional time step is allowed, the following schedule is feasible:
\begin{equation}
\begin{bmatrix}B&0&A\\ C&0&0\end{bmatrix}
\rightarrow
\begin{bmatrix}B&0&A\\ 0&C&0\end{bmatrix}
\rightarrow
\begin{bmatrix}0&A&0\\ B&C&0\end{bmatrix}
\rightarrow
\begin{bmatrix}A&0&0\\ 0&B&C\end{bmatrix}
\rightarrow
\begin{bmatrix}A&B&0\\ 0&0&C\end{bmatrix}
\rightarrow
\begin{bmatrix}A&B&0\\ 0&C&0\end{bmatrix}.
\end{equation}
Its per-step costs are $3,2,3,2,$ and $1$, respectively, giving a total SoC of $11$. In each displayed arrow, all seat actions---including any waiting actions---are executed simultaneously. The third transition uses straight simultaneous following: seat $B$ enters the current vertex of seat $C$ while $C$ moves one vertex forward in the same direction. A shortest-path calculation on the full $120$-state transition graph shows that no schedule, of any length, has SoC below $11$. Hence, the globally minimum SoC is $11$, but it cannot be achieved under the globally minimum makespan $T^*=4$. This example motivates the lexicographic objective used in this study: first minimize the operationally critical decoupling time and then minimize SoC among schedules attaining that time.

\section{Modeling Framework}\label{sec:milp}

This section develops an exact time-expanded mixed-integer programming model for the fixed-makespan MB seat movement problem defined in Section~\ref{sec:problem_des_form}. The problem data inherited from Section~\ref{sec:problem_des_form} are the physical track graph $\mathcal G_{N,m}=(\mathcal V_{N,m},\mathcal E_{N,m})$, the color set $\mathcal C$, the initial and target state matrices $X^0$ and $X^{\mathcal G}$, the action-arc set $\mathcal A_{N,m}$, the target-cell sets $\mathcal G_c$, and the seat-operation costs $\omega_{uv}^c$.

The following notation is introduced specifically for the time-expanded MILP. For each $v\in\mathcal V_{N,m}$, define the outgoing and incoming action-arc sets as
\begin{equation}
\delta_{N,m}^{+}(v)=\{(v,w)\in\mathcal A_{N,m}\},\ \delta_{N,m}^{-}(v)=\{(u,v)\in\mathcal A_{N,m}\}.
\end{equation}

For a fixed makespan $T$, define
\begin{equation}
\mathcal T(T)=\{0,1,\ldots,T\},\ \mathcal T^{-}(T)=\{0,1,\ldots,T-1\}.
\end{equation}

When $T$ is clear from context, we write $\mathcal T$ and $\mathcal T^{-}$. The model determines whether $X^0$ can be transformed into $X^{\mathcal G}$ in exactly $T$ time steps and, if feasible, minimizes SoC among all such schedules.

The decision variables are
\begin{equation}
x_{vt}^c\in\{0,1\},\ \forall c\in\mathcal C,\ v\in\mathcal V_{N,m},\ t\in\mathcal T,
\end{equation}
where $x_{vt}^c=1$ if vertex $v$ is occupied by a color-$c$ seat at time $t$. Let
\begin{equation}
y_{uvt}^c\in\{0,1\},\ \forall c\in\mathcal C,\ (u,v)\in\mathcal A_{N,m},\ t\in\mathcal T^{-},
\end{equation}
where $y_{uvt}^c=1$ if a color-$c$ seat takes action $(u,v)$ from time $t$ to $t+1$.

For a fixed $T$, the SoC-minimization model is
\begin{equation}
\operatorname{SoC}^{*}(T)=\min\sum_{c\in\mathcal C}\sum_{t\in\mathcal T^{-}}\sum_{(u,v)\in\mathcal A_{N,m}}\omega_{uv}^c y_{uvt}^c
\end{equation}

subject to
\begin{equation}
x_{v0}^c=\mathbf 1\{X^0[v]=c\},\ \forall c\in\mathcal C,\ v\in\mathcal V_{N,m},
\end{equation}
\begin{equation}
x_{vT}^c=\mathbf 1\{X^{\mathcal G}[v]=c\},\ \forall c\in\mathcal C,\ v\in\mathcal V_{N,m},
\end{equation}
\begin{equation}
\sum_{(v,w)\in\delta_{N,m}^{+}(v)}y_{vwt}^c=x_{vt}^c,\ \forall c\in\mathcal C,\ v\in\mathcal V_{N,m},\ t\in\mathcal T^{-},
\end{equation}
\begin{equation}
\sum_{(u,v)\in\delta_{N,m}^{-}(v)}y_{uvt}^c=x_{v,t+1}^c,\ \forall c\in\mathcal C,\ v\in\mathcal V_{N,m},\ t\in\mathcal T^{-},
\end{equation}
\begin{equation}
\sum_{c\in\mathcal C}x_{vt}^c\le1,\ \forall v\in\mathcal V_{N,m},\ t\in\mathcal T,
\end{equation}
\begin{equation}
\sum_{c\in\mathcal C}\left(y_{uvt}^c+y_{vut}^c\right)\le1,\ \forall\{u,v\}\in\mathcal E_{N,m},\ t\in\mathcal T^{-}.
\end{equation}

The objective function assigns unit cost to every physical movement.
A waiting action $(v,v)\in\mathcal A_{N,m}^{\mathrm{wait}}$ has zero cost only when $v\in\mathcal G_c$, i.e., when the color-$c$ seat is already waiting at a target vertex of its own color. Waiting at any non-target vertex incurs unit cost. Therefore, the SoC is not merely the total number of physical movements; it is the cumulative seat-operation cost that penalizes both physical movements and non-target waiting before decoupling.

The first two constraints impose the initial state $X^0$ and the target state $X^\mathcal{G}$. The next two constraints are flow conservation constraints on the time-expanded network. 

Specifically, a color-$c$ seat occupying vertex $v$ at time $t$ must choose exactly one outgoing action from $v$, either a physical movement action or a waiting action. Similarly, any color-$c$ seat occupying vertex $v$ at time $t+1$ must be reached by exactly one incoming action. The vertex capacity constraint ensures that at most one seat occupies each vertex at each time. The undirected edge capacity constraint ensures that at most one seat uses a physical track edge in one time step, regardless of direction; therefore, vertex conflict and edge conflict are prohibited.

We next impose the straight-following rule. Define the aggregate occupancy and action variables
\begin{equation}
\bar{x}_{vt}=\sum_{c\in\mathcal C}x_{vt}^c,\ \forall v\in\mathcal V_{N,m},\ t\in\mathcal T,
\end{equation}
and
\begin{equation}
\bar{y}_{uvt}=\sum_{c\in\mathcal C}y_{uvt}^c,\ \forall (u,v)\in\mathcal A_{N,m},\ t\in\mathcal T^{-}.
\end{equation}

Recall from Section~\ref{sec:problem_des_form} that, for each directed physical movement arc $(u,v)\in\mathcal A_{N,m}^{\mathrm{move}}$, $F(u,v)=2v-u$ is the vertex immediately ahead of $v$ in the movement direction from $u$ to $v$. Define the arcs for which the required forward action is physically feasible as
\begin{equation}
\mathcal A_{N,m}^{F}=\left\{(u,v)\in\mathcal A_{N,m}^{\mathrm{move}}:
(v,F(u,v))\in\mathcal A_{N,m}^{\mathrm{move}}\right\}.
\end{equation}

The straight-following rule is enforced by
\begin{equation}
\bar y_{uvt}+\bar x_{vt}\le1+\bar y_{v,F(u,v),t},\ \forall (u,v)\in\mathcal A_{N,m}^{F},\ t\in\mathcal T^{-},
\end{equation}
and
\begin{equation}
\bar y_{uvt}+\bar x_{vt}\le1,\ \forall (u,v)\in \mathcal A_{N,m}^{\mathrm{move}}\setminus\mathcal A_{N,m}^{F},\ t\in\mathcal T^{-}.
\end{equation}

These constraints encode the following logic. If a seat moves from $u$ to $v$, then $\bar{y}_{uvt}=1$. If $v$ is occupied at the beginning of time step $t$, then $\bar{x}_{vt}=1$. In this case, the first inequality forces $\bar{y}_{v,F(u,v),t}=1$, meaning that the seat originally occupying $v$ must simultaneously move one vertex forward in the same direction. If $F(u,v)$ is outside the physical track graph or if the arc $(v,F(u,v))$ is not physically feasible, the second inequality prohibits any seat from entering the occupied vertex $v$. Therefore, a trailing seat can enter the current position of a leading seat only when the leading seat moves one cell forward in the same direction.

For any fixed $T$, the above model either proves that no schedule with makespan $T$ exists, or returns a schedule with the minimum SoC among all schedules that reach $X^\mathcal{G}$ exactly at time $T$. Let $\mathcal{M}(T)$ denote the feasible region of this fixed-$T$ model. The makespan-optimal decoupling time is obtained by the outer search
\begin{equation}
\operatorname{SoC}^{*}=\operatorname{SoC}^{*}(T^{*})=\min_{(x,y)\in\mathfrak F(T^{*})}
\sum_{c\in\mathcal C}\sum_{t\in\mathcal T^{-}(T^{*})}\sum_{(u,v)\in\mathcal A_{N,m}}\omega_{uv}^c y_{uvt}^c.
\end{equation}

Because waiting arcs are included, if the model is feasible for $T$, then it is also feasible for any $T'>T$. In particular, once the target state $X^\mathcal G$ is reached, all seats can remain at their target vertices with zero additional cost. Therefore, $T^*$ can be found by a standard linear or binary search over $T$, once a valid upper bound is available. After $T^*$ is identified, the final model solved in this study is the fixed-makespan SoC minimization problem
\begin{equation}
\operatorname{SoC}^{*}
=\operatorname{SoC}^{*}(T^{*})
=\min_{\mathcal M(T^{*})}
\sum_{c\in\mathcal C}
\sum_{t\in\mathcal T^{-}(T^{*})}
\sum_{(u,v)\in\mathcal A_{N,m}}
\omega_{uv}^c y_{uvt}^c.
\end{equation}

This produces a optimal seat movement schedule: the modular
bus units decouple at the earliest possible time $T^*$, and among all schedules that achieve this earliest decoupling time, the selected schedule minimizes the SoC. This SoC penalizes both physical movements
and waiting at non-target vertices, while waiting at a correct target vertex is cost-free. Therefore, the secondary objective encourages seats to settle at their target positions as early as possible without delaying decoupling, and minimize the number of seat movements required to maximize passenger comfort.

\section{Optimal Boundary-Sequence Dynamic Programming Algorithm}\label{sec:bsdp}

This section develops an exact boundary-sequence dynamic programming (BSDP) algorithm for the fixed-makespan problem. Adjacent modular units interact through only one physical boundary edge. Once the crossing event on every boundary is specified for every time step, each module can evaluate its own feasible movements and local cost. The resulting local transfer costs are then combined over consistent shared boundary schedules.

\begin{figure}[!htbp]
    \centering
    \includegraphics[width=\linewidth]{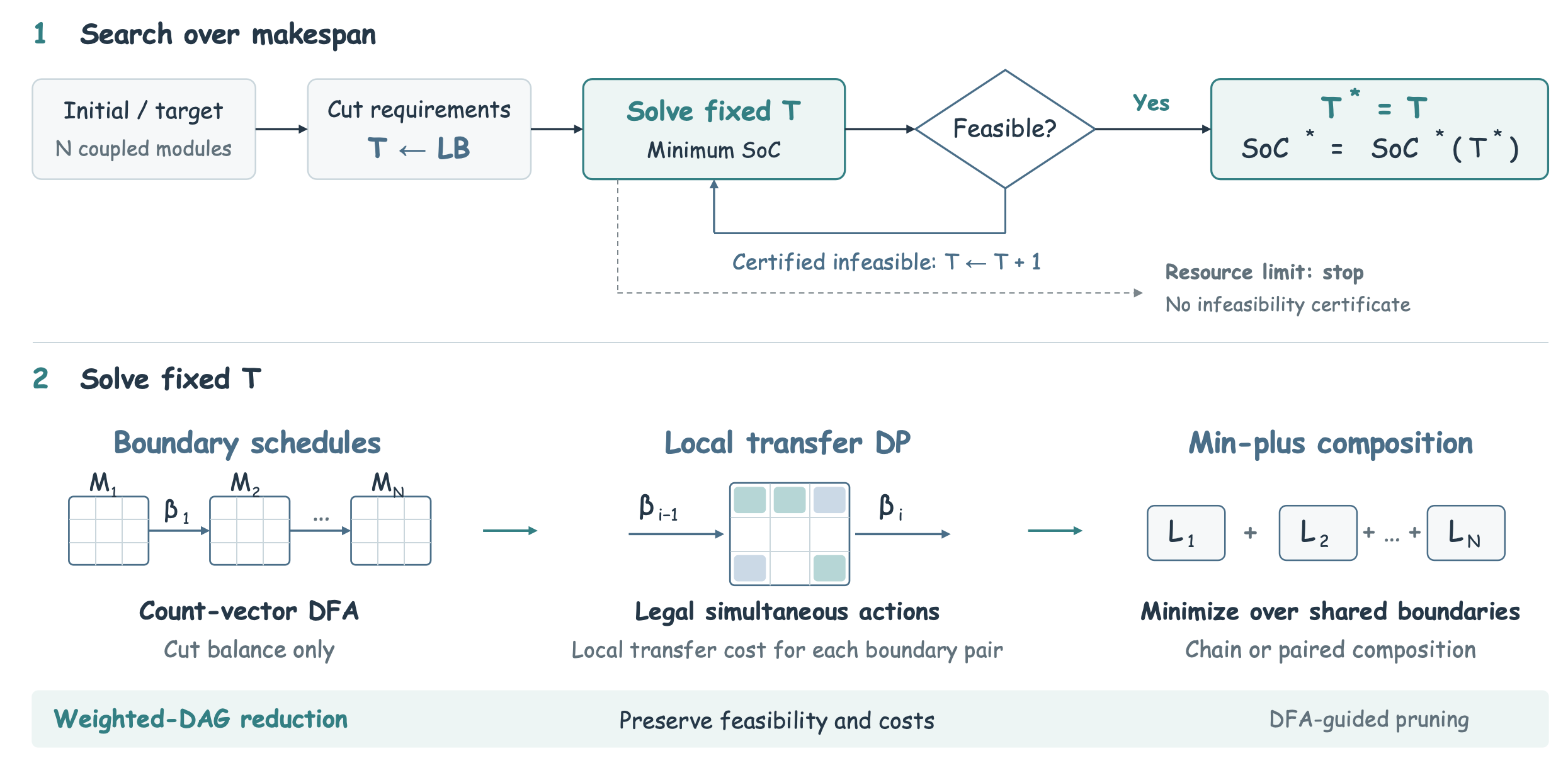}
    \caption{Overview of the BSDP algorithm.}
    \label{fig:BSDP_diagram}
\end{figure}

Figure~\ref{fig:BSDP_diagram} summarizes how the fixed-$T$ algorithm supports lexicographic makespan--SoC optimization. The outer search starts from a valid integer lower bound on the minimum makespan, denoted by $\mathrm{LB}$ in the figure. It tests successive values of $T$ and increases $T$ only after infeasibility at the current value has been established. The first feasible value is $T^*$. BSDP then minimizes SoC at $T^*$ to obtain $\operatorname{SoC}^*(T^*)$. The fixed-$T$ procedure may be used in feasibility mode, which can stop after finding a feasible schedule, or in optimization mode, which computes the minimum SoC. The ``Minimum SoC'' label in the figure describes the latter mode; finding a feasible schedule alone does not certify its SoC optimality. If optimization has already been completed at the first feasible $T$, its result is retained without a second run. A resource limit before feasibility is resolved does not constitute an infeasibility certificate and cannot justify increasing $T$.

The lower part of Figure~\ref{fig:BSDP_diagram} expands the fixed-$T$ computation. Boundary schedules encode the interactions between adjacent modules, as defined in Section~\ref{sec:boundary_schedule}. Local actions and local transfer dynamic programming determine the feasibility and minimum cost of each boundary-schedule pair in Sections~\ref{sec:local_boundary} and~\ref{sec:local_transfer_dp}. Section~\ref{sec:dfa_compression} introduces DFA-guided pruning and weighted-DAG reduction to represent these schedules compactly. Sections~\ref{sec:chain_composition} and~\ref{sec:paired_composition} then combine the local transfer costs by two equivalent min-plus evaluation orders. The labels $L_i$ in the overview abbreviate the fixed-$T$ local transfer costs $L_i^T$; their boundary arguments are omitted there and defined below.

\subsection{Finite-Automaton Preliminaries}\label{sec:dfa_prelim}

Finite automata provide a standard way to recognize sequences that satisfy constraints expressible through a finite amount of memory~\citep{rabin1959finite}. We briefly define the required terminology before applying it to the MB problem.

\begin{definition}[Alphabet, word, and language]
An \emph{alphabet} $\Sigma$ is a finite set of symbols. A length-$T$ \emph{word} over $\Sigma$ is an ordered sequence $w=(a_0,a_1,\ldots,a_{T-1})\in\Sigma^T$. Let $\Sigma^*=\bigcup_{T\ge0}\Sigma^T$ be the set of all finite words over $\Sigma$. A \emph{language} is a set $\mathcal L\subseteq\Sigma^*$. In this paper, every makespan-$T$ boundary language is a fixed-length language contained in $\Sigma^T$.
\end{definition}

\begin{definition}[Deterministic finite automaton]
A deterministic finite automaton (DFA) is a five-tuple $\mathscr A=(\mathscr Q,\Sigma,\eta,q^0,\mathscr F)$, where $\mathscr Q$ is a finite state set, $\Sigma$ is an input alphabet, $\eta:\mathscr Q\times\Sigma\rightarrow\mathscr Q$ is a deterministic transition function, $q^0\in\mathscr Q$ is the initial state, and $\mathscr F\subseteq\mathscr Q$ is the accepting-state set. Starting from $q^0$, the DFA reads a word one symbol at a time and updates its state through $\eta$. The word is accepted if the state after the last symbol belongs to $\mathscr F$; otherwise, it is rejected. The set of accepted words is the language recognized by the DFA. An invalid prefix may be sent to a nonaccepting sink state that transitions to itself on every later input.
\end{definition}

In the present problem, one symbol describes what happens at one inter-module boundary during one time step; a word describes the complete length-$T$ boundary-crossing schedule; and the language contains all schedules satisfying the required net seat transfers. For the cut-balance constraints, whether a boundary prefix can be completed depends only on the number of symbols already read and the cumulative color-specific net transfers. Prefixes with the same summary can therefore share a count-vector DFA state. Physical realizability inside a module additionally depends on its local configurations, which are incorporated in Section~\ref{sec:dfa_compression}.

For optimization, the representation must also retain the relevant local states and their minimum accumulated costs. BSDP uses local cost maps and a weighted finite-state representation~\citep{mohri1997finite} for this purpose. Equality of the transfer counts alone is insufficient for merging these cost-bearing states. In BSDP, the automaton is a computational data structure used to avoid explicit enumeration of all $(1+2|\mathcal C|)^T$ boundary words; it does not change the feasible schedules or the objective.

\subsection{Boundary Events, Schedules, and Cut Requirements}\label{sec:boundary_schedule}

For $i=1,2,\ldots,N-1$, define the right portal of module $\mathcal M_i$ and the left portal of module $\mathcal M_{i+1}$ as
\begin{equation}
    p_i^{R}=(2,mi),\ p_{i+1}^{L}=(2,mi+1).
\end{equation}

Their unique physical boundary edge is
\begin{equation}
e_i^{\mathrm{bd}}=\{p_i^{R},p_{i+1}^{L}\}.
\label{eq:bsdp_boundary_edge}
\end{equation}
We reserve $e_i^{\mathrm{bd}}$ for the physical edge and use $\mathcal B_i^T$ below for the cut-balanced boundary language. Membership in this language is a necessary condition for global feasibility; local physical realizability is checked separately.

The boundary-event alphabet is
\begin{equation}
\Sigma=\{0\}\cup\{+c,-c:c\in\mathcal C\}.
\label{eq:bsdp_boundary_alphabet}
\end{equation}
Table~\ref{tab:boundary_symbols} gives the physical meaning of each symbol. The signs refer to the global left-to-right orientation of the module chain.
\begin{table}[htbp]
\centering
\caption{Interpretation of a boundary-event symbol on $e_i^{\mathrm{bd}}$.}
\label{tab:boundary_symbols}
\begin{tabular}{cl}
\toprule
Symbol & Event during one time step \\
\midrule
$0$ & No seat crosses $e_i^{\mathrm{bd}}$. \\
$+c$ & One color-$c$ seat moves from $\mathcal M_i$ to $\mathcal M_{i+1}$. \\
$-c$ & One color-$c$ seat moves from $\mathcal M_{i+1}$ to $\mathcal M_i$. \\
\bottomrule
\end{tabular}
\end{table}

For a fixed makespan $T$, a boundary schedule is the word
\begin{equation}
\beta_i=(\sigma_{i,0},\sigma_{i,1},\ldots,\sigma_{i,T-1})\in\Sigma^T,
\label{eq:bsdp_boundary_word}
\end{equation}
where $\sigma_{i,t}$ is the event on $e_i^{\mathrm{bd}}$ from time $t$ to $t+1$. Because $e_i^{\mathrm{bd}}$ is a single edge, at most one crossing event can occur on it in a time step.

Define the prefix vertex set
\begin{equation}
P_i=\bigcup_{h=1}^{i}\mathcal W_h,\ i=1,2,\ldots,N-1.
\end{equation}
For color $c$, the net left-to-right transfer encoded by $\beta_i$ is
\begin{equation}
\Delta_i^c(\beta_i)
=\sum_{t=0}^{T-1}\mathbf 1\{\sigma_{i,t}=+c\}
-\sum_{t=0}^{T-1}\mathbf 1\{\sigma_{i,t}=-c\}.
\end{equation}
The transfer required by the initial and target states is
\begin{equation}
d_i^c
=\sum_{v\in P_i}\mathbf 1\{X^0[v]=c\}
-\sum_{v\in P_i}\mathbf 1\{X^{\mathcal G}[v]=c\}.
\label{eq:bsdp_cut_requirement}
\end{equation}
Thus, $d_i^c>0$ means that the left prefix must send $d_i^c$ color-$c$ seats to the right, whereas $d_i^c<0$ means that it must receive $-d_i^c$ such seats. We collect these requirements in the vector $\mathbf d_i=(d_i^c)_{c\in\mathcal C}$. In the figures, $M_i$ denotes module $\mathcal M_i$, and the prefix $P_i$ comprises all vertices in modules $\mathcal M_1,\ldots,\mathcal M_i$.

The cut-balanced boundary language is
\begin{equation}
\mathcal B_i^T=\left\{\beta_i\in\Sigma^T:\Delta_i^c(\beta_i)=d_i^c,\ \forall c\in\mathcal C\right\}.
\label{eq:bsdp_boundary_language}
\end{equation}
For the two exterior boundaries, define
$\mathcal B_0^T=\mathcal B_N^T=\{\mathbf 0_T\}$, where $\mathbf 0_T=(0,\ldots,0)$. For notational uniformity, write $\sigma_{0,t}=\sigma_{N,t}=0$ for every $t=0,1,\ldots,T-1$.

\begin{lemma}[Boundary cut balance]\label{lemma:boundary}
Every feasible global schedule induces a boundary word $\beta_i\in\mathcal B_i^T$ for every $i=1,2,\ldots,N-1$.
\end{lemma}
\begin{proof}
The number of color-$c$ seats in $P_i$ changes only when a color-$c$ seat crosses $e_i^{\mathrm{bd}}$. A $+c$ event decreases this count by one, whereas a $-c$ event increases it by one. Since $X^T=X^{\mathcal G}$,
\[
\Delta_i^c(\beta_i)
=\sum_{v\in P_i}\mathbf 1\{X^0[v]=c\}
-\sum_{v\in P_i}\mathbf 1\{X^{\mathcal G}[v]=c\}
=d_i^c.
\]
Hence $\beta_i\in\mathcal B_i^T$.
\end{proof}

The cut requirements specify net transfers, whereas a boundary word also specifies their timing and order. Distinct words can satisfy the same cut requirements but differ in local physical feasibility or SoC. These differences motivate the local transition model below.

\subsection{Local Actions and Boundary Consistency} \label{sec:local_boundary}

For module $\mathcal M_i$, let
\begin{equation}
\mathcal S_i=(\mathcal C\cup\{0\})^{\mathcal W_i}
\end{equation}
be its local state space, and define
\begin{equation}
S_i^0=X^0|_{\mathcal W_i},\ S_i^{\mathcal G}=X^{\mathcal G}|_{\mathcal W_i}.
\end{equation}

For $i>1$, let $q_i^L=(2,(i-1)m)$ be the vertex immediately outside the left boundary of $\mathcal M_i$, and let $p_i^L=(2,(i-1)m+1)$ be its left portal. For $i<N$, let $p_i^R=(2,im)$ be its right portal and $q_i^R=(2,im+1)$ the vertex immediately outside its right boundary.

Let $\mathcal A_i^{\mathrm{loc}}=\{(u,v)\in\mathcal A_{N,m}:u\in\mathcal W_i\text{ or }v\in\mathcal W_i\}$ be the action arcs incident to module $\mathcal M_i$. A candidate local simultaneous action set is a set $U\subseteq\mathcal C\times\mathcal A_i^{\mathrm{loc}}$ of color-labeled actions. It may contain waiting and internal actions, outbound boundary actions whose tail lies in $\mathcal W_i$, and inbound boundary actions whose head lies in $\mathcal W_i$.

For a candidate $U$ containing at most one crossing action on each physical boundary and for $i>1$, define its left-boundary projection by
\begin{equation}
\partial_i^L(U)=
\begin{cases}
+c, & (c,q_i^L,p_i^L)\in U,\\
-c, & (c,p_i^L,q_i^L)\in U,\\
0,  & \text{no action in $U$ crosses the left boundary},
\end{cases}
\end{equation}

For $i<N$, define its right-boundary projection by
\begin{equation}
\partial_i^R(U)=
\begin{cases}
+c, & (c,p_i^R,q_i^R)\in U,\\
-c, & (c,q_i^R,p_i^R)\in U,\\
0,  & \text{no action in $U$ crosses the right boundary}.
\end{cases}
\end{equation}
For the exterior sides, set $\partial_1^L(U)=0$ and $\partial_N^R(U)=0$.

Table~\ref{tab:local_boundary_actions} summarizes the local interpretation for a nonexterior side. The same global symbol has different inbound/outbound meanings on the two sides of a module because every $+c$ event is oriented from left to right globally.
\begin{table}[htbp]
\centering
\caption{Local action represented by a nonzero boundary symbol.}
\label{tab:local_boundary_actions}
\begin{tabular}{ccl}
\toprule
Module side & Symbol & Required color-labeled action \\
\midrule
Left  & $+c$ & $(c,q_i^L,p_i^L)$: enter from the left \\
Left  & $-c$ & $(c,p_i^L,q_i^L)$: leave to the left \\
Right & $+c$ & $(c,p_i^R,q_i^R)$: leave to the right \\
Right & $-c$ & $(c,q_i^R,p_i^R)$: enter from the right \\
\bottomrule
\end{tabular}
\end{table}

Given boundary schedules $\beta_{i-1}$ and $\beta_i$, define the actual left and right symbols observed by module $\mathcal M_i$ at time $t$ as
\begin{equation}
\lambda_{i,t}=\sigma_{i-1,t},\ \rho_{i,t}=\sigma_{i,t}.
\label{eq:bsdp_local_symbols}
\end{equation}
The phrase ``$U$ respects the left and right boundary symbols'' now has the precise meaning
\begin{equation}
\partial_i^L(U)=\lambda_{i,t},
\qquad
\partial_i^R(U)=\rho_{i,t}.
\end{equation}
For example, if $(\lambda_{i,t},\rho_{i,t})=(+c,0)$, then $U$ must contain the inbound left-boundary action $(c,q_i^L,p_i^L)$ and must contain no right-boundary crossing. An action set with an outbound left crossing or any right crossing is inconsistent with these prescribed symbols.

The global one-step definition cannot be imposed verbatim on a single module, because an inbound action has its tail in a neighboring module. We therefore impose the local counterpart explicitly.

For generic one-step boundary symbols $a^L,a^R\in\Sigma$ and local states $S,S'\in\mathcal S_i$, let
\begin{equation}
\mathcal U_i(S,S';a^L,a^R)
\end{equation}
be the collection of sets $U\subseteq\mathcal C\times\mathcal A_i^{\mathrm{loc}}$ satisfying all of the following conditions:
\begin{enumerate}
    \item \textbf{Local tail consistency and completeness.} Every action $(c,u,v)\in U$ with $u\in\mathcal W_i$ satisfies $S[u]=c$, and every occupied $u\in\mathcal W_i$ is the tail of exactly one action in $U$. An inbound action has its tail outside $\mathcal W_i$; its color is prescribed by $a^L$ or $a^R$.
    \item \textbf{No local vertex conflict.} Every $v\in\mathcal W_i$ is the head of at most one action in $U$.
    \item \textbf{No edge conflict.} No two actions in $U$ traverse the same undirected edge in opposite directions.
    \item \textbf{Straight simultaneous following at local vertices.} If $(c,u,v)\in U$, $u\neq v$, $v\in\mathcal W_i$, and $S[v]=c'\neq0$, then $(c',v,F(u,v))\in U$. Following into a vertex outside $\mathcal W_i$ is checked by the neighboring module, for which that vertex is local.
    \item \textbf{Local simultaneous update.} All actions are applied simultaneously, and, for every $v\in\mathcal W_i$, $S'[v]=c\quad\Longleftrightarrow\quad(c,u,v)\in U\text{ for some }u$.
    \item \textbf{Boundary consistency.}
    $\partial_i^L(U)=a^L$ and $\partial_i^R(U)=a^R$.
\end{enumerate}
This generic relation has no time index because the physical one-step rules are time-homogeneous. Its actual use for module $i$ at time $t$ is $\mathcal U_i(S,S';\lambda_{i,t},\rho_{i,t})$.

These conditions are the exact local counterparts of Definition~\ref{def:one-step}. In particular, a boundary action is represented in both incident modules, but its color and direction are fixed by their shared boundary symbol. Conditions involving its local endpoint are checked by the module containing that endpoint. The tail-ownership convention below then charges the action only once.

The local one-step cost is counted by tail ownership:
\begin{equation}
\operatorname{cost}_i(U)=\sum_{\substack{(c,u,v)\in U:\ u\in\mathcal W_i}}\omega_{uv}^c.
\label{eq:bsdp_tail_cost}
\end{equation}
Thus, an internal or outbound action is charged to the module containing its tail. An inbound boundary action is charged to the neighboring sending module, so every global action is counted exactly once.

Define
\begin{equation}
\kappa_i(S,S';a^L,a^R)=\min_{U\in\mathcal U_i(S,S';a^L,a^R)}\operatorname{cost}_i(U),
\label{eq:bsdp_local_step_cost}
\end{equation}
with value $+\infty$ when the feasible-action set is empty.

\subsection{Local Transfer Dynamic Programming}\label{sec:local_transfer_dp}

For fixed boundary schedules $\beta_{i-1}$ and $\beta_i$, let
$H_{i,t}^{\beta_{i-1},\beta_i}(S)$ be the minimum local SoC required for module $\mathcal M_i$ to reach local state $S\in\mathcal S_i$ at time $t$. The initialization is
\begin{equation}
H_{i,0}^{\beta_{i-1},\beta_i}(S)=
\begin{cases}
0, & S=S_i^0,\\
+\infty, & \text{otherwise}.
\end{cases}
\label{eq:bsdp_local_initial}
\end{equation}

For $t=0,1,\ldots,T-1$, the recurrence is
\begin{equation}
H_{i,t+1}^{\beta_{i-1},\beta_i}(S')=\min_{S\in\mathcal S_i}\left\{
H_{i,t}^{\beta_{i-1},\beta_i}(S)+\kappa_i(S,S';\lambda_{i,t},\rho_{i,t})\right\}.
\label{eq:bsdp_local_recursion}
\end{equation}

The fixed-$T$ local transfer cost is
\begin{equation}
L_i^T(\beta_{i-1},\beta_i)=H_{i,T}^{\beta_{i-1},\beta_i}(S_i^{\mathcal G}).
\label{eq:bsdp_local_transfer}
\end{equation}

A finite value of $L_i^T(\beta_{i-1},\beta_i)$ means that the module can reach its target configuration within the fixed horizon while respecting both prescribed boundary words. Otherwise, the value is $+\infty$. Equation~\eqref{eq:bsdp_local_recursion} retains a separate minimum cost for each reachable local state, because different states can admit different feasible continuations.

\begin{figure}[!htbp]
    \centering
    \includegraphics[width=\linewidth]{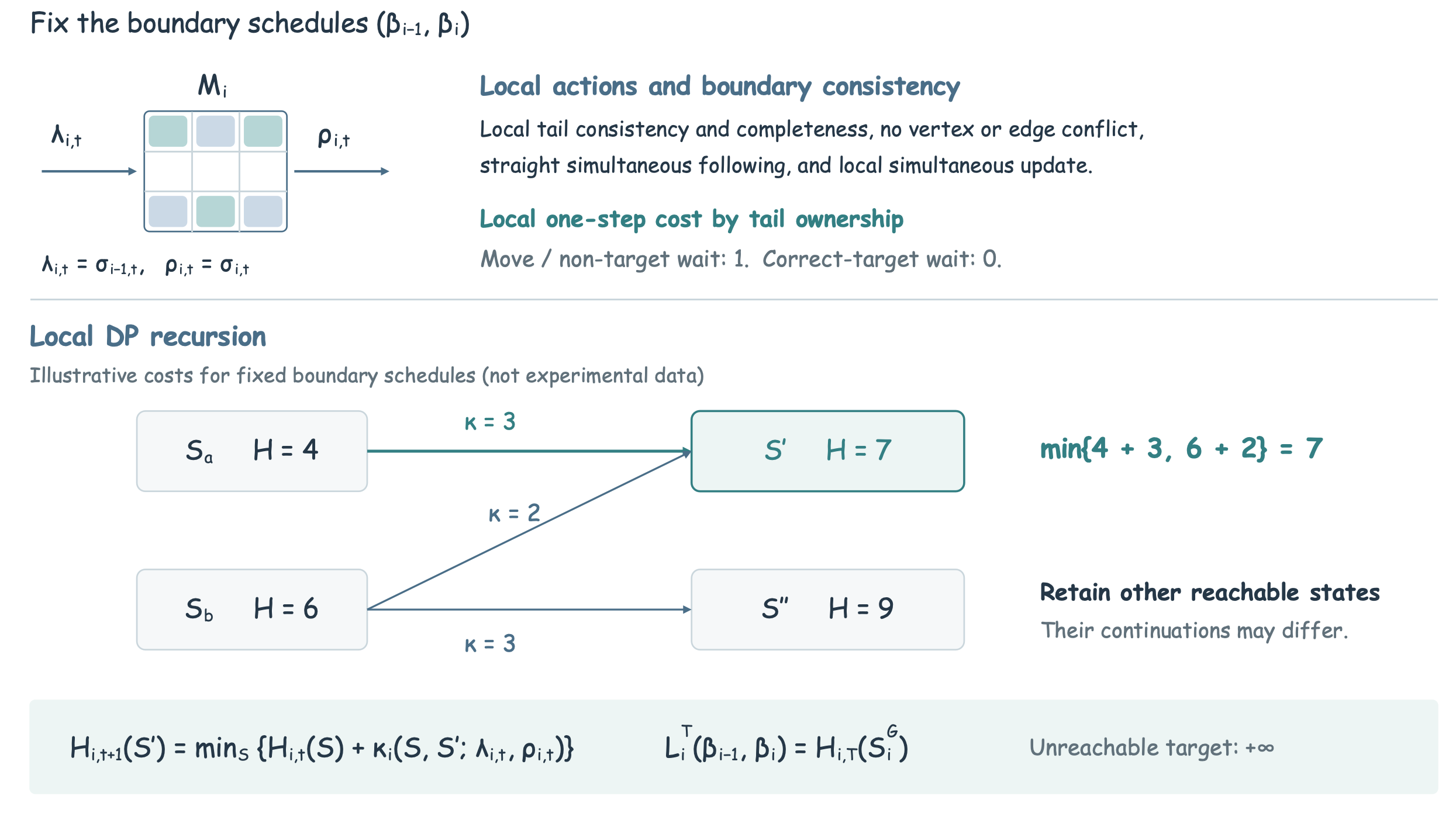}
    \caption{Local transfer dynamic programming.}
    \label{fig:bu_diagram}
\end{figure}

Figure~\ref{fig:bu_diagram} illustrates the local action model and one application of this recurrence. In its upper part, $\lambda_{i,t}$ and $\rho_{i,t}$ are the two boundary-event symbols defined in~\eqref{eq:bsdp_local_symbols}. The adjacent arrows indicate the locations of the boundary interactions and the positive crossing orientation. They do not restrict either symbol to $+c$: each symbol may be $0$, $+c$, or $-c$, with the interpretation in Table~\ref{tab:local_boundary_actions}. The pictured module is schematic and does not specify the local configurations used in the numerical example below. The one-step cost follows the tail-ownership rule in~\eqref{eq:bsdp_tail_cost}: movements and waiting at non-target vertices have unit cost, whereas waiting at a correct target vertex has zero cost.

In the lower part of Figure~\ref{fig:bu_diagram}, the module index $i$ and boundary schedules $(\beta_{i-1},\beta_i)$ are fixed. The states $S_a,S_b\in\mathcal S_i$ belong to time $t$, whereas $S',S''\in\mathcal S_i$ belong to time $t+1$. The label $H$ at each node abbreviates the corresponding value of $H_{i,t}^{\beta_{i-1},\beta_i}$ or $H_{i,t+1}^{\beta_{i-1},\beta_i}$. An edge label $\kappa$ abbreviates the local one-step cost $\kappa_i(S,S';\lambda_{i,t},\rho_{i,t})$ for that edge's endpoints. In particular, the two predecessor labels mean
\[
H_{i,t}^{\beta_{i-1},\beta_i}(S_a)=4,
\qquad
H_{i,t}^{\beta_{i-1},\beta_i}(S_b)=6.
\]
For the illustrated incoming transitions to $S'$, the one-step costs are $3$ and $2$, respectively. The update in~\eqref{eq:bsdp_local_recursion} therefore gives
\[
H_{i,t+1}^{\beta_{i-1},\beta_i}(S')
=\min\{4+3,\,6+2\}=7.
\]
The other illustrated successor has value
$H_{i,t+1}^{\beta_{i-1},\beta_i}(S'')=6+3=9$.
It must still be retained: a lower cost at a different state does not imply a better feasible continuation. These numbers illustrate the recurrence and are not experimental results or costs derived from the pictured seat configuration. The diagram shows selected states and transitions, with the displayed incoming alternatives determining the stated values in this example. In the formula strip, $H_{i,t}(S)$ suppresses the fixed boundary-schedule superscripts, and $S_i^G$ denotes the target state $S_i^{\mathcal G}$ used in~\eqref{eq:bsdp_local_transfer}.

\subsection{DFA-Guided Boundary-Schedule Compression}
\label{sec:dfa_compression}

Direct enumeration of all length-$T$ boundary words requires examining $|\Sigma|^T=(1+2|\mathcal C|)^T$ candidates for each inter-module boundary. We therefore use a deterministic finite automaton (DFA) to recognize the cut-balanced boundary language and then augment the automaton with local-state and cost information. The DFA removes boundary prefixes that can no longer satisfy the required transfers, while the augmented construction removes words that cannot be realized by the physical movements of the corresponding module.

For a boundary symbol $a\in\Sigma$, define its color-specific net-transfer increment as
\begin{equation}
\varepsilon^c(a)=
\begin{cases}
 1, & a=+c,\\
-1, & a=-c,\\
 0, & \text{otherwise},
\end{cases}
\qquad c\in\mathcal C,
\end{equation}
and let
\begin{equation}
\boldsymbol\varepsilon(a)
=\bigl(\varepsilon^c(a)\bigr)_{c\in\mathcal C}
\in\mathbb Z^{\mathcal C}.
\end{equation}
Thus, reading $+c$ increases the left-to-right net transfer of color $c$ by one, reading $-c$ decreases it by one, and reading $0$ leaves the transfer vector unchanged.

For boundary $i$, define the count-vector DFA
\begin{equation}
\mathscr A_i^T=\bigl(\mathscr Q_i^T,\Sigma,\eta_i^T,q_i^0,
\mathscr F_i^T\bigr).
\end{equation}
A nonsink state is a pair $q=(\tau,\mathbf z)$, where $\tau\in\{0,1,\ldots,T\}$ is the number of symbols already read and $\mathbf z=(z^c)_{c\in\mathcal C}$ is the cumulative net-transfer vector. Formally,
\begin{equation}
\mathscr Q_i^T
={}
\Biggl\{
(\tau,\mathbf z):
0\le\tau\le T,\ 
\exists(a_0,\ldots,a_{\tau-1})\in\Sigma^\tau
\text{ such that }
\mathbf z
=\sum_{\ell=0}^{\tau-1}
\boldsymbol\varepsilon(a_\ell)
\Biggr\}
\cup\{q_i^{\bot}\},
\label{eq:bsdp_count_states}
\end{equation}
where $q_i^{\bot}$ is a nonaccepting sink state. This state set is finite because $0\le\tau\le T$ and every component of a reachable vector satisfies $|z^c|\le\tau$.

The initial state is
\begin{equation}
q_i^0=(0,\mathbf0).
\label{eq:bsdp_dfa_initial}
\end{equation}
Consider a nonsink state $(\tau,\mathbf z)$ with $\tau<T$ and an input symbol $a\in\Sigma$. Let $\mathbf z'=\mathbf z+\boldsymbol\varepsilon(a)$.

Because at most one seat can cross a boundary in each remaining time step, the required transfer vector $\mathbf d_i=(d_i^c)_{c\in\mathcal C}$ can still be reached only if
\begin{equation}
\sum_{c\in\mathcal C}
\left|d_i^c-z'^c\right|
\le T-\tau-1.
\label{eq:dfa_residual_condition}
\end{equation}
Accordingly, define
\begin{equation}
\eta_i^T\bigl((\tau,\mathbf z),a\bigr)
=
\begin{cases}
(\tau+1,\mathbf z'),
&
\text{if \eqref{eq:dfa_residual_condition} holds},\\
q_i^{\bot},
&
\text{otherwise}.
\end{cases}
\label{eq:bsdp_dfa_transition}
\end{equation}
All inputs from $q_i^{\bot}$ return to $q_i^{\bot}$, and every input from a nonsink state with $\tau=T$ also enters $q_i^{\bot}$. The accepting-state set is
\begin{equation}
\mathscr F_i^T=\mathscr Q_i^T\cap\{(T,\mathbf d_i)\}.
\label{eq:bsdp_dfa_accepting}
\end{equation}

\begin{figure}[!htbp]
    \centering
    \includegraphics[width=\linewidth]{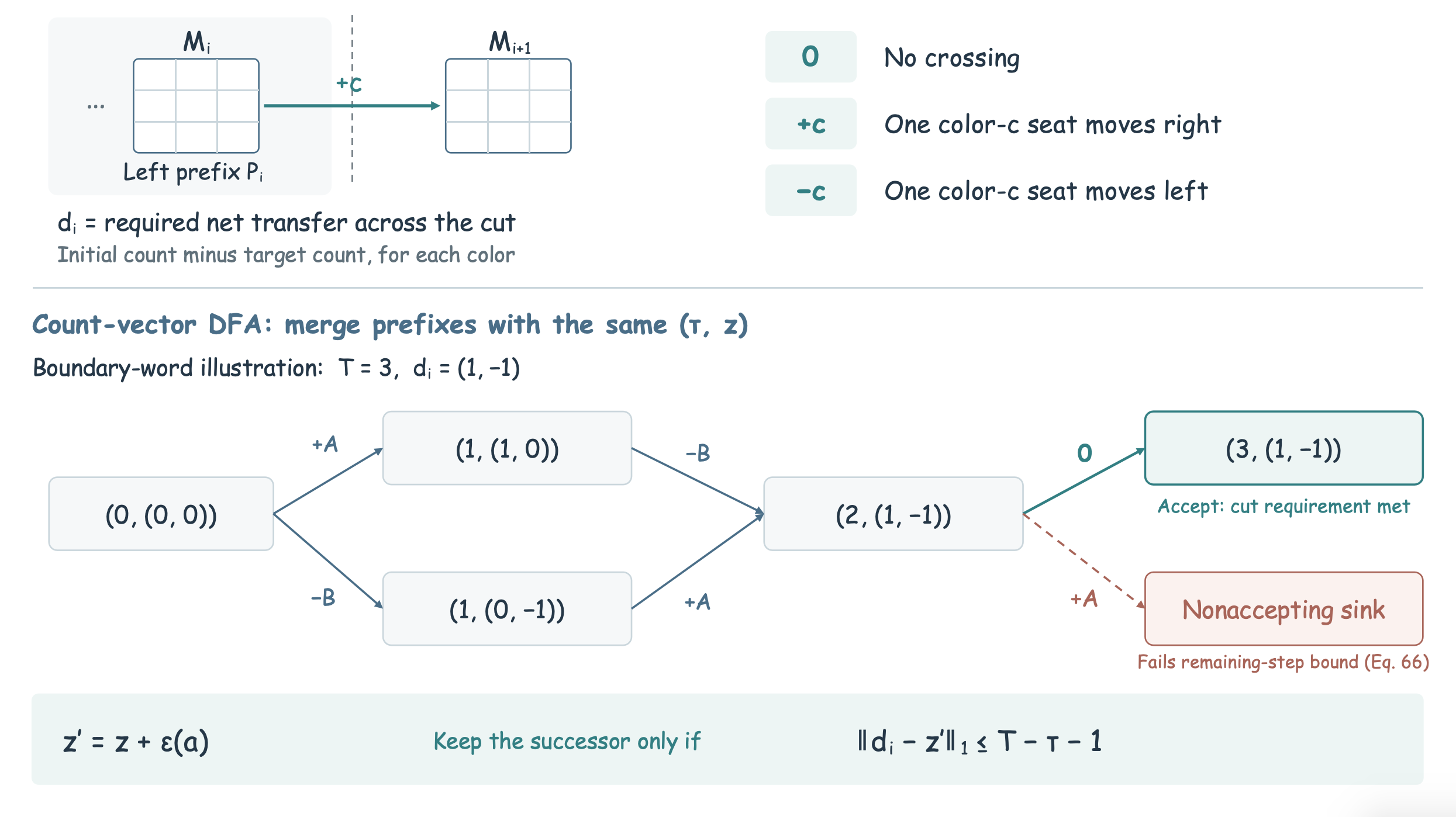}
    \caption{Count-vector DFA for boundary schedules.}
    \label{fig:bsc_diagram}
\end{figure}

A length-$T$ word is accepted if and only if its final cumulative transfer vector equals the cut requirement $\mathbf d_i$. Therefore,
\begin{equation}
\mathcal L(\mathscr A_i^T)=\mathcal B_i^T.
\label{eq:bsdp_dfa_language}
\end{equation}
Figure~\ref{fig:bsc_diagram} connects the physical boundary events in Section~\ref{sec:boundary_schedule} to the count-vector DFA defined above. Its upper part shows the cut between $\mathcal M_i$ and $\mathcal M_{i+1}$, the left prefix $P_i$, and the symbols $0,+c,-c$. The labels $d_i$, $z$, and $\varepsilon(a)$ in the figure denote the vectors $\mathbf d_i$, $\mathbf z$, and $\boldsymbol\varepsilon(a)$, respectively. The norm notation in its pruning condition is shorthand for
\[
\|\mathbf d_i-\mathbf z'\|_1
=\sum_{c\in\mathcal C}|d_i^c-z'^c|,
\]
so the condition is exactly~\eqref{eq:dfa_residual_condition}.

For the numerical boundary-word example in the lower part of Figure~\ref{fig:bsc_diagram}, take $\mathcal C=\{A,B\}$, use the coordinate order $(A,B)$, and set $T=3$ and $\mathbf d_i=(1,-1)$. Here $A$ and $B$ denote color classes. A node labeled $(\tau,(z^A,z^B))$ is the state $q=(\tau,\mathbf z)$ in~\eqref{eq:bsdp_count_states}: $\tau$ counts the boundary-event symbols already read, and $z^A,z^B$ are the cumulative net left-to-right transfers of the two colors. In particular, $(0,(0,0))$ is $q_i^0=(0,\mathbf0)$; no symbols have been read and no net transfer has occurred.

Reading $+A$ from the initial state gives $(1,(1,0))$, while reading $-B$ gives $(1,(0,-1))$. The two prefixes $(+A,-B)$ and $(-B,+A)$ both reach $(2,(1,-1))$. They therefore share a count-vector DFA state even though their event orders differ. From this common state, reading $0$ produces $(3,(1,-1))=(T,\mathbf d_i)$, which is accepting by~\eqref{eq:bsdp_dfa_accepting}. By contrast, reading $+A$ would give the candidate vector $\mathbf z'=(2,-1)$, for which
\[
\sum_{c\in\{A,B\}}|d_i^c-z'^c|=1
>0=T-2-1.
\]
Equation~\eqref{eq:bsdp_dfa_transition} therefore sends this input to $q_i^\bot$, shown as the nonaccepting sink. The figure displays selected transitions; other transitions and the sink self-loops are omitted.

This example concerns boundary words rather than complete physical seat schedules. Acceptance by the count-vector DFA certifies cut balance only. In particular, the two prefixes that share $(2,(1,-1))$ need not induce the same reachable local configurations or the same local costs. The local-state and cost information required to distinguish them is introduced next.

Consider a module $\mathcal M_i$ treated as a side subproblem with one exposed inter-module boundary. Let $\chi\in\{L,R\}$ identify whether the exposed boundary is on the left or right. For an input symbol $a\in\Sigma$, define the corresponding pair of local boundary symbols by
\begin{equation}
(a_\chi^L,a_\chi^R)=
\begin{cases}
(a,0), & \chi=L,\\
(0,a), & \chi=R.
\end{cases}
\end{equation}
Thus, the nonexposed exterior boundary is assigned the zero symbol. In particular, $\chi=R$ for the leftmost module $\mathcal M_1$, whereas $\chi=L$ for the rightmost module $\mathcal M_N$.

Let
\begin{equation}
b=
\begin{cases}
i-1, & \chi=L,\\
i,   & \chi=R,
\end{cases}
\end{equation}
be the index of the exposed physical boundary, and let $\mathbf d_b=(d_b^c)_{c\in\mathcal C}$ be its required net-transfer vector.

The fixed-boundary DP in Section~\ref{sec:local_transfer_dp} evaluates a prescribed pair of boundary words. The side automaton instead reads an exposed boundary word incrementally and retains the local states and costs compatible with each prefix. At layer $\tau$, different boundary prefixes may induce different local cost maps. To distinguish them explicitly, let $\alpha=(a_0,a_1,\ldots,a_{\tau-1})\in\Sigma^\tau$ be a boundary prefix and define
\begin{equation}
R_{i,\chi,\tau}^{\alpha}:
\mathcal S_i\times\mathbb Z^{\mathcal C}
\longrightarrow
\mathbb R_{\ge0}\cup\{+\infty\}.
\label{eq:bsdp_side_cost_map}
\end{equation}

For $S\in\mathcal S_i$ and $\mathbf z\in\mathbb Z^{\mathcal C}$, $R_{i,\chi,\tau}^{\alpha}(S,\mathbf z)$ is the minimum local SoC among all length-$\tau$ local schedules of module $\mathcal M_i$ that are consistent with the exposed-side boundary prefix $\alpha$, reach local state $S$, and accumulate net-transfer vector $\mathbf z$. Pairs with value $+\infty$ need not be stored.

For the empty prefix $\emptyset$, the initial cost map is
\begin{equation}
R_{i,\chi,0}^{\emptyset}(S,\mathbf z)=\begin{cases}
0,&S=S_i^0\text{ and }\mathbf z=\mathbf 0,
\\+\infty,&\text{otherwise}.
\end{cases}
\label{eq:side_automaton_initial}
\end{equation}

Suppose that the automaton has read a prefix $\alpha\in\Sigma^\tau$ and next reads symbol $a\in\Sigma$. Let $\alpha a\in\Sigma^{\tau+1}$ denote their concatenation. The successor cost map is defined by
\begin{equation}
R_{i,\chi,\tau+1}^{\alpha a}=\Theta_{i,\chi,\tau}
\left(R_{i,\chi,\tau}^{\alpha},a\right).
\end{equation}
Specifically, for every $S'\in\mathcal S_i$ and
$\mathbf z'\in\mathbb Z^{\mathcal C}$,
\begin{equation}
\begin{aligned}
&R_{i,\chi,\tau+1}^{\alpha a}(S',\mathbf z')\\
&\quad=
\begin{cases}
\displaystyle
\min_{\substack{
S\in\mathcal S_i:\\
\mathcal U_i
(S,S';a_\chi^L,a_\chi^R)
\neq\emptyset}}
\left\{R_{i,\chi,\tau}^{\alpha}
\left(S,\mathbf z'-\boldsymbol\varepsilon(a)\right)
+\kappa_i (S,S';a_\chi^L,a_\chi^R)\right\},&
\displaystyle \sum_{c\in\mathcal C}\left|d_b^c-z'^c\right|\le T-\tau-1,
\\[3ex]
+\infty,&\text{otherwise}.
\end{cases}
\end{aligned}
\label{eq:side_automaton_update}
\end{equation}
As usual, the minimum over an empty set is $+\infty$.
The transition is deterministic because the complete successor map is
uniquely determined by the current cost map and the next input symbol.

Let
\begin{equation}
\mathfrak R_{i,\chi,\tau}=\left\{R_{i,\chi,\tau}^{\alpha}:\alpha\in\Sigma^\tau,\ 
R_{i,\chi,\tau}^{\alpha}\not\equiv+\infty
\right\}
\end{equation}
be the set of distinct reachable cost maps at layer $\tau$.
If two different boundary prefixes induce identical cost maps, they correspond to the same element of $\mathfrak R_{i,\chi,\tau}$ and are therefore merged into one automaton node. Equality here applies to the entire map over $(S,\mathbf z)$, including infeasible entries with value $+\infty$. It is stronger than equality of the count-vector states illustrated in Figure~\ref{fig:bsc_diagram}.

After reading a complete boundary word $\beta_b=(\sigma_{b,0},\ldots,\sigma_{b,T-1})\in\Sigma^T$, the resulting cost map is $R_{i,\chi,T}^{\beta_b}$. The word is accepted by the side automaton
if
\begin{equation}
R_{i,\chi,T}^{\beta_b}
\left(S_i^{\mathcal G},\mathbf d_b\right)
<+\infty.
\end{equation}
Its accepted weight is
\begin{equation}
W_{i,\chi}^T(\beta_b)
=
R_{i,\chi,T}^{\beta_b}
\left(S_i^{\mathcal G},\mathbf d_b\right).
\end{equation}
Consequently, the side automaton accepts exactly those boundary words
that are both cut-balanced and locally realizable, and assigns each
accepted word its minimum local SoC.

The distinct reachable cost maps are stored as nodes of a layered
weighted directed acyclic graph. Layer $\tau$ contains the maps in
$\mathfrak R_{i,\chi,\tau}$. For
$R_{i,\chi,\tau}^{\alpha}\in\mathfrak R_{i,\chi,\tau}$ and
$a\in\Sigma$, an arc labeled by $a$ connects
$R_{i,\chi,\tau}^{\alpha}$ to
$R_{i,\chi,\tau+1}^{\alpha a}$ whenever the successor map is not
identically $+\infty$. A root-to-terminal path therefore spells one
length-$T$ boundary word, while the terminal output gives its minimum
local SoC. This is a weighted finite-state representation in the sense
of \citet{mohri1997finite}.

To reduce this layered graph without changing the represented weighted language, let a node $p$ at layer $\tau$ represent the weighted continuation function
\begin{equation}
\phi_p:\Sigma^{T-\tau}\longrightarrow \mathbb R_{\ge0}\cup\{+\infty\},
\end{equation}
where $\phi_p(\omega)$ is the minimum final local SoC obtained by reading suffix $\omega$ from node $p$, and $\phi_p(\omega)=+\infty$ if that suffix cannot lead to an accepting terminal state. Two nodes $p$ and $q$ at the same layer are equivalent if
\begin{equation}
\phi_p(\omega)=\phi_q(\omega),
\ \forall\omega\in\Sigma^{T-\tau}.
\label{eq:bsdp_weighted_equivalence}
\end{equation}
Equivalent nodes are merged.

This reduction is exact because equivalent nodes agree on the feasibility and minimum cost of every remaining boundary suffix. The feasibility filtering and weighted-DAG reduction described above therefore preserve the realizable boundary words and their minimum local costs. They need not retain every individual local schedule realizing a boundary word.

Incumbent-based pruning is a separate optimization mechanism. Let $U$ be the cost of a retained feasible global schedule, and let $\underline{F}(p)$ be a lower bound on the total SoC of every complete schedule extending a partial search state $p$. The state $p$ can be discarded if $\underline{F}(p)\ge U$, because none of its completions can improve the incumbent. Such pruning may discard feasible schedules, but preserves the optimal objective value provided that the lower bound is valid and the incumbent is retained.

\subsection{Modules Chain Composition}\label{sec:chain_composition}

For a fixed makespan $T$, let $D_i^T(\beta_i)$ be the minimum cumulative SoC for modules $\mathcal M_1,\ldots,\mathcal M_i$ when their exposed right-boundary schedule is $\beta_i\in\mathcal B_i^T$. Initialize
\begin{equation}
D_0^T(\mathbf0_T)=0.
\label{eq:bsdp_chain_initial}
\end{equation}
For $i=1,2,\ldots,N$, the chain recurrence is
\begin{equation}
D_i^T(\beta_i)=\min_{\beta_{i-1}\in\mathcal B_{i-1}^T}
\left\{D_{i-1}^T(\beta_{i-1})+L_i^T(\beta_{i-1},\beta_i)\right\},
\ \beta_i\in\mathcal B_i^T.
\label{eq:bsdp_chain_recursion}
\end{equation}

Because $\mathcal B_N^T=\{\mathbf0_T\}$, the terminal value is $D_N^T(\mathbf0_T)$. It is finite if and only if a feasible length-$T$ global schedule exists. The recurrence is a min-plus composition: $L_i^T$ is the transfer relation contributed by module $\mathcal M_i$, and $\beta_i$ is the complete interface shared by the processed prefix and the remaining modules.

The chain evaluation is illustrated in the lower-left part of Figure~\ref{fig:mcc_diagram}, alongside the paired evaluation introduced next. Each step adds the next module's local transfer cost and minimizes over its shared boundary word. It retains the resulting cost separately for every exposed right-boundary word $\beta_i$.

\subsection{Paired Boundary-Language Composition} \label{sec:paired_composition}

The chain recurrence may be evaluated left to right. It may also be evaluated by recursively composing contiguous module intervals, which is useful when the boundary languages are stored as compressed weighted DAGs. For an interval $[a,b]$, define
\begin{equation}
\Phi_{a:b}^T(\beta_{a-1},\beta_b)=\min_{\substack{\beta_j\in\mathcal B_j^T\\j=a,a+1,\ldots,b-1}}\sum_{i=a}^{b}L_i^T(\beta_{i-1},\beta_i).
\label{eq:bsdp_interval_transfer}
\end{equation}

For a single module,
\begin{equation}
\Phi_{i:i}^T(\beta_{i-1},\beta_i)=L_i^T(\beta_{i-1},\beta_i).
\label{eq:bsdp_single_transfer}
\end{equation}

For any split index $k$ satisfying $a\le k<b$,
\begin{equation}
\Phi_{a:b}^T(\beta_{a-1},\beta_b)=\min_{\gamma\in\mathcal B_k^T}\left\{\Phi_{a:k}^T(\beta_{a-1},\gamma)+\Phi_{k+1:b}^T(\gamma,\beta_b)\right\}.
\label{eq:bsdp_paired_recursion}
\end{equation}
This operation is a min-plus join over the shared boundary word $\gamma$. Because the compressed weighted DAGs preserve the exact cost of every represented boundary word, evaluating the join on the DAG representation gives the same value as explicit enumeration.

\begin{figure}[!htbp]
    \centering
    \includegraphics[width=\linewidth]{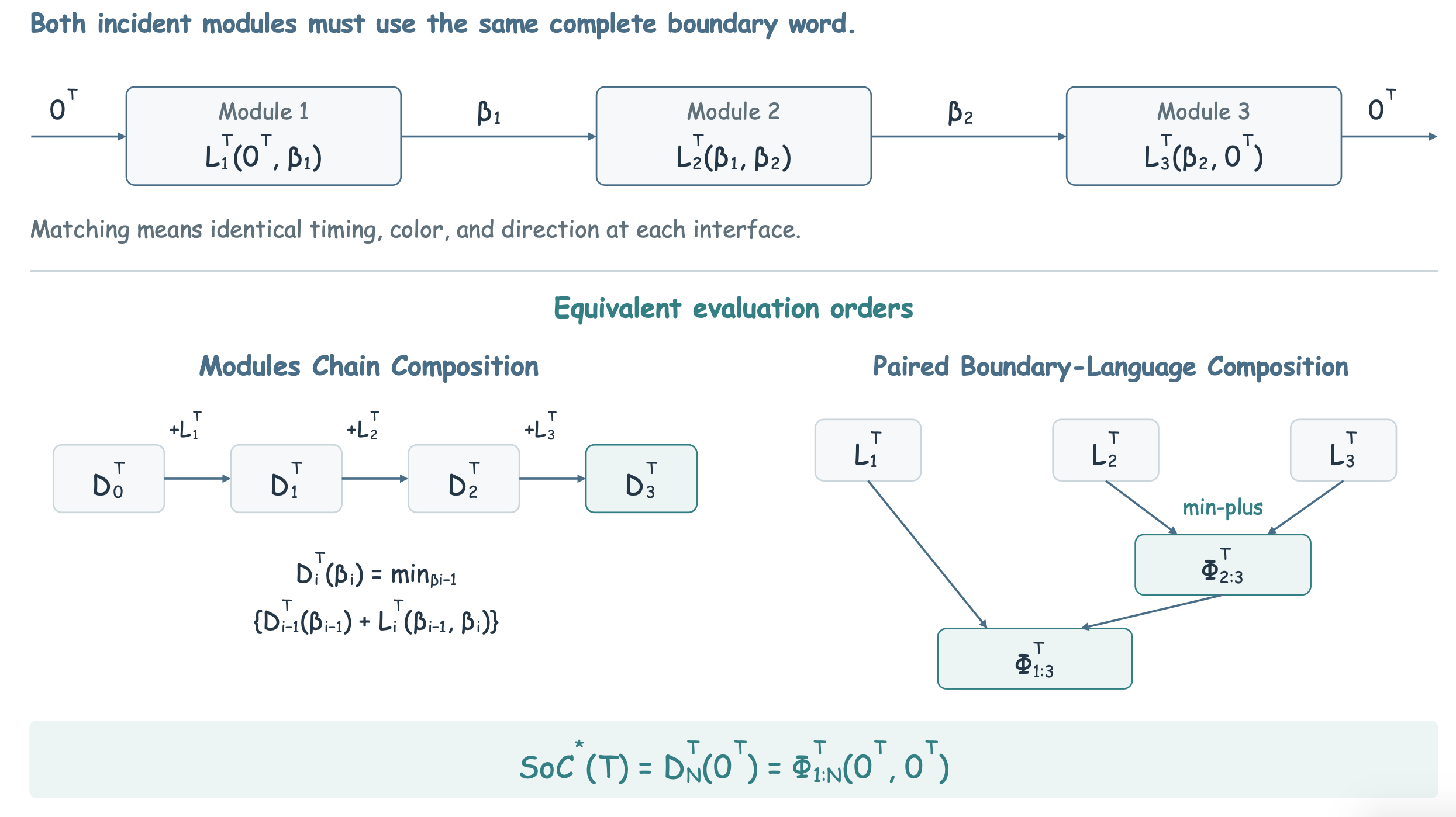}
    \caption{Chain and paired composition of local transfer costs.}
    \label{fig:mcc_diagram}
\end{figure}

For the entire vehicle,
\begin{equation}
\operatorname{SoC}^*(T)=\Phi_{1:N}^T(\mathbf0_T,\mathbf0_T)=D_N^T(\mathbf0_T).
\label{eq:bsdp_composition_value}
\end{equation}
For example, when $N=3$,
\begin{equation}
\operatorname{SoC}^*(T)=\min_{\substack{\beta_1\in\mathcal B_1^T\\
                   \beta_2\in\mathcal B_2^T}}
\left\{
L_1^T(\mathbf0_T,\beta_1)
+L_2^T(\beta_1,\beta_2)
+L_3^T(\beta_2,\mathbf0_T)
\right\}.
\label{eq:bsdp_three_modules}
\end{equation}
Figure~\ref{fig:mcc_diagram} illustrates~\eqref{eq:bsdp_three_modules} for $N=3$. The upper part shows that the same complete word $\beta_1$ is used by modules $\mathcal M_1$ and $\mathcal M_2$, and that the same word $\beta_2$ is used by modules $\mathcal M_2$ and $\mathcal M_3$. Matching requires identical events at every time step, including zero events. Equality of the final net-transfer vectors alone is insufficient. The notation $0^T$ in the figure denotes the all-zero boundary word $\mathbf0_T$ defined in Section~\ref{sec:boundary_schedule}; it is a length-$T$ word, not a scalar power.

In the lower-left chain diagram, the nodes labeled $D_0^T,D_1^T,D_2^T,D_3^T$ abbreviate $D_0^T(\mathbf0_T)$, $D_1^T(\beta_1)$, $D_2^T(\beta_2)$, and $D_3^T(\mathbf0_T)$, respectively. Each arrow labeled $+L_i^T$ includes the minimization over $\beta_{i-1}\in\mathcal B_{i-1}^T$ in~\eqref{eq:bsdp_chain_recursion}. Thus, the internal nodes represent functions of their exposed boundary words, rather than a single scalar cost obtained by summation alone.

In the lower-right paired diagram, the leaves abbreviate the three local transfer functions shown in the upper part. Modules $\mathcal M_2$ and $\mathcal M_3$ are combined first:
\[
\Phi_{2:3}^T(\beta_1,\mathbf0_T)
=\min_{\beta_2\in\mathcal B_2^T}
\left\{
L_2^T(\beta_1,\beta_2)+L_3^T(\beta_2,\mathbf0_T)
\right\}.
\]
The resulting interval transfer function is then combined with module $\mathcal M_1$:
\[
\Phi_{1:3}^T(\mathbf0_T,\mathbf0_T)
=\min_{\beta_1\in\mathcal B_1^T}
\left\{
L_1^T(\mathbf0_T,\beta_1)+\Phi_{2:3}^T(\beta_1,\mathbf0_T)
\right\}.
\]
The tree nodes $\Phi_{2:3}^T$ and $\Phi_{1:3}^T$ suppress these external boundary arguments. The two diagrams show alternative evaluation orders of the same min-plus composition, not consecutive stages of the algorithm. Their arrows represent composition dependencies rather than physical seat movements. Both evaluations return the value in~\eqref{eq:bsdp_composition_value}, as summarized in the figure's bottom strip.

For this three-module case, the middle-module search advances a product state $(S,\nu_L,\nu_R)$, where $S\in\mathcal S_2$ is the local module state and $\nu_L$ and $\nu_R$ are nodes of the two boundary-language DAGs. An outgoing product transition is allowed only when its two arc labels form a locally feasible pair of boundary symbols for the same time step.

\begin{theorem}[Exactness of paired composition]
For every interval $[a,b]$ and fixed makespan $T$, the paired recurrence computes the exact transfer cost $\Phi_{a:b}^T$. Consequently,
\begin{equation}
\Phi_{1:N}^T(\mathbf0_T,\mathbf0_T)=D_N^T(\mathbf0_T)=\operatorname{SoC}^*(T).
\end{equation}
\end{theorem}

\begin{proof}
The claim holds for a single-module interval because $\Phi_{i:i}^T=L_i^T$. For a longer interval $[a,b]$, choose any $k$ with $a\le k<b$. Every feasible interval schedule induces one shared boundary word $\gamma\in\mathcal B_k^T$ and separates into feasible schedules on $[a,k]$ and $[k+1,b]$. Conversely, two feasible subinterval schedules with the same $\gamma$ stitch into a feasible schedule on $[a,b]$. Their tail-owned costs add because the two intervals have disjoint vertex sets. Minimizing over $\gamma$ gives the stated recurrence. Induction on the interval length proves exactness. The weighted-DAG representation does not change the result because its merging rule preserves the feasibility and minimum continuation cost of every suffix.
\end{proof}

\subsection{Optimality Guarantee}\label{sec:bsdp_optimality}

\begin{theorem}[Optimality of BSDP]\label{thm:bsdp_exact}
For every fixed makespan $T$,
\begin{equation}
D_N^T(\mathbf0_T)
=\min_{\Pi\in\mathcal P_T(X^0,X^{\mathcal G})}
\operatorname{SoC}(\Pi)
=\operatorname{SoC}^*(T),
\label{eq:bsdp_fixed_T_optimum}
\end{equation}
where the minimum over an empty set is $+\infty$. Consequently, $D_N^T(\mathbf0_T)<+\infty$ if and only if a feasible length-$T$ schedule exists.
\end{theorem}

\begin{proof}
First consider a feasible global schedule $\Pi\in\mathcal P_T(X^0,X^{\mathcal G})$. Recording the color and direction of the crossing on each boundary $e_i^{\mathrm{bd}}$ at each time step produces a unique word $\beta_i$. By Lemma~\ref{lemma:boundary}, $\beta_i\in\mathcal B_i^T$. Restricting every state and simultaneous action set of $\Pi$ to module $\mathcal M_i$ produces a local schedule consistent with $\beta_{i-1}$ and $\beta_i$. It satisfies the six local conditions in Section~\ref{sec:local_boundary}; hence, its local cost is at least $L_i^T(\beta_{i-1},\beta_i)$. Because $\mathcal W_1,\ldots,\mathcal W_N$ partition $\mathcal V_{N,m}$, every global action has its tail in exactly one module. The tail-ownership rule therefore gives
\begin{equation}
\operatorname{SoC}(\Pi)
\ge\sum_{i=1}^{N}L_i^T(\beta_{i-1},\beta_i).
\end{equation}

Conversely, fix boundary words $\beta_i\in\mathcal B_i^T$ and, for every module, a local schedule attaining the finite value $L_i^T(\beta_{i-1},\beta_i)$. Adjacent modules prescribe the same symbol $\sigma_{i,t}$ on their shared edge. Thus, an outbound boundary action in one module is identical in time, direction, and color to the inbound action seen by the neighboring module. The local tail-completeness and update conditions reconstruct one global seat action for every occupied vertex; the local vertex and following conditions check each destination in the module containing that destination; and the shared boundary symbol enforces the boundary-edge rule. Therefore, the local schedules stitch into a feasible global schedule on $\mathcal G_{N,m}$. By tail ownership, its SoC is exactly
\begin{equation}
\sum_{i=1}^{N}L_i^T(\beta_{i-1},\beta_i).
\end{equation}

It remains to minimize this sum over the internal boundary words. The recurrence for $D_i^T$ performs precisely this minimization, one module at a time. Induction on $i$ shows that $D_i^T(\beta_i)$ is the minimum cost of all locally compatible schedules for modules $\mathcal M_1,\ldots,\mathcal M_i$ exposing $\beta_i$. At $i=N$, the exterior schedule is $\mathbf0_T$, so $D_N^T(\mathbf0_T)=\operatorname{SoC}^*(T)$.

Finally, the count-vector DFA recognizes exactly $\mathcal B_i^T$, and the weighted-DAG merging criterion preserves the realizability and minimum cost of each represented boundary word. Thus, these representations preserve the value of the exact recurrence.

If incumbent-based pruning is additionally used, every discarded search state must have a valid lower bound on the total completion cost that is no smaller than the cost of a retained feasible incumbent. Consequently, no discarded state can yield a strictly better solution. Once the remaining search is completed, the retained incumbent therefore attains the same optimum as the unpruned recurrence.
\end{proof}

The theorem establishes exactness for a fixed $T$ when the required minimizations are completed. If the outer search starts from a valid lower bound and certifies infeasibility at every smaller tested horizon, its first feasible horizon is $T^*$. Completing the fixed-$T^*$ optimization then yields the lexicographically optimal objective pair $(T^*,\operatorname{SoC}^*(T^*))$. A feasible incumbent found before optimization is complete provides an upper bound on SoC, but does not by itself establish SoC optimality. Likewise, interruption without a feasibility certificate does not establish infeasibility.

An optimal schedule can be recovered by retaining minimizing boundary choices in the composition recurrences, predecessor states in~\eqref{eq:bsdp_local_recursion}, and local action sets attaining~\eqref{eq:bsdp_local_step_cost}. Backtracking first recovers the shared boundary words and then each module's local actions. The boundary-consistency and tail-ownership arguments above ensure that these local schedules stitch into a feasible global schedule with total cost $\operatorname{SoC}^*(T)$.

\section{Experiments}\label{sec:exp}

In this section, we compare the proposed BSDP Algorithm with a mixed-integer linear programming formulation solved by Gurobi. Both methods optimize the same lexicographic objective: first minimizing the makespan and then minimizing the sum-of-costs under the optimal makespan. All experiments were conducted on a MacBook Pro equipped with an Apple M3 Pro processor with 11 CPU cores (5 performance cores and 6 efficiency cores) and 18 GB of unified memory. The operating system was macOS 26.3. The proposed algorithm was implemented in C++ and compiled using GCC 15.2.0 with OpenMP enabled. Gurobi Optimizer 13.0.0 was used as the MILP solver.

We do not report a comparison between BSDP and exact solvers for colored MAPF. Existing exact colored MAPF methods, such as CBM~\citep{ma2016optimal} and ITA-CBS~\citep{tang2023solving}, are fundamentally based on conflict-based search. In the present problem, the narrow central aisle and the limited inter-module connections create a large number of corridor and bottleneck conflicts. Resolving these conflicts may require extensive branching at the high level of a conflict-based search, causing the conflict tree to grow exponentially in the worst case. Consequently, such methods are not computationally competitive with BSDP for the problem structure considered in this study, and their inclusion would provide little informative value as a numerical baseline. Nevertheless, to assess their applicability, we conducted preliminary
experiments using CBM under the same computational environment and time limit as the other methods. CBM failed to solve even the smallest test instances within the 1800 s time limit. We therefore do not include conflict-based colored-MAPF solvers as baselines in the reported computational results.

\subsection{Data Generation}\label{sec:data_generation}

For a prescribed pair $(N,m)$, we generate benchmark instances using the $3\times mN$ modular-bus layout defined in Section~\ref{sec:problem_des_form}. The state matrix has vertex set $\mathcal V_{N,m}$, color set $\mathcal C=\{1,2,\cdots,N\}$, and $2mN$ movable seats. Each color occurs exactly $2m$ times. In both the initial and target state matrices, movable seats are placed only in the non-aisle rows, while all middle-row vertices are empty: $X[2,j]=0,\ j=1,2,\cdots,mN$. The middle row therefore provides the $mN$ empty track positions used for seat movement.

The target state matrix $X^{\mathcal G}$ is deterministic. For each color $c\in\mathcal C$, color-$c$ seats occupy the top and bottom non-aisle positions of module $\mathcal M_c$. Thus,
\begin{equation}
X^{\mathcal G}[v]=
\begin{cases}
c, & v\in\mathcal G_c,\ c\in\mathcal C,\\
0, & v=(2,j),\ j=1,2,\ldots,mN.
\end{cases}
\end{equation}

Let $\mathcal V_{N,m}^{\mathrm{seat}}=\{(r,j)\in\mathcal V_{N,m}:r\in\{1,3\}\}$ be the set of non-aisle vertices. We first create $N$ unlabeled groups indexed by $\mathcal K=\{1,2,\ldots,N\}$, with each group containing exactly $2m$ seats. Using a fixed random seed, we randomly permute the multiset containing $2m$ copies of each group index. The first $mN$ entries of the resulting sequence are placed on the top row from left to right, and the remaining $mN$ entries are placed on the bottom row from left to right. This defines a group-placement map $h:\mathcal V_{N,m}^{\mathrm{seat}}\rightarrow\mathcal K$. For each group $k\in\mathcal K$, let $\mathcal R_k=\{v\in\mathcal V_{N,m}^{\mathrm{seat}}:h(v)=k\}$. By construction, $|\mathcal R_k|=2m$ for every $k\in\mathcal K$.

The random shuffling step fixes the spatial positions of the $N$ groups but does not yet assign final colors. For each group $k\in\mathcal K$ and each target color $c\in\mathcal C$, define
\begin{equation}
\Gamma_{kc}
=\min_{\mu\in\operatorname{Bij}(\mathcal R_k,\mathcal G_c)}
\sum_{u\in\mathcal R_k}
\operatorname{dist}_{\mathcal G_{N,m}}\bigl(u,\mu(u)\bigr),
\end{equation}
where $\operatorname{Bij}(\mathcal R_k,\mathcal G_c)$ is the set of bijections from the $2m$ current positions of group $k$ to the $2m$ target vertices of color $c$, and $\operatorname{dist}_{\mathcal G_{N,m}}(u,v)$ is the shortest-path distance on $\mathcal G_{N,m}$.

Let $\mathfrak S_N$ denote the set of one-to-one assignments from $\mathcal K$ to $\mathcal C$. The target-aware color assignment is
\begin{equation}
\pi^\star\in\arg\min_{\pi\in\mathfrak S_N}
\sum_{k\in\mathcal K}\Gamma_{k,\pi(k)}.
\end{equation}
Given $\pi^\star$, the initial state matrix is
\begin{equation}
X^0[v]=
\begin{cases}
\pi^\star(h(v)), & v\in\mathcal V_{N,m}^{\mathrm{seat}},\\
0, & v=(2,j),\ j=1,2,\cdots,mN.
\end{cases}
\end{equation}
If this assignment produces $X^0=X^{\mathcal G}$, we instead select the best non-target assignment:
\begin{equation}
\pi^\star\in
\arg\min_{\substack{\pi\in\mathfrak S_N:\ X^0_\pi\neq X^{\mathcal G}}}
\sum_{k\in\mathcal K}\Gamma_{k,\pi(k)},
\end{equation}
where
\begin{equation}
X^0_\pi[v]=
\begin{cases}
\pi(h(v)), & v\in\mathcal V_{N,m}^{\mathrm{seat}},\\
0, & v=(2,j),\ j=1,2,\cdots,mN.
\end{cases}
\end{equation}
This procedure preserves the $3\times mN$ grid structure, the count of $2m$ seats per color, and the empty middle aisle for every tested pair $(N,m)$.

\subsection{Main Results}

We compare the our BSDP with the fixed-makespan mixed-integer programming formulation solved by Gurobi (Hereinafter referred to as Gurobi). The benchmark instances are generated according to the data-generation procedure described in Section~\ref{sec:data_generation}. Most existing MB studies treat the number of modular units in a coupled vehicle as a capacity decision rather than as a fixed constant. Typical operational settings impose an upper bound of approximately five to six units per coupled vehicle, while some conceptual fleet-size studies allow platoons with up to ten modules. In contrast, the present study focuses on the exact optimization of
in-vehicle seat reconfiguration after modules are coupled. This problem is combinatorially more demanding than capacity-level dispatching. For the main benchmark, we fix the module-width parameter at $m=3$ and vary the number of coupled modules. We use $N=2$ and $N=3$ as controlled benchmark classes: $N=2$ captures the canonical two-module transfer process, whereas $N=3$ is the smallest setting with multiple inter-module boundaries. For each class, we generate 50 independent instances on the $3\times3N$ physical track layout, with six seats per color and an empty middle aisle.

For each instance and each solver, we perform 10 repeated runs under a time limit of 1800 seconds per run. We report the makespan $T^*$, the $\operatorname{SoC}^*$, and the runtime. Runtime is reported as the empirical mean together with the 95\% confidence interval computed over the 10 runs. For instance $i$, let $\bar{\tau}^{\mathrm{BSDP}}_i$ and $\bar{\tau}^{\mathrm{Gurobi}}_i$ denote the mean runtimes of BSDP and Gurobi, respectively. The per-instance speedup is defined as $\mathrm{Speedup}_i = \bar{\tau}^{\mathrm{Gurobi}}_i/\bar{\tau}^{\mathrm{BSDP}}_i$. Thus, a speedup greater than one indicates that BSDP is faster than Gurobi. When BSDP and Gurobi reach the time limit, the corresponding runtime is counted as 1800 seconds in the runtime summary.

\begin{figure}[h!]
    \centering
    \begin{subfigure}[t]{0.9\textwidth}
        \centering
        \includegraphics[width=\textwidth]{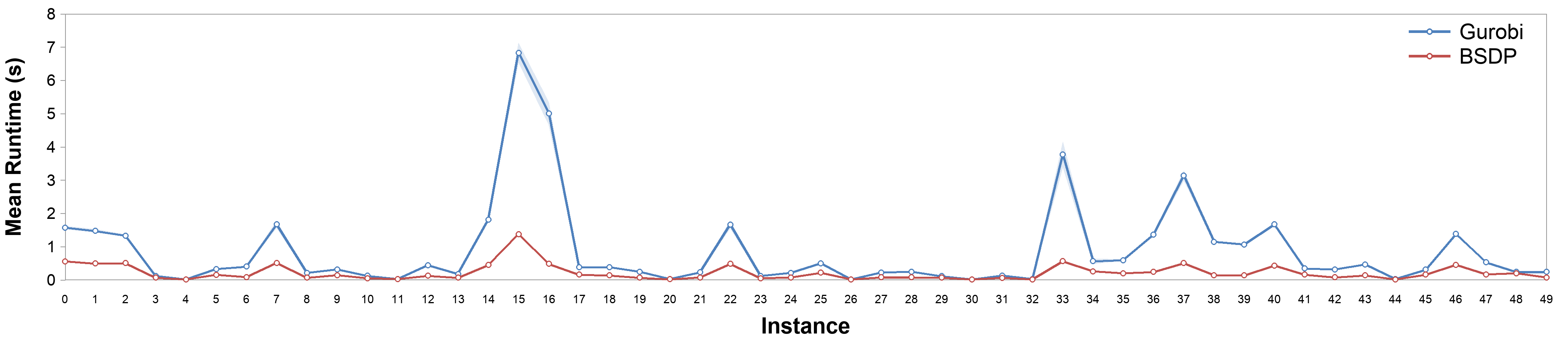}
        \caption{2 modules.}
        \label{fig:runtime_2modules}
    \end{subfigure}
    \begin{subfigure}[t]{0.9\textwidth}
        \centering
        \includegraphics[width=\textwidth]{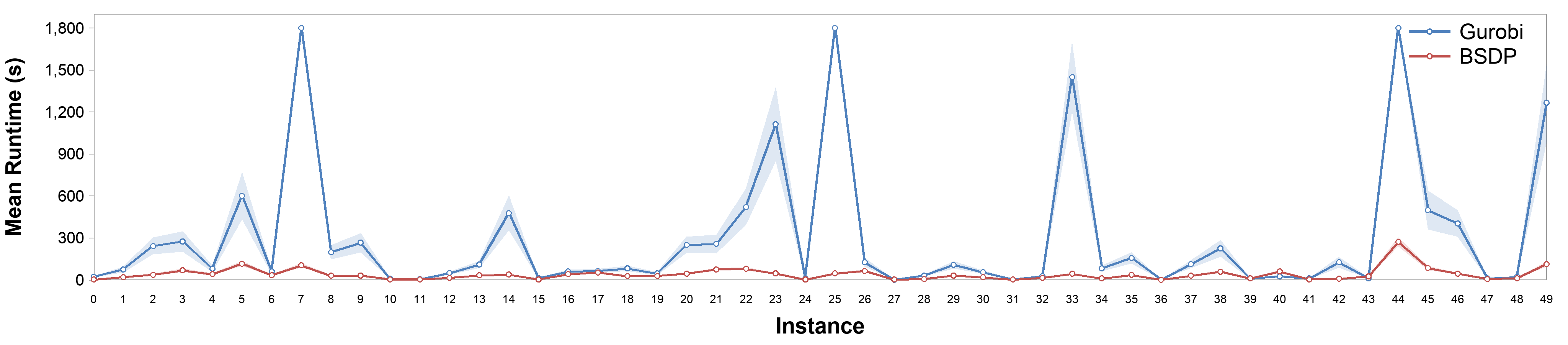}
        \caption{3 modules.}
        \label{fig:runtime_3modules}
    \end{subfigure}

    \caption{Runtime comparison between BSDP and Gurobi.}
    \label{fig:runtime_comparison}
\end{figure}

Figure~\ref{fig:runtime_comparison} reports the runtime comparison between BSDP and Gurobi for the two benchmark classes. For the $N=2$ instances, Figure~\ref{fig:runtime_2modules} shows that for the $N=2$ modules, BSDP solves the instances with an average runtime of 0.213s, whereas Gurobi requires 0.869s on average. The average per-instance speedup is $3.24\times$. Figure~\ref{fig:runtime_3modules} shows that for the $N=3$ modules, BSDP requires 39.238s on average, whereas Gurobi requires 300.503s on average when time-limit runs are counted as 1800s. The corresponding average per-instance speedup is $6.30\times$. The detailed results shown in the Appendix~\ref{app:detailed_res_2} and Appendix~\ref{app:detailed_res_3}.

\begin{table}[h!]
\centering
\small
\caption{Runtime performance of Gurobi and BSDP on the two- and three-module benchmarks ($m=3$).}
\label{tab:runtime_summary}
\begin{tabular}{llrrrrr}
\toprule
Method & Modules & Mean Runtime (s) & Median Runtime (s) & Max Runtime (s) & CV & Timeout Rate \\
\midrule
Gurobi & 2 modules & 0.869 & 0.335 & 6.828 & 1.516 & 0.0\% \\
BSDP   & 2 modules & 0.213 & 0.138 & 1.375 & 1.137 & 0.0\% \\
Gurobi & 3 modules & 300.503 & 81.972 & 1800.000 & 1.645 & 6.0\% \\
BSDP   & 3 modules & 39.238 & 31.024 & 270.900 & 1.136 & 0.0\% \\
\bottomrule
\end{tabular}
\end{table}

Table~\ref{tab:runtime_summary} summarizes the main runtime results for the two- and three-module benchmarks. For the two-module instances, both methods solve all cases within the 1800-second time limit. BSDP reduces the mean runtime from 0.869~s to 0.213~s and achieves a mean per-instance speedup of $3.24\times$. The advantage is more pronounced on the three-module instances, where BSDP reduces the mean runtime from 300.503~s to 39.238~s and achieves a mean per-instance speedup of $6.30\times$.

BSDP solves all three-module instances within the time limit, whereas Gurobi reaches the limit on $6.0\%$ of the instances. Runs reaching the time limit are recorded as 1800~s when computing the reported statistics. BSDP also has lower median runtimes and coefficients of variation in both benchmark classes, indicating shorter typical runtimes and less relative dispersion in the recorded runtimes across instances.

\subsection{Cases with $N\ (N\geq 4)$ modules.}

To assess performance beyond the three-module setting, we conduct fixed-$T$ improvement experiments with $N\in\{4,5,6,7\}$ and the baseline module width $m=3$. Each size class contains 50 instances, giving 200 instances in total. For each instance, a greedy procedure generates a feasible initial schedule, whose horizon $T$ is then fixed throughout the comparison. BSDP and Gurobi receive identical initial trajectories and are each allocated 11 threads and a 300-s time limit. These experiments therefore evaluate their ability to improve SoC and strengthen lower bounds under the same initial solution, fixed horizon, and computational budget. Since the greedy horizon is not necessarily makespan-optimal, the results assess the fixed-horizon SoC optimization component rather than certifying the full lexicographic optimum. Complete instance-level results are reported in Appendices~\ref{app:detailed_res_4}-\ref{app:detailed_res_7}.

Table~\ref{tab:large_modules_summary} summarizes the final solution quality and lower bounds. For four modules, BSDP obtains a lower final SoC than Gurobi on 47 instances and matches Gurobi on the remaining three. For each of the five-, six-, and seven-module classes, BSDP obtains a lower final SoC on all 50 instances. Overall, BSDP improves upon Gurobi on 197 of the 200 instances, with three ties and no worse final solutions. Its mean final SoC is 173.28, 307.32, 447.66, and 611.84 for four through seven modules, respectively, compared with 211.14, 716.14, 1249.18, and 1978.44 for Gurobi. The corresponding reductions relative to Gurobi are 17.93\%, 57.09\%, 64.16\%, and 69.07\%. These percentages are calculated as the difference between the two mean final SoC values divided by the Gurobi mean, rather than as the mean of instance-level percentage reductions.

\begin{table}[H]
\centering
\small
\setlength{\tabcolsep}{5pt}
\caption{Fixed-horizon results for four to seven modules.
Each row summarizes 50 instances at the 300-s time limit.
Reduction denotes the percentage decrease in mean final SoC
relative to Gurobi. LB denotes the reported lower bound.}
\label{tab:large_modules_summary}
\begin{tabular}{crrrrrr}
\toprule
$N$ & Mean initial
    & \multicolumn{2}{c}{Mean final SoC}
    & Reduction
    & \multicolumn{2}{c}{Mean final LB} \\
\cmidrule(lr){3-4}
\cmidrule(lr){6-7}
    & SoC & BSDP & Gurobi & (\%) & BSDP & Gurobi \\
\midrule
4 &  356.10 & 173.28 &  211.14 & 17.93 & 144.50 & 136.74 \\
5 &  908.54 & 307.32 &  716.14 & 57.09 & 206.24 & 197.46 \\
6 & 1409.90 & 447.66 & 1249.18 & 64.16 & 268.22 & 262.94 \\
7 & 2082.54 & 611.84 & 1978.44 & 69.07 & 335.74 & 339.40 \\
\bottomrule
\end{tabular}
\end{table}

The comparison with the common greedy initial schedules further highlights the difference in improvement capability. For four through seven modules, BSDP reduces the aggregate initial SoC by 51.34\%, 66.17\%, 68.25\%, and 70.62\%, respectively. The corresponding reductions for Gurobi are 40.71\%, 21.18\%, 11.40\%, and 5.00\%. Here, the aggregate reduction is calculated from the total initial and final SoC over the 50 instances in each size class. Thus, within the prescribed time budget, BSDP maintains substantial improvements over the initial schedules across all four classes, whereas Gurobi achieves progressively smaller aggregate reductions on the larger tested classes.

The lower-bound comparison exhibits a different pattern. BSDP reports a higher final lower bound than Gurobi on 45, 50, and 47 instances for four, five, and six modules, respectively. Their mean final lower bounds are 144.50 versus 136.74, 206.24 versus 197.46, and 268.22 versus 262.94. For seven modules, however, Gurobi reports a higher final lower bound on 39 instances, BSDP reports a higher bound on nine, and the bounds are equal on two. The corresponding means are 339.40 for Gurobi and 335.74 for BSDP. Hence, BSDP's advantage in final solution quality does not imply uniformly stronger lower bounds. In particular, the seven-module results show that BSDP can produce substantially better feasible schedules even when Gurobi provides slightly stronger bounds on average.

\begin{figure*}[p]
\centering



\begin{subfigure}[t]{0.48\textwidth}
\centering
\includegraphics[width=\linewidth,height=0.25\textheight,keepaspectratio]{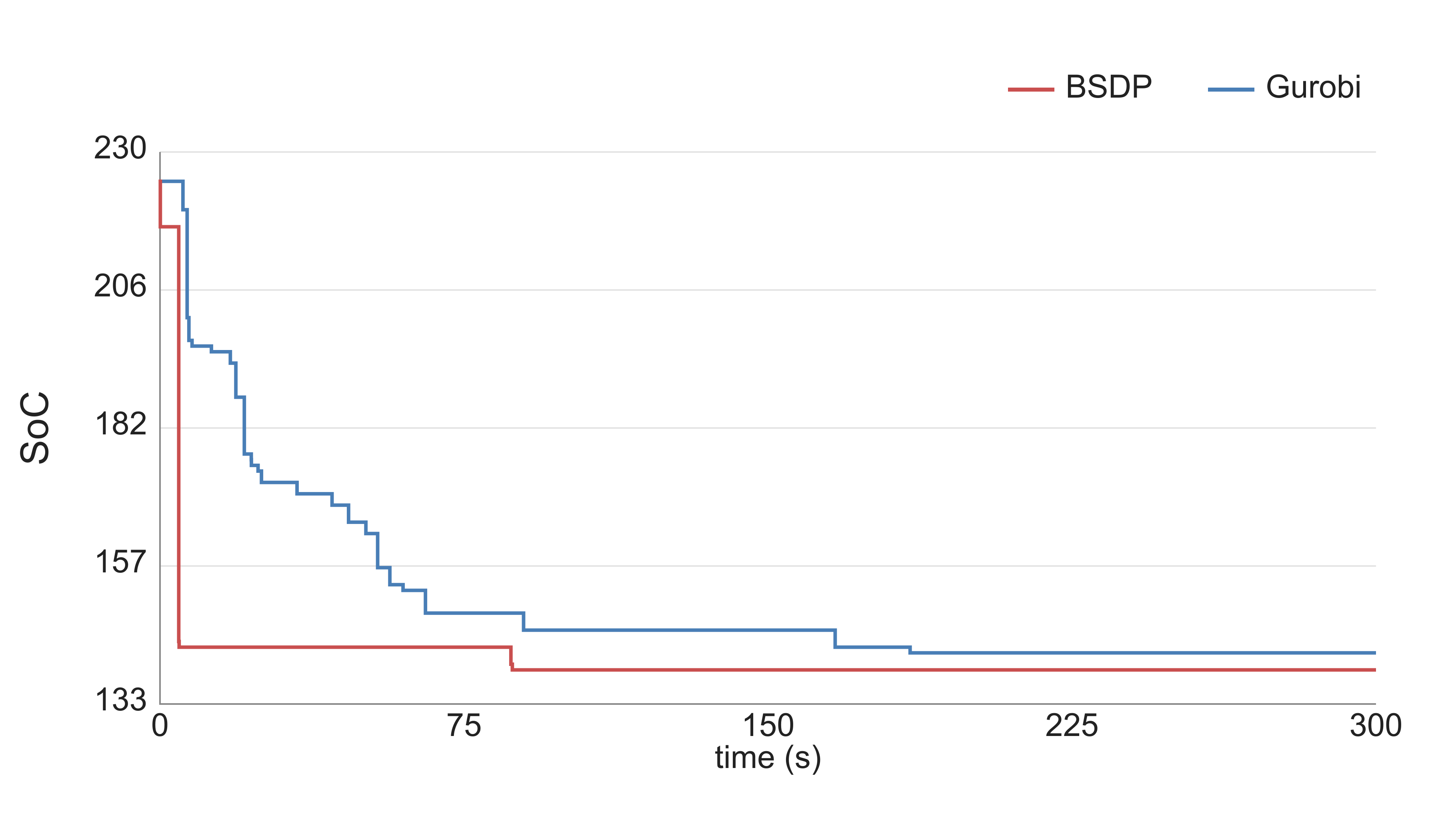}
\caption{Instance~19: SoC}
\end{subfigure}\hfill
\begin{subfigure}[t]{0.48\textwidth}
\centering
\includegraphics[width=\linewidth,height=0.25\textheight,keepaspectratio]{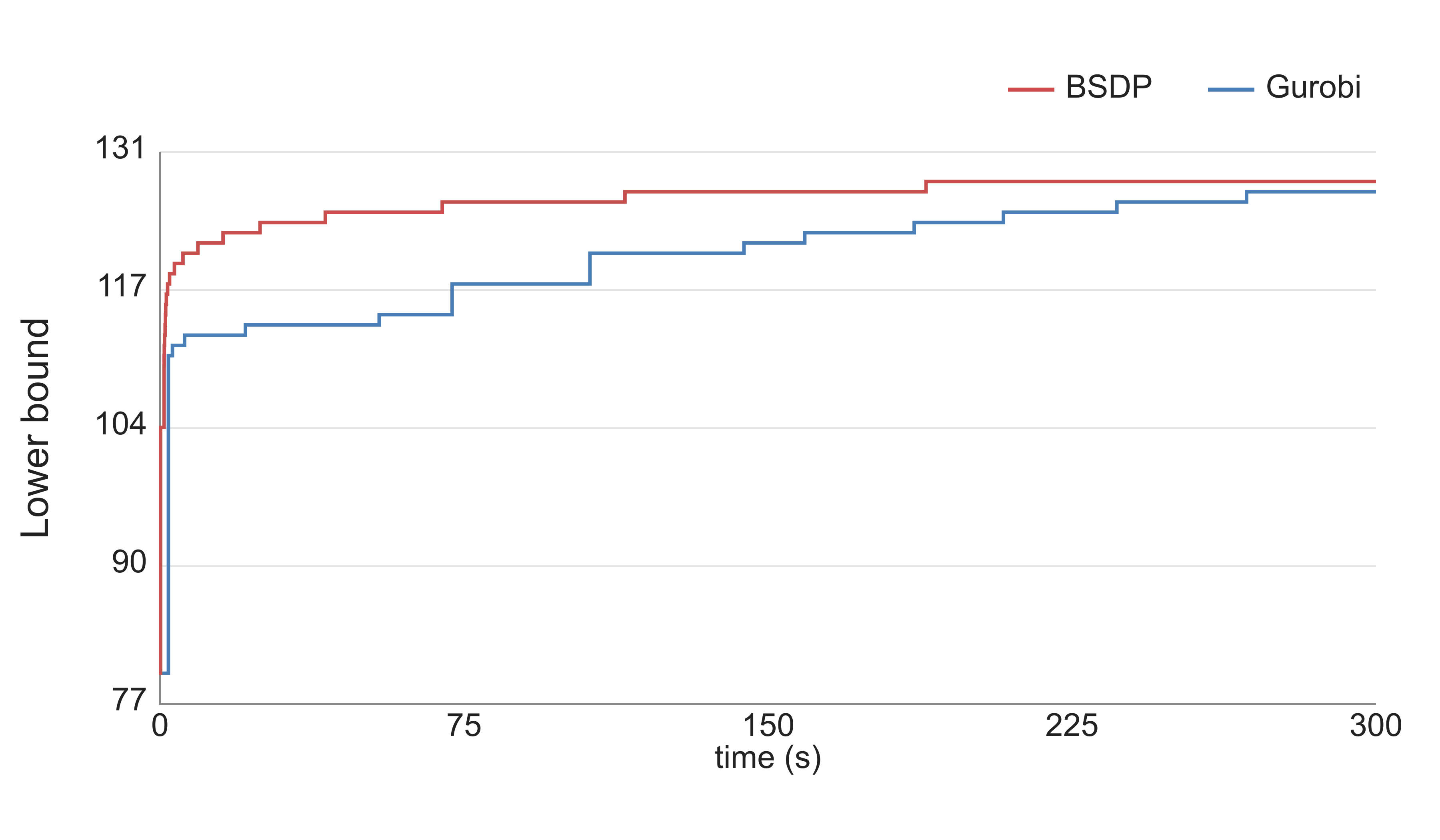}
\caption{Instance~19: Lower bound}
\end{subfigure}
\par\medskip

\begin{subfigure}[t]{0.48\textwidth}
\centering
\includegraphics[width=\linewidth,height=0.25\textheight,keepaspectratio]{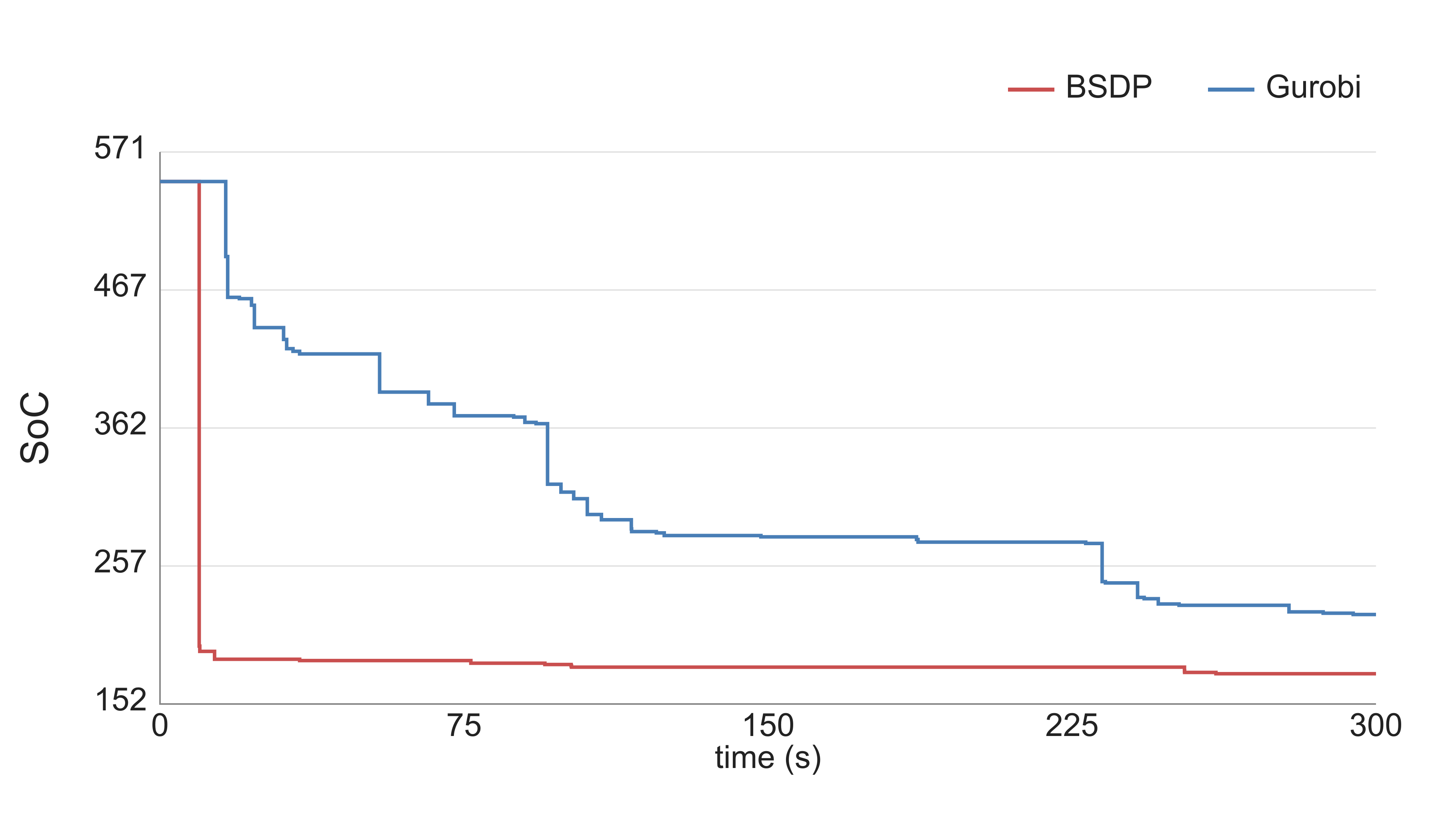}
\caption{Instance~20: SoC}
\end{subfigure}\hfill
\begin{subfigure}[t]{0.48\textwidth}
\centering
\includegraphics[width=\linewidth,height=0.25\textheight,keepaspectratio]{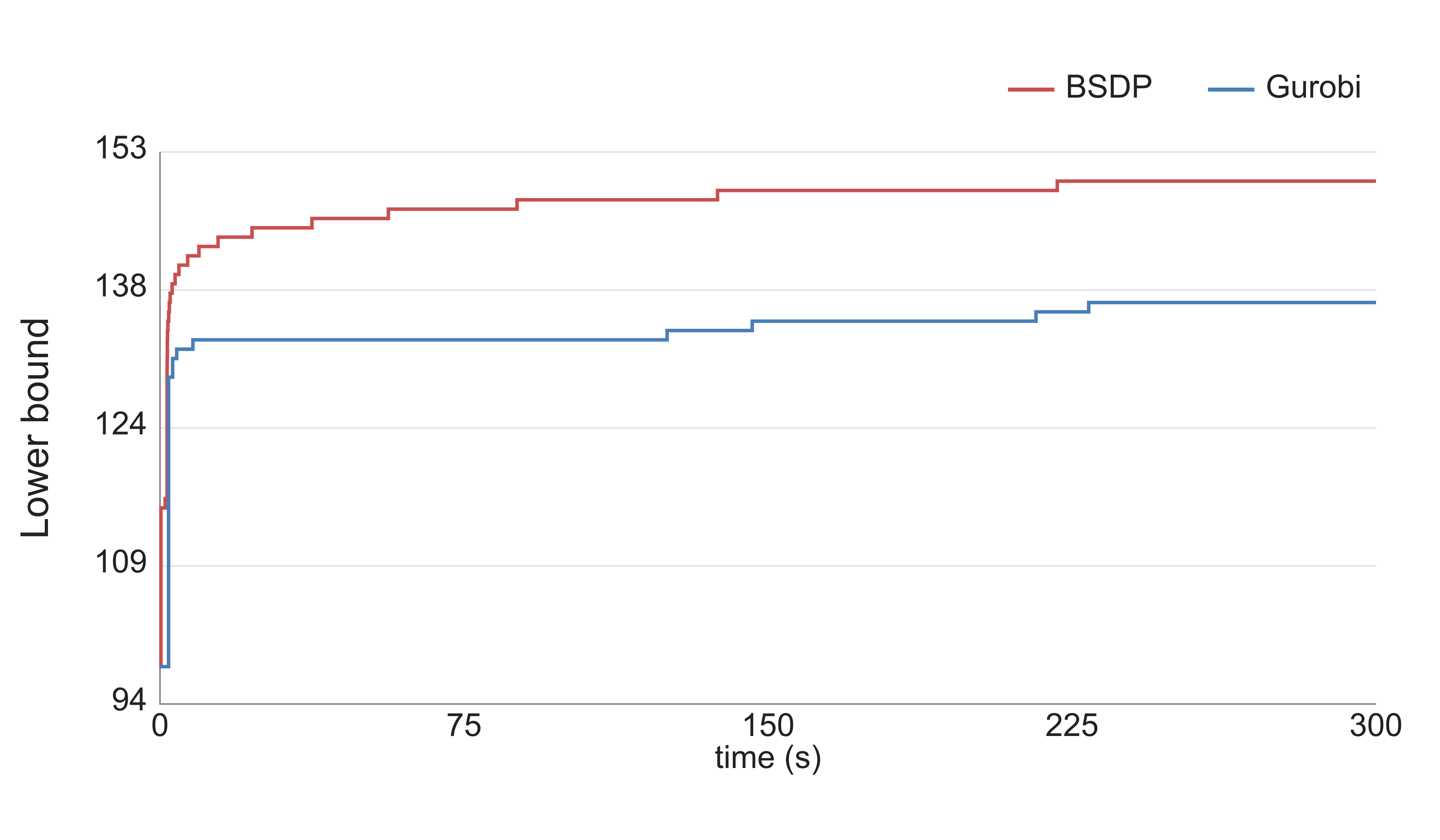}
\caption{Instance~20: Lower bound}
\end{subfigure}
\par\medskip

\begin{subfigure}[t]{0.48\textwidth}
\centering
\includegraphics[width=\linewidth,height=0.25\textheight,keepaspectratio]{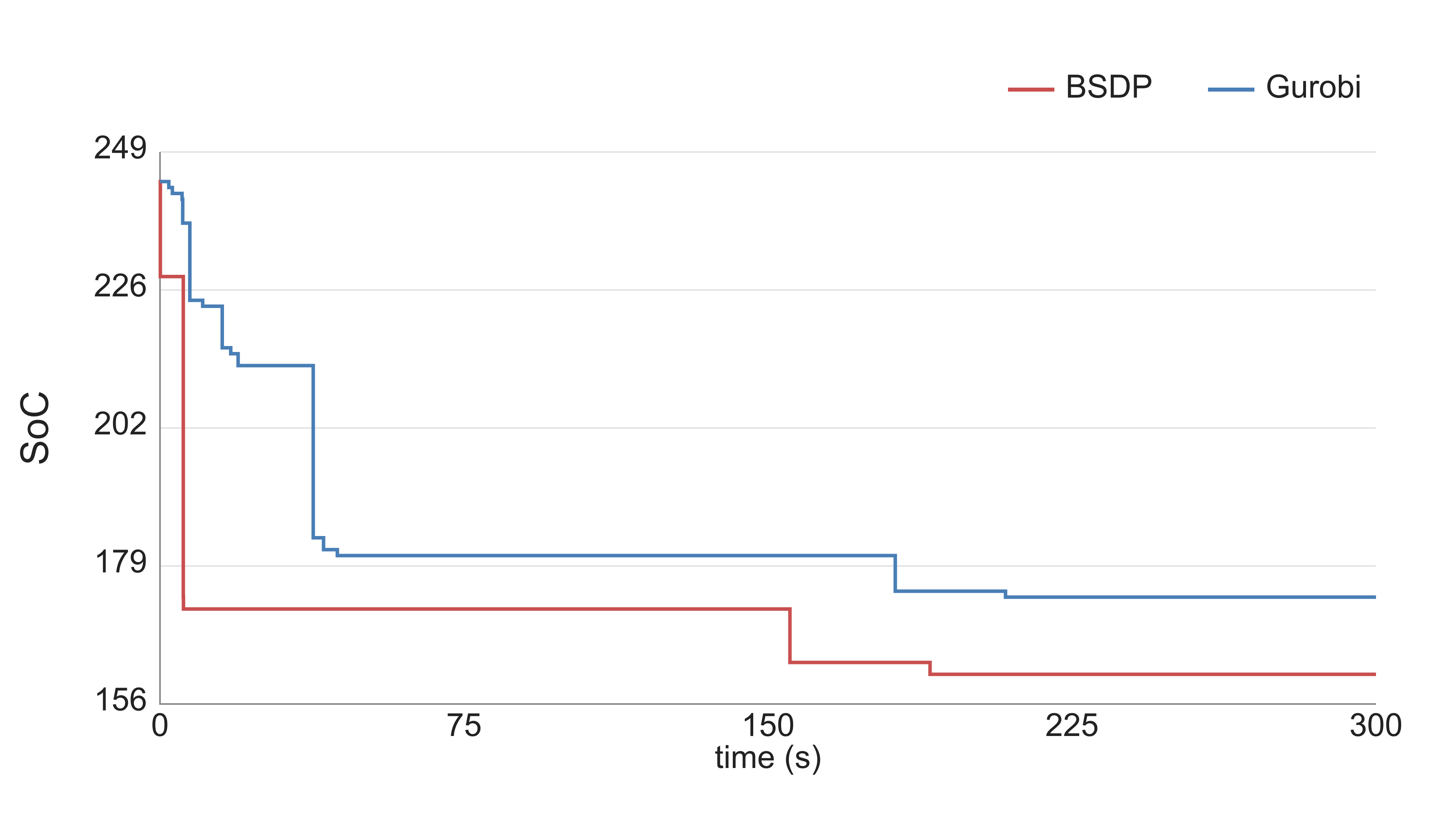}
\caption{Instance~22: SoC}
\end{subfigure}\hfill
\begin{subfigure}[t]{0.48\textwidth}
\centering
\includegraphics[width=\linewidth,height=0.25\textheight,keepaspectratio]{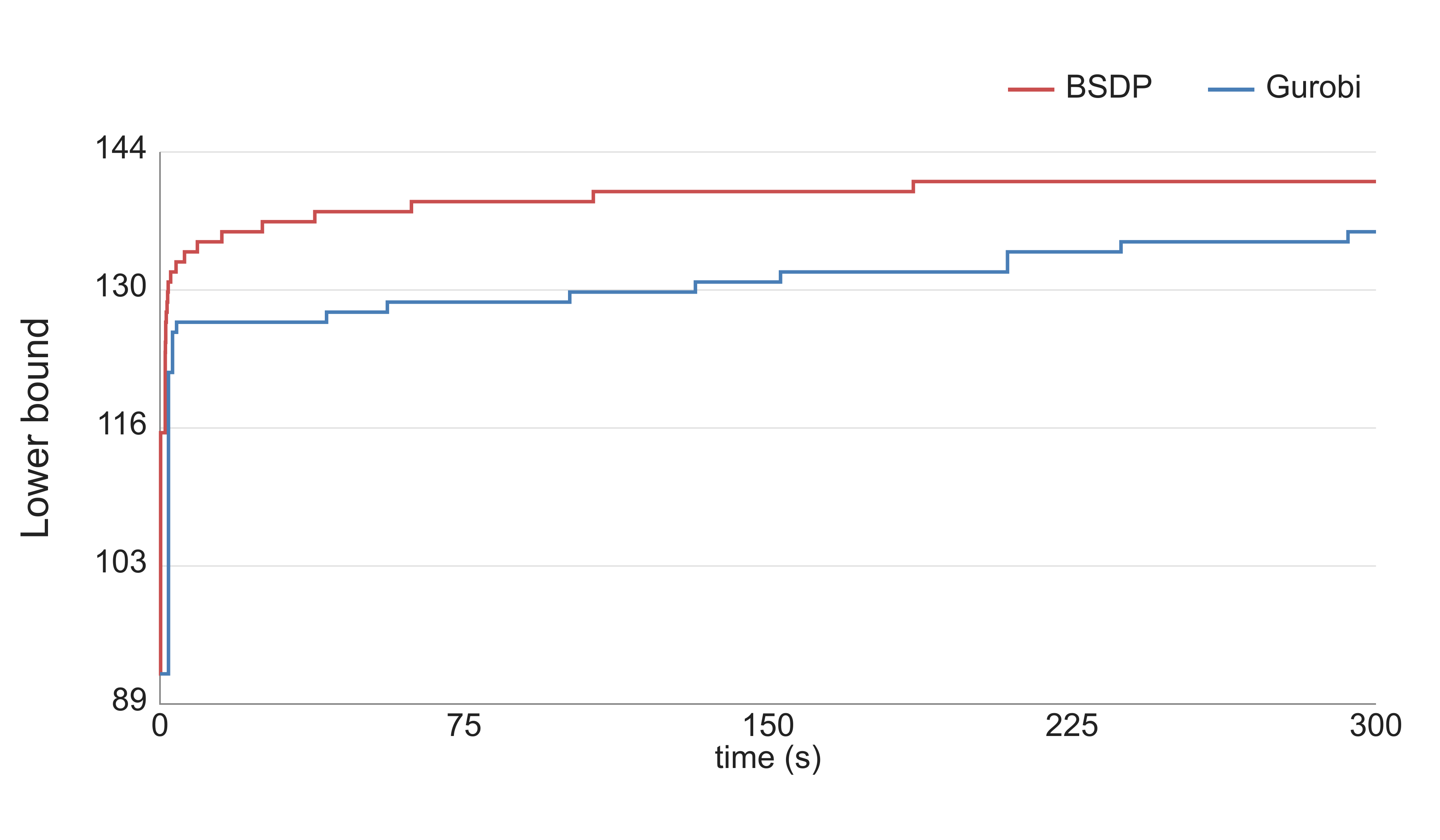}
\caption{Instance~22: Lower bound}
\end{subfigure}

\caption{Convergence of the incumbent SoC (left) and lower bound (right) for three selected four-module instances. For each instance, BSDP and Gurobi use the same greedy initial solution and fixed horizon, with 11 threads and a 300-s time limit. Gurobi lower bounds are conservatively extracted from solver logs.}
\label{fig:four_module_convergence}
\end{figure*}

Figure~\ref{fig:four_module_convergence} illustrates the search process using three selected four-module instances, namely instances 19, 20, and 22. The left column shows the incumbent SoC, and the right column shows the lower bound. In these examples, BSDP achieves substantial SoC reductions early in the search and subsequently improves its incumbent solutions and lower bounds. These trajectories illustrate its ability to provide improved feasible schedules before the time limit, complementing the final-value comparisons across the full dataset.

Taken together, the results demonstrate a consistent advantage for BSDP in fixed-horizon solution improvement on the evaluated four- to seven-module instances. The relative advantage in mean final SoC increases across the tested size classes. Nevertheless, the remaining differences between incumbent costs and lower bounds show that certifying optimality remains challenging under the 300-s budget, particularly for the larger instances.

\subsection{Sensitivity to the Module-Width Parameter $m$}
The generalized formulation treats the number of columns per modular unit, $m$, as a vehicle-design parameter. A $3\times m$ unit provides $2m$ seat positions and $m$ central-aisle positions. Consequently, $m$ affects both vehicle capacity and computational complexity: increasing $m$ adds two seats per unit but also enlarges the local state space and the number of feasible intra-module movement patterns.

We evaluate $m\in\{2,3,4\}$. The main benchmark uses the baseline width $m=3$, while this subsection reports the additional $m=2$ and $m=4$ cases. Specifically, the additional benchmarks contain two- and three-module instances for $m=2$ and two-module instances for $m=4$. The tested $(N,m)$ combinations are intended to demonstrate the applicability of the generalized formulation and the sensitivity of computational performance to module geometry; they do not constitute a full factorial experiment over all combinations of $N$ and $m$.

For each additional benchmark setting, we generate 50 random instances using the parameterized procedure in Section~\ref{sec:data_generation}. BSDP and Gurobi use the same physical layout, initial and target states, lexicographic objective, thread count, and time limit. Both methods obtain the same optimal objective values on all instances solved to optimality, so the comparison focuses on computational efficiency.

\begin{figure}[h!]
    \centering
    \begin{subfigure}[t]{0.9\textwidth}
        \centering
        \includegraphics[width=\textwidth]{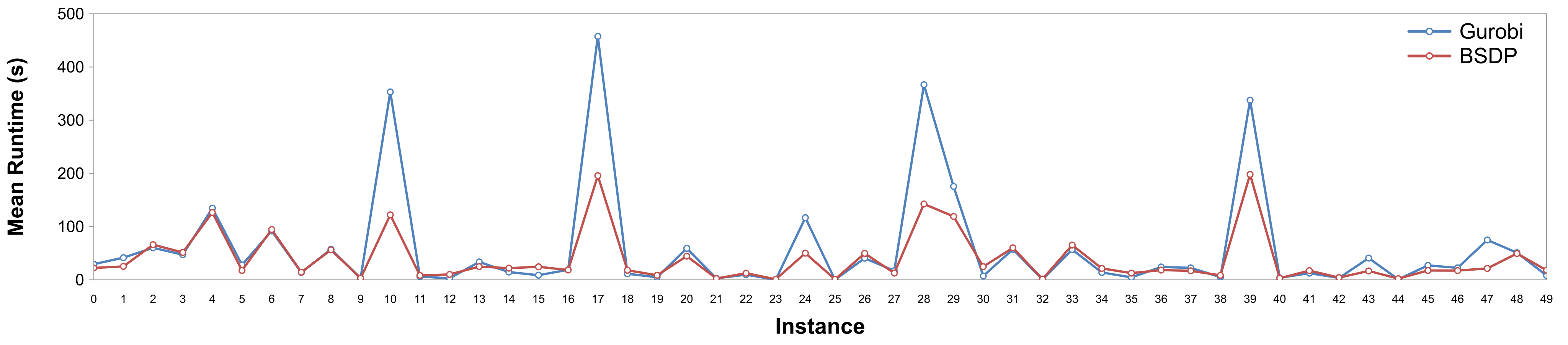}
        \caption{$3\times4$ module geometry with two modules.}
        \label{fig:2modules_3x4_new}
    \end{subfigure}
    \begin{subfigure}[t]{0.9\textwidth}
        \centering
        \includegraphics[width=\textwidth]{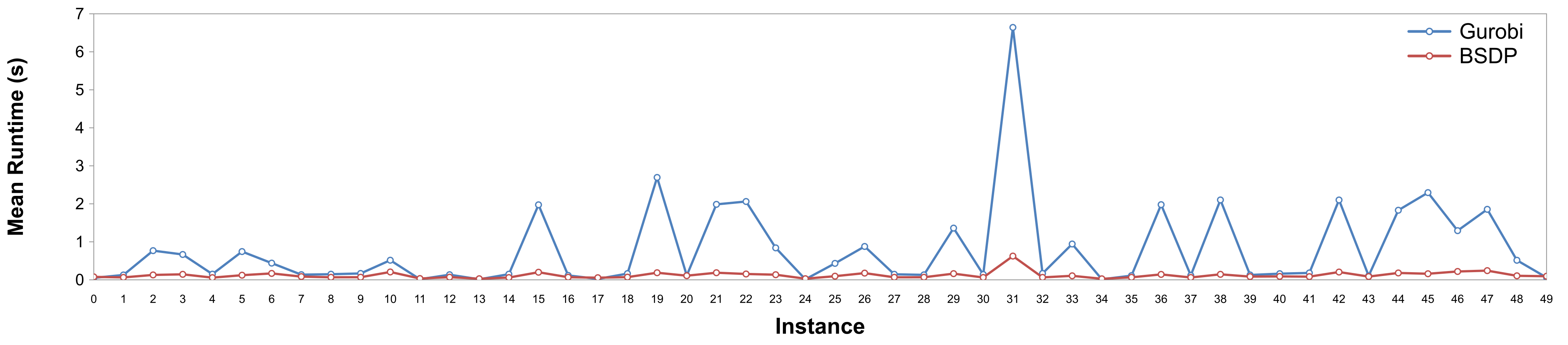}
        \caption{$3\times2$ module geometry with two modules.}
        \label{fig:2modules_3x2_new}
    \end{subfigure}
    \begin{subfigure}[t]{0.9\textwidth}
        \centering
        \includegraphics[width=\textwidth]{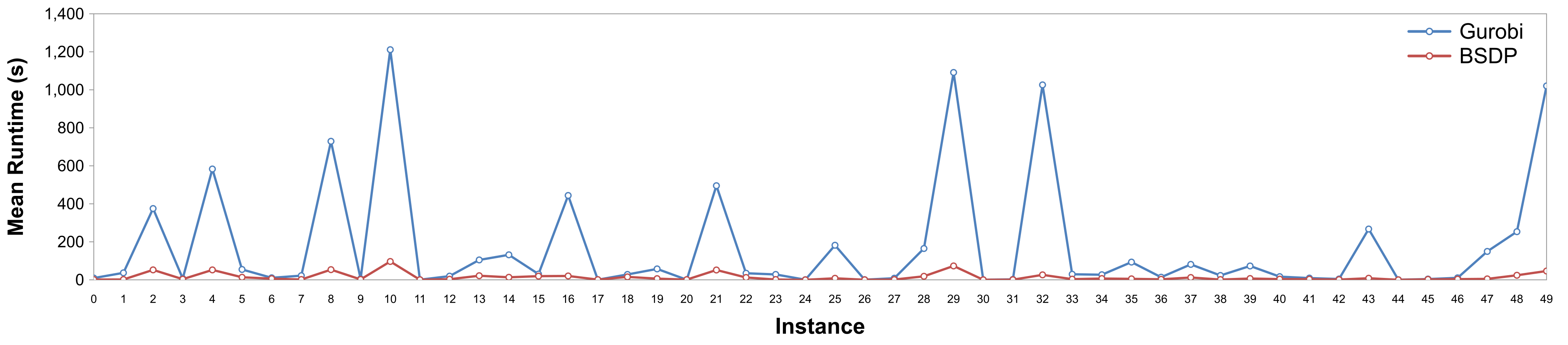}
        \caption{$3\times2$ module geometry with three modules.}
        \label{fig:3modules_3x2_new}
    \end{subfigure}
    \caption{Per-instance runtime comparison between BSDP and Gurobi under different rectangular module geometries.}
    \label{fig:geometry_case_runtime}
\end{figure}

\begin{table}[htbp]
\centering
\caption{Performance comparison between BSDP and Gurobi across different rectangular module settings.}
\label{tab:dp_gurobi_module_settings}
\begin{tabular}{llrr}
\toprule
Setting & Metric & BSDP & Gurobi \\
\midrule
2 modules, $3\times4$ & Total runtime & 1960.9 s & 2983.9 s \\
                      & Mean runtime & 39.2 s & 59.7 s \\
                      & Median runtime & 18.4 s & 22.5 s \\
                      & Aggregate speedup & 1.52$\times$ & -- \\
\midrule
2 modules, $3\times2$ & Total runtime & 6.0 s & 39.9 s \\
                      & Mean runtime & 0.120 s & 0.798 s \\
                      & Median runtime & 0.093 s & 0.168 s \\
                      & Aggregate speedup & 6.64$\times$ & -- \\
\midrule
3 modules, $3\times2$ & Total runtime & 719.1 s & 8972.8 s \\
                      & Mean runtime & 14.4 s & 179.5 s \\
                      & Median runtime & 5.1 s & 28.9 s \\
                      & Aggregate speedup & 12.48$\times$ & -- \\
\bottomrule
\end{tabular}
\end{table}

Figure~\ref{fig:geometry_case_runtime} shows the per-instance runtime profiles of BSDP and Gurobi under the three module settings, and Table~\ref{tab:dp_gurobi_module_settings} summarizes the corresponding aggregate statistics. In the $3\times4$ two-module setting, BSDP and Gurobi are relatively close on many easy instances, but Gurobi exhibits several pronounced runtime spikes on harder cases. As a result, BSDP achieves a lower total runtime of 1960.9 s, compared with 2983.9 s for Gurobi. This corresponds to an aggregate speedup of 1.52$\times$, or equivalently a 34.3\% reduction in total computational time. BSDP also reduces the mean runtime from 59.7 s to 39.2 s and the median runtime from 22.5 s to 18.4 s. These results indicate that, even when the module is enlarged to a $3\times4$ rectangular geometry with richer internal path combinations, BSDP remains competitive and provides a clear advantage in aggregate runtime.

The advantage of BSDP becomes more substantial on the $3\times2$ two-module benchmark. Although this setting contains fewer cells per module than the $3\times4$ case, its narrow geometry produces a denser interaction structure with less room for path separation. In this setting, BSDP requires only 6.0 s in total, whereas Gurobi requires 39.9 s. The resulting aggregate speedup is 6.64$\times$, which is much larger than that observed in the $3\times4$ two-module case. The mean runtime is reduced from 0.798 s to 0.120 s, and the median runtime is reduced from 0.168 s to 0.093 s. This suggests that the compact boundary-state representation used by BSDP is especially effective when local feasibility is dominated by tight spatial interactions.

The strongest improvement is observed in the $3\times2$ three-module setting. When the number of modules increases from two to three, the interactions among local decisions propagate over a longer module sequence, making the problem significantly more challenging for the global Gurobi formulation. Gurobi requires 8972.8 s in total and 179.5 s on average, whereas BSDP requires only 719.1 s in total and 14.4 s on average. This gives an aggregate speedup of 12.48$\times$, corresponding to an approximately 92.0\% reduction in total runtime. The median runtime is also reduced from 28.9 s to 5.1 s, showing that BSDP improves not only the aggregate computational cost but also the typical-instance performance.

Overall, the results confirm that BSDP is not restricted to the baseline width $m=3$. Its computational advantage persists for both the narrower $m=2$ layout and the wider $m=4$ layout. More importantly, the results show that $m$ is not merely a geometric label but a consequential vehicle-design parameter. Increasing $m$ raises the seating capacity of each modular unit, but it also expands the local configuration space and can substantially increase solution time. A smaller $m$ reduces capacity and creates tighter movement interactions, although its smaller local state space can be exploited effectively by BSDP. The choice of module width should therefore balance passenger capacity, maneuvering flexibility, and the real-time computational effort required for seat reconfiguration.

\subsection{Ablation Study}
\begin{figure}[h!]
    \centering
    \begin{subfigure}[t]{0.9\textwidth}
        \centering
        \includegraphics[width=\textwidth]{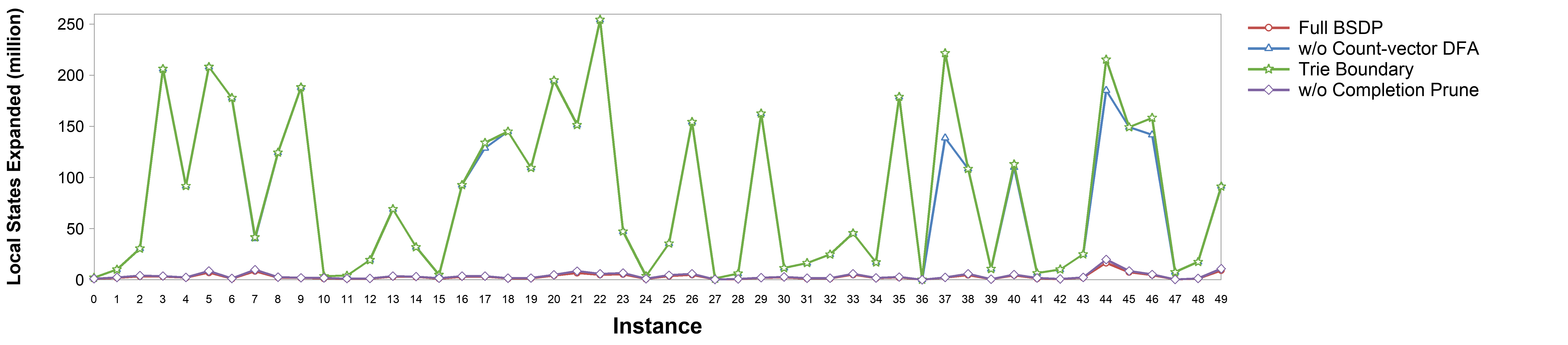}
        \caption{Number of local states expanded.}
        \label{fig:ablation_local_expanded}
    \end{subfigure}
    \begin{subfigure}[t]{0.9\textwidth}
        \centering
        \includegraphics[width=\textwidth]{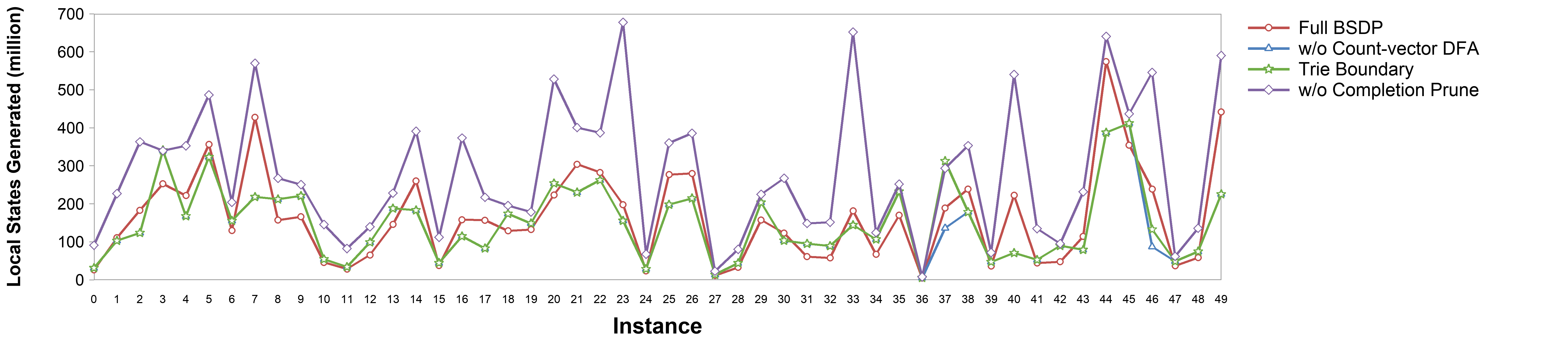}
        \caption{Number of local states generated.}
        \label{fig:ablation_local_generated}
    \end{subfigure}
    \begin{subfigure}[t]{0.9\textwidth}
        \centering
        \includegraphics[width=\textwidth]{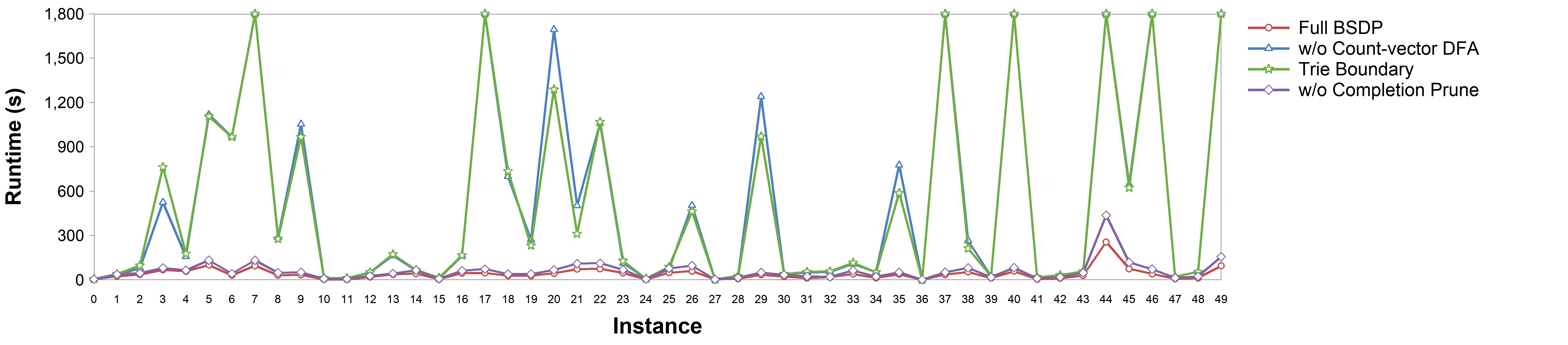}
        \caption{Runtime.}
        \label{fig:ablation_runtime}
    \end{subfigure}
    \caption{Ablation study of BSDP on the 3-module benchmark with 50 instances.}
    \label{fig:ablation_study}
\end{figure}

To quantify the contribution of the major algorithmic components of BSDP, we conduct an ablation study on the 3-module benchmark consisting of 50 instances. All variants are evaluated under the same exact-safe setting with an 1800-second time limit. The full BSDP is compared against three degraded variants: (i) BSDP without the count-vector DFA compression for boundary languages, (ii) BSDP using a basic trie-based boundary representation instead of the DFA-compressed representation, and (iii) BSDP without feasibility-based side-completion pruning. The numbers of local states expanded and generated, together with the runtime results, are reported in Figures~\ref{fig:ablation_local_expanded}--\ref{fig:ablation_runtime}.

Figure~\ref{fig:ablation_local_expanded} shows that the DFA-compressed boundary representation substantially reduces the effective search space. The full BSDP expands only 3.08 million local states on average. In contrast, removing the count-vector DFA compression increases the average number of expanded states to 79.94 million, while replacing the compressed DFA boundary with a basic trie boundary increases it to 82.71 million. These two variants therefore expand roughly 26 times more local states than the full method. This demonstrates that the performance gain of BSDP is not merely due to implementation-level efficiency, but is primarily driven by a structural reduction of the boundary state space.

The results on generated local states, shown in Figure~\ref{fig:ablation_local_generated}, further clarify the role of side-completion pruning. Removing this pruning mechanism leads to the largest increase in generated states, from 164.84 million in the full BSDP to 281.63 million on average. This indicates that side-completion pruning effectively eliminates many unnecessary successor states before they enter later stages of the dynamic program. Nevertheless, this variant still avoids timeouts and remains relatively close to the full method in runtime, suggesting that side-completion pruning mainly improves efficiency by reducing redundant state generation, whereas DFA-based boundary compression is more critical for controlling the number of states that must actually be expanded.

The runtime comparison in Figure~\ref{fig:ablation_runtime} is consistent with the above observations. The full BSDP is the fastest and most stable variant, with an average runtime of 38.996 seconds and no timeout over all 50 instances. Disabling the count-vector DFA compression increases the mean runtime to 514.040 seconds and results in a 14.0\% timeout rate. Similarly, using the trie boundary representation leads to an average runtime of 495.946 seconds and the same 14.0\% timeout rate. By comparison, removing side-completion pruning increases the average runtime to 59.001 seconds, which is about 1.5 times that of the full BSDP, but does not cause any timeout. These results confirm that the compact DFA representation is the dominant factor behind the scalability and robustness of BSDP, especially on hard instances.

Overall, the ablation study shows that both components contribute to the efficiency of BSDP, but they play different roles. The count-vector DFA compression is essential for scalability because it prevents the boundary state space from growing explosively and significantly reduces the number of expanded local states. The side-completion pruning provides an additional efficiency gain by suppressing redundant successor generation. Together, these mechanisms make BSDP faster, more stable, and less susceptible to hard-instance timeouts on the 3-module benchmark.

\section{Conclusions and Future Research}\label{sec:conclusion}

This study addresses the physical seat-reconfiguration problem that arises when passengers must be redistributed among coupled modular bus units before decoupling. Given seat-to-module assignments, the proposed model represents seats as colored tiles moving simultaneously along a constrained track graph. It explicitly accounts for occupancy restrictions, opposing edge movements,
and corner-following conflicts. The lexicographic objective distinguishes two operational requirements: completing reconfiguration at the earliest feasible time and minimizing cumulative seat-operation cost among schedules attaining that
time. This distinction matters because schedules with the same minimum makespan can require substantially different amounts of movement and non-target waiting.

We develop a time-expanded mixed-integer linear programming formulation and an exact Boundary-Sequence Dynamic Programming algorithm. The central methodological insight is that the single-edge connection between adjacent modules provides a
sufficient interface for coordinating their local schedules. Encoding the timing, direction, and color of boundary crossings allows the global problem to be decomposed into module-level subproblems and solved through min-plus composition.
Finite-automaton compression and weighted boundary representations reduce the search space while preserving feasibility and optimal costs. Combined with a search over the time horizon, the fixed-horizon algorithm provides an exact solution approach to the lexicographic problem.

The computational results demonstrate benefits in both exact solution efficiency and feasible-solution improvement. On the baseline two- and three-module benchmarks, BSDP solves all instances within the 1800-s limit and achieves mean
per-instance runtime speedups of 3.24 and 6.30 over Gurobi, respectively, with timeout runtimes capped at the time limit. For four to seven modules, the experiments instead fix the feasible horizons supplied by common greedy schedules. Under identical initial solutions and 300-s budgets, BSDP obtains a lower final SoC on 197 of 200 instances and ties on the remaining three. The reductions in mean final SoC relative to Gurobi range from 17.93\% for four modules to 69.07\% for seven modules. These results establish an advantage in fixed-horizon improvement, rather than a claim that the larger instances have been solved to lexicographic optimality. The seven-module results also show that better incumbent solutions do not necessarily imply stronger lower bounds, highlighting the remaining difficulty of optimality certification.

The additional experiments clarify the sources and scope of these gains. Runtime advantages persist in the tested narrower and wider module geometries, indicating that BSDP is not restricted to the baseline layout. The ablation study identifies
boundary-language compression as a major contributor to efficiency, while side-completion pruning further reduces redundant state generation. From an operational perspective, the results support treating seat reconfiguration as an explicit component of modular bus planning. Minimizing makespan alone does not control the cumulative movement and waiting incurred before decoupling. The secondary objective provides a means of reducing this burden without sacrificing the earliest feasible completion time.

Several extensions remain necessary for broader application. The present model assumes predetermined seat-to-module assignments, a fixed track topology, and synchronized movements with uniform action durations. Jointly optimizing assignments, seat movements, and vehicle coupling decisions could reveal additional opportunities to reduce reconfiguration effort. Incorporating heterogeneous movement durations, acceleration constraints, and execution uncertainty would bring the scheduling model closer to physical operation. Moreover, SoC is an operational proxy rather than a direct measurement of passenger comfort. Physical trials and passenger studies are therefore needed to calibrate its cost components and assess the relationship between optimized schedules and experienced disturbance. Finally, stronger lower bounds and methods that provide explicit solution-quality guarantees under short computational budgets would extend the approach to larger formations and support its integration with online transit control.

\bibliographystyle{apalike} 
\bibliography{reference}

\newpage
\appendix

\section{Detailed Results}

\subsection{Results of 2 modules}\label{app:detailed_res_2}

\begin{longtable}{c l r r p{4.4cm} r}
\caption{Results for 2 modules}\label{tab:results-2modules}\\
\toprule
Instance & Method & Makespan & SoC & Runtime (s) & Speedup \\
\midrule
\endfirsthead
\toprule
Instance & Method & Makespan & SoC & Runtime (s) & Speedup \\
\midrule
\endhead
\midrule
\multicolumn{6}{r}{Continued on next page} \\
\endfoot
\bottomrule
\endlastfoot
0 & BSDP & 11 & 52 & 0.559 [0.554, 0.564] & 2.812 \\
 & Gurobi &  &  & 1.572 [1.519, 1.625] &  \\
1 & BSDP & 11 & 51 & 0.491 [0.487, 0.495] & 3 \\
 & Gurobi &  &  & 1.472 [1.418, 1.526] &  \\
2 & BSDP & 11 & 50 & 0.507 [0.503, 0.510] & 2.625 \\
 & Gurobi &  &  & 1.331 [1.296, 1.365] &  \\
3 & BSDP & 8 & 30 & 0.068 [0.067, 0.069] & 1.756 \\
 & Gurobi &  &  & 0.119 [0.115, 0.124] &  \\
4 & BSDP & 5 & 18 & 0.012 [0.012, 0.012] & 1.286 \\
 & Gurobi &  &  & 0.015 [0.015, 0.016] &  \\
5 & BSDP & 9 & 29 & 0.155 [0.154, 0.157] & 2.105 \\
 & Gurobi &  &  & 0.327 [0.317, 0.338] &  \\
6 & BSDP & 9 & 30 & 0.083 [0.082, 0.083] & 4.842 \\
 & Gurobi &  &  & 0.401 [0.384, 0.417] &  \\
7 & BSDP & 11 & 50 & 0.511 [0.507, 0.514] & 3.268 \\
 & Gurobi &  &  & 1.669 [1.567, 1.770] &  \\
8 & BSDP & 8 & 26 & 0.064 [0.064, 0.065] & 3.252 \\
 & Gurobi &  &  & 0.209 [0.200, 0.218] &  \\
9 & BSDP & 9 & 29 & 0.140 [0.139, 0.140] & 2.242 \\
 & Gurobi &  &  & 0.314 [0.296, 0.331] &  \\
10 & BSDP & 8 & 33 & 0.048 [0.048, 0.049] & 2.55 \\
 & Gurobi &  &  & 0.123 [0.118, 0.128] &  \\
11 & BSDP & 6 & 11 & 0.026 [0.025, 0.027] & 0.923 \\
 & Gurobi &  &  & 0.024 [0.023, 0.025] &  \\
12 & BSDP & 9 & 39 & 0.123 [0.121, 0.125] & 3.59 \\
 & Gurobi &  &  & 0.443 [0.429, 0.456] &  \\
13 & BSDP & 8 & 26 & 0.065 [0.064, 0.066] & 2.694 \\
 & Gurobi &  &  & 0.175 [0.166, 0.183] &  \\
14 & BSDP & 11 & 55 & 0.448 [0.446, 0.449] & 4.036 \\
 & Gurobi &  &  & 1.806 [1.767, 1.845] &  \\
15 & BSDP & 12 & 57 & 1.375 [1.365, 1.384] & 4.967 \\
 & Gurobi &  &  & 6.828 [6.538, 7.118] &  \\
16 & BSDP & 11 & 49 & 0.480 [0.479, 0.481] & 10.431 \\
 & Gurobi &  &  & 5.004 [4.660, 5.347] &  \\
17 & BSDP & 9 & 30 & 0.156 [0.156, 0.157] & 2.428 \\
 & Gurobi &  &  & 0.380 [0.362, 0.397] &  \\
18 & BSDP & 9 & 39 & 0.136 [0.136, 0.137] & 2.805 \\
 & Gurobi &  &  & 0.382 [0.356, 0.408] &  \\
19 & BSDP & 8 & 36 & 0.066 [0.065, 0.066] & 3.736 \\
 & Gurobi &  &  & 0.245 [0.225, 0.264] &  \\
20 & BSDP & 6 & 11 & 0.018 [0.017, 0.018] & 1.508 \\
 & Gurobi &  &  & 0.027 [0.025, 0.028] &  \\
21 & BSDP & 8 & 29 & 0.074 [0.074, 0.075] & 3.044 \\
 & Gurobi &  &  & 0.227 [0.218, 0.235] &  \\
22 & BSDP & 11 & 52 & 0.481 [0.476, 0.486] & 3.457 \\
 & Gurobi &  &  & 1.663 [1.551, 1.775] &  \\
23 & BSDP & 8 & 33 & 0.048 [0.047, 0.048] & 2.433 \\
 & Gurobi &  &  & 0.116 [0.110, 0.122] &  \\
24 & BSDP & 8 & 33 & 0.070 [0.068, 0.071] & 3.045 \\
 & Gurobi &  &  & 0.212 [0.203, 0.220] &  \\
25 & BSDP & 9 & 29 & 0.217 [0.214, 0.221] & 2.283 \\
 & Gurobi &  &  & 0.496 [0.460, 0.533] &  \\
26 & BSDP & 5 & 18 & 0.013 [0.013, 0.014] & 1.156 \\
 & Gurobi &  &  & 0.016 [0.015, 0.017] &  \\
27 & BSDP & 8 & 27 & 0.079 [0.078, 0.080] & 2.829 \\
 & Gurobi &  &  & 0.224 [0.214, 0.233] &  \\
28 & BSDP & 8 & 27 & 0.079 [0.078, 0.080] & 3.129 \\
 & Gurobi &  &  & 0.247 [0.238, 0.257] &  \\
29 & BSDP & 8 & 27 & 0.065 [0.064, 0.066] & 1.682 \\
 & Gurobi &  &  & 0.110 [0.105, 0.114] &  \\
30 & BSDP & 5 & 18 & 0.013 [0.013, 0.014] & 1.185 \\
 & Gurobi &  &  & 0.016 [0.014, 0.018] &  \\
31 & BSDP & 8 & 31 & 0.062 [0.062, 0.063] & 2.09 \\
 & Gurobi &  &  & 0.131 [0.124, 0.137] &  \\
32 & BSDP & 5 & 13 & 0.014 [0.014, 0.014] & 1.736 \\
 & Gurobi &  &  & 0.024 [0.023, 0.026] &  \\
33 & BSDP & 11 & 51 & 0.563 [0.559, 0.567] & 6.69 \\
 & Gurobi &  &  & 3.764 [3.390, 4.139] &  \\
34 & BSDP & 10 & 35 & 0.259 [0.257, 0.261] & 2.171 \\
 & Gurobi &  &  & 0.562 [0.483, 0.641] &  \\
35 & BSDP & 9 & 29 & 0.198 [0.195, 0.200] & 2.984 \\
 & Gurobi &  &  & 0.590 [0.565, 0.615] &  \\
36 & BSDP & 10 & 36 & 0.241 [0.239, 0.243] & 5.642 \\
 & Gurobi &  &  & 1.359 [1.311, 1.406] &  \\
37 & BSDP & 11 & 50 & 0.503 [0.497, 0.509] & 6.228 \\
 & Gurobi &  &  & 3.134 [2.990, 3.278] &  \\
38 & BSDP & 9 & 39 & 0.143 [0.130, 0.156] & 8.015 \\
 & Gurobi &  &  & 1.146 [1.114, 1.178] &  \\
39 & BSDP & 9 & 38 & 0.138 [0.137, 0.140] & 7.673 \\
 & Gurobi &  &  & 1.062 [1.028, 1.096] &  \\
40 & BSDP & 11 & 62 & 0.428 [0.423, 0.434] & 3.895 \\
 & Gurobi &  &  & 1.669 [1.632, 1.706] &  \\
41 & BSDP & 9 & 29 & 0.158 [0.155, 0.160] & 2.174 \\
 & Gurobi &  &  & 0.342 [0.338, 0.347] &  \\
42 & BSDP & 9 & 30 & 0.079 [0.079, 0.080] & 3.956 \\
 & Gurobi &  &  & 0.313 [0.308, 0.317] &  \\
43 & BSDP & 9 & 38 & 0.138 [0.137, 0.139] & 3.369 \\
 & Gurobi &  &  & 0.464 [0.447, 0.481] &  \\
44 & BSDP & 5 & 13 & 0.014 [0.013, 0.014] & 1.71 \\
 & Gurobi &  &  & 0.024 [0.023, 0.024] &  \\
45 & BSDP & 9 & 30 & 0.160 [0.160, 0.161] & 1.916 \\
 & Gurobi &  &  & 0.307 [0.299, 0.315] &  \\
46 & BSDP & 11 & 53 & 0.453 [0.450, 0.455] & 3.047 \\
 & Gurobi &  &  & 1.379 [1.327, 1.431] &  \\
47 & BSDP & 9 & 29 & 0.163 [0.162, 0.164] & 3.246 \\
 & Gurobi &  &  & 0.528 [0.517, 0.539] &  \\
48 & BSDP & 9 & 30 & 0.198 [0.195, 0.201] & 1.205 \\
 & Gurobi &  &  & 0.238 [0.233, 0.244] &  \\
49 & BSDP & 8 & 29 & 0.071 [0.070, 0.072] & 3.365 \\
 & Gurobi &  &  & 0.240 [0.237, 0.242] &  \\
\end{longtable}

\subsection{Results of 2 modules}\label{app:detailed_res_3}

\begin{longtable}{c l r r p{4.4cm} r}
\caption{Results for 3 modules}\label{tab:results-3modules}\\
\toprule
Instance & Method & Makespan & SoC & Runtime (s) & Speedup \\
\midrule
\endfirsthead
\toprule
Instance & Method & Makespan & SoC & Runtime (s) & Speedup \\
\midrule
\endhead
\midrule
\multicolumn{6}{r}{Continued on next page} \\
\endfoot
\bottomrule
\endlastfoot
0 & BSDP & 11 & 101 & 2.086 [1.483, 2.688] & 10.675 \\
 & Gurobi &  &  & 22.264 [17.465, 27.064] &  \\
1 & BSDP & 12 & 90 & 19.669 [13.823, 25.515] & 3.785 \\
 & Gurobi &  &  & 74.453 [54.564, 94.342] &  \\
2 & BSDP & 13 & 103 & 36.511 [29.156, 43.866] & 6.629 \\
 & Gurobi &  &  & 242.0 [181.1, 303.0] &  \\
3 & BSDP & 14 & 98 & 68.274 [56.057, 80.492] & 4.012 \\
 & Gurobi &  &  & 273.9 [201.3, 346.5] &  \\
4 & BSDP & 14 & 125 & 41.032 [33.459, 48.604] & 1.948 \\
 & Gurobi &  &  & 79.909 [57.308, 102.510] &  \\
5 & BSDP & 14 & 111 & 115.3 [98.7, 131.9] & 5.206 \\
 & Gurobi &  &  & 600.3 [435.0, 765.6] &  \\
6 & BSDP & 14 & 91 & 33.124 [26.865, 39.384] & 1.788 \\
 & Gurobi &  &  & 59.221 [42.899, 75.543] &  \\
7 & BSDP & 15 & 150 & 103.5 [88.3, 118.7] & 17.392 \\
 & Gurobi & \textemdash{} & \textemdash{} & 1800.0 [1800.0, 1800.0] &  \\
8 & BSDP & 13 & 103 & 29.233 [24.937, 33.530] & 6.783 \\
 & Gurobi &  &  & 198.3 [147.5, 249.1] &  \\
9 & BSDP & 13 & 112 & 31.084 [26.302, 35.867] & 8.505 \\
 & Gurobi &  &  & 264.4 [196.1, 332.7] &  \\
10 & BSDP & 11 & 87 & 3.389 [2.765, 4.012] & 2.214 \\
 & Gurobi &  &  & 7.502 [6.086, 8.918] &  \\
11 & BSDP & 11 & 75 & 2.943 [2.415, 3.472] & 1.739 \\
 & Gurobi &  &  & 5.118 [4.197, 6.040] &  \\
12 & BSDP & 12 & 83 & 13.871 [11.752, 15.989] & 3.487 \\
 & Gurobi &  &  & 48.364 [37.423, 59.306] &  \\
13 & BSDP & 14 & 111 & 32.026 [27.137, 36.915] & 3.398 \\
 & Gurobi &  &  & 108.8 [83.7, 133.9] &  \\
14 & BSDP & 14 & 146 & 38.137 [32.546, 43.729] & 12.497 \\
 & Gurobi &  &  & 476.6 [353.3, 599.9] &  \\
15 & BSDP & 11 & 84 & 3.255 [2.697, 3.813] & 3.67 \\
 & Gurobi &  &  & 11.946 [9.734, 14.158] &  \\
16 & BSDP & 13 & 85 & 40.018 [33.219, 46.816] & 1.494 \\
 & Gurobi &  &  & 59.788 [46.010, 73.567] &  \\
17 & BSDP & 14 & 103 & 52.173 [42.912, 61.434] & 1.222 \\
 & Gurobi &  &  & 63.743 [49.428, 78.058] &  \\
18 & BSDP & 13 & 91 & 26.717 [22.498, 30.937] & 3.047 \\
 & Gurobi &  &  & 81.414 [61.768, 101.060] &  \\
19 & BSDP & 13 & 79 & 28.557 [23.399, 33.715] & 1.615 \\
 & Gurobi &  &  & 46.122 [35.355, 56.889] &  \\
20 & BSDP & 13 & 94 & 44.251 [37.146, 51.357] & 5.656 \\
 & Gurobi &  &  & 250.3 [191.7, 308.9] &  \\
21 & BSDP & 14 & 112 & 75.099 [65.454, 84.744] & 3.42 \\
 & Gurobi &  &  & 256.8 [191.6, 322.1] &  \\
22 & BSDP & 14 & 92 & 79.329 [70.461, 88.197] & 6.572 \\
 & Gurobi &  &  & 521.3 [393.1, 649.6] &  \\
23 & BSDP & 13 & 122 & 45.411 [38.827, 51.994] & 24.472 \\
 & Gurobi &  &  & 1111.3 [850.0, 1372.6] &  \\
24 & BSDP & 11 & 82 & 2.504 [2.246, 2.761] & 3.361 \\
 & Gurobi &  &  & 8.414 [6.868, 9.960] &  \\
25 & BSDP & 14 & 145 & 45.694 [38.517, 52.870] & 39.393 \\
 & Gurobi & \textemdash{} & \textemdash{} & 1800.0 [1800.0, 1800.0] &  \\
26 & BSDP & 14 & 136 & 63.025 [53.578, 72.473] & 1.997 \\
 & Gurobi &  &  & 125.9 [94.8, 156.9] &  \\
27 & BSDP & 10 & 58 & 1.787 [1.634, 1.941] & 0.659 \\
 & Gurobi &  &  & 1.179 [1.009, 1.348] &  \\
28 & BSDP & 12 & 95 & 5.940 [4.928, 6.952] & 5.215 \\
 & Gurobi &  &  & 30.975 [24.141, 37.809] &  \\
29 & BSDP & 13 & 93 & 30.964 [25.256, 36.673] & 3.451 \\
 & Gurobi &  &  & 106.9 [80.9, 132.9] &  \\
30 & BSDP & 12 & 98 & 18.130 [14.918, 21.342] & 2.991 \\
 & Gurobi &  &  & 54.219 [41.958, 66.480] &  \\
31 & BSDP & 11 & 101 & 1.739 [1.514, 2.104] & 1.702 \\
 & Gurobi &  &  & 2.960 [2.484, 3.436] &  \\
32 & BSDP & 12 & 80 & 14.196 [11.770, 16.622] & 1.841 \\
 & Gurobi &  &  & 26.134 [20.068, 32.201] &  \\
33 & BSDP & 13 & 114 & 43.280 [35.684, 50.875] & 33.487 \\
 & Gurobi &  &  & 1449.3 [1206.6, 1692.0] &  \\
34 & BSDP & 13 & 122 & 10.213 [8.706, 11.720] & 8.081 \\
 & Gurobi &  &  & 82.529 [61.312, 103.746] &  \\
35 & BSDP & 14 & 113 & 34.681 [28.783, 40.579] & 4.508 \\
 & Gurobi &  &  & 156.3 [117.3, 195.3] &  \\
36 & BSDP & 9 & 59 & 0.660 [0.607, 0.712] & 2.074 \\
 & Gurobi &  &  & 1.368 [1.125, 1.611] &  \\
37 & BSDP & 13 & 82 & 29.101 [23.645, 34.558] & 3.823 \\
 & Gurobi &  &  & 111.3 [83.1, 139.4] &  \\
38 & BSDP & 14 & 143 & 58.165 [48.168, 68.163] & 3.867 \\
 & Gurobi &  &  & 224.9 [166.6, 283.2] &  \\
39 & BSDP & 12 & 70 & 10.074 [8.385, 11.763] & 0.987 \\
 & Gurobi &  &  & 9.942 [7.678, 12.206] &  \\
40 & BSDP & 13 & 81 & 59.090 [48.386, 69.795] & 0.436 \\
 & Gurobi &  &  & 25.734 [19.360, 32.108] &  \\
41 & BSDP & 11 & 84 & 3.845 [3.150, 4.540] & 2.6 \\
 & Gurobi &  &  & 9.995 [7.835, 12.155] &  \\
42 & BSDP & 12 & 89 & 7.565 [6.286, 8.844] & 16.537 \\
 & Gurobi &  &  & 125.1 [87.6, 162.7] &  \\
43 & BSDP & 12 & 82 & 25.033 [20.519, 29.547] & 0.485 \\
 & Gurobi &  &  & 12.150 [9.322, 14.979] &  \\
44 & BSDP & 16 & 128 & 270.9 [239.7, 302.2] & 6.643 \\
 & Gurobi & \textemdash{} & \textemdash{} & 1800.0 [1800.0, 1800.0] &  \\
45 & BSDP & 15 & 104 & 84.451 [70.609, 98.293] & 5.89 \\
 & Gurobi &  &  & 497.4 [359.5, 635.3] &  \\
46 & BSDP & 13 & 103 & 44.573 [37.282, 51.864] & 9.034 \\
 & Gurobi &  &  & 402.6 [308.4, 496.9] &  \\
47 & BSDP & 12 & 83 & 6.103 [5.375, 6.831] & 1.794 \\
 & Gurobi &  &  & 10.951 [8.560, 13.342] &  \\
48 & BSDP & 12 & 95 & 12.103 [10.192, 14.014] & 1.781 \\
 & Gurobi &  &  & 21.552 [16.503, 26.601] &  \\
49 & BSDP & 15 & 122 & 113.1 [95.8, 130.4] & 11.168 \\
 & Gurobi &  &  & 1263.5 [990.3, 1536.8] &  \\
\end{longtable}

\subsection{Results of 4 modules}\label{app:detailed_res_4}
\begingroup
\small
\setlength{\tabcolsep}{7pt}
\begin{longtable}{rrrrrrr}
\caption{Comparison of BSDP and Gurobi for four $3\times3$ modules with identical greedy initial schedules, fixed horizons, 11 threads, and a 300~s time limit.}
\label{tab:four_modules_fixed_T}\\
\toprule
Instance & $T$ & Initial SoC & BSDP SoC & Gurobi SoC & BSDP LB & Gurobi LB \\
\midrule
\endfirsthead
\toprule
Instance & $T$ & Initial SoC & BSDP SoC & Gurobi SoC & BSDP LB & Gurobi LB \\
\midrule
\endhead
\midrule
\endfoot
\bottomrule
\endlastfoot
0  & 30 & 265 & 200 & 207 & 145 & 141 \\
1  & 33 & 171 & 110 & 111 & 106 & 104 \\
2  & 47 & 673 & 190 & 256 & 157 & 143 \\
3  & 34 & 225 & 151 & 157 & 135 & 128 \\
4  & 34 & 220 & 178 & 188 & 137 & 131 \\
5  & 26 & 233 & 196 & 205 & 153 & 146 \\
6  & 41 & 331 & 170 & 178 & 149 & 139 \\
7  & 26 & 195 & 171 & 184 & 144 & 143 \\
8  & 37 & 283 & 186 & 186 & 142 & 142 \\
9  & 36 & 275 & 173 & 184 & 142 & 138 \\
10 & 40 & 503 & 183 & 229 & 161 & 147 \\
11 & 49 & 508 & 190 & 313 & 158 & 147 \\
12 & 37 & 329 & 219 & 232 & 161 & 155 \\
13 & 25 & 167 & 132 & 143 & 116 & 107 \\
14 & 44 & 496 & 146 & 224 & 129 & 117 \\
15 & 38 & 263 & 172 & 177 & 149 & 140 \\
16 & 24 & 181 & 138 & 143 & 122 & 125 \\
17 & 41 & 497 & 175 & 190 & 149 & 142 \\
18 & 50 & 687 & 219 & 299 & 146 & 135 \\
19 & 28 & 225 & 139 & 142 & 128 & 127 \\
20 & 49 & 549 & 175 & 220 & 150 & 137 \\
21 & 54 & 766 & 198 & 505 & 167 & 155 \\
22 & 37 & 244 & 161 & 174 & 141 & 136 \\
23 & 31 & 232 & 172 & 178 & 141 & 136 \\
24 & 55 & 838 & 225 & 618 & 179 & 168 \\
25 & 44 & 451 & 173 & 216 & 149 & 135 \\
26 & 43 & 409 & 175 & 200 & 144 & 132 \\
27 & 31 & 288 & 180 & 203 & 155 & 147 \\
28 & 36 & 334 & 178 & 200 & 152 & 145 \\
29 & 52 & 582 & 173 & 397 & 153 & 139 \\
30 & 39 & 346 & 152 & 152 & 131 & 120 \\
31 & 47 & 315 & 165 & 184 & 144 & 137 \\
32 & 37 & 411 & 167 & 170 & 138 & 132 \\
33 & 25 & 239 & 160 & 164 & 142 & 129 \\
34 & 37 & 273 & 159 & 165 & 141 & 135 \\
35 & 34 & 321 & 184 & 197 & 154 & 145 \\
36 & 52 & 520 & 213 & 238 & 149 & 138 \\
37 & 48 & 600 & 209 & 212 & 177 & 162 \\
38 & 44 & 529 & 188 & 276 & 146 & 136 \\
39 & 29 & 204 & 140 & 141 & 127 & 129 \\
40 & 29 & 196 & 153 & 155 & 130 & 126 \\
41 & 27 & 198 & 149 & 151 & 124 & 128 \\
42 & 38 & 335 & 178 & 188 & 156 & 139 \\
43 & 36 & 343 & 162 & 171 & 150 & 141 \\
44 & 21 & 118 & 102 & 102 & 101 & 102 \\
45 & 33 & 249 & 191 & 206 & 149 & 143 \\
46 & 37 & 348 & 212 & 262 & 158 & 150 \\
47 & 40 & 277 & 165 & 175 & 137 & 128 \\
48 & 37 & 304 & 191 & 203 & 156 & 149 \\
49 & 31 & 259 & 176 & 186 & 155 & 141 \\
\end{longtable}
\endgroup

\subsection{Results of 5 modules}\label{app:detailed_res_5}
\begingroup
\small
\setlength{\tabcolsep}{7pt}
\begin{longtable}{rrrrrrr}
\caption{Comparison of BSDP and Gurobi for five $3\times3$ modules with identical greedy initial schedules, fixed horizons, 11 threads, and a 300~s time limit.}
\label{tab:five_modules_fixed_T}\\
\toprule
Instance & $T$ & Initial SoC & BSDP SoC & Gurobi SoC & BSDP LB & Gurobi LB \\
\midrule
\endfirsthead
\toprule
Instance & $T$ & Initial SoC & BSDP SoC & Gurobi SoC & BSDP LB & Gurobi LB \\
\midrule
\endhead
\midrule
\endfoot
\bottomrule
\endlastfoot
0  & 56 & 866  & 329 & 576  & 213 & 204 \\
1  & 58 & 844  & 277 & 642  & 192 & 181 \\
2  & 56 & 966  & 307 & 859  & 219 & 208 \\
3  & 66 & 1087 & 321 & 1024 & 224 & 214 \\
4  & 59 & 818  & 282 & 561  & 193 & 186 \\
5  & 79 & 1248 & 401 & 1071 & 244 & 236 \\
6  & 35 & 363  & 287 & 313  & 217 & 209 \\
7  & 63 & 980  & 325 & 792  & 227 & 220 \\
8  & 62 & 1101 & 304 & 825  & 215 & 201 \\
9  & 56 & 767  & 342 & 609  & 204 & 196 \\
10 & 59 & 864  & 356 & 693  & 225 & 215 \\
11 & 38 & 459  & 267 & 288  & 199 & 189 \\
12 & 64 & 1011 & 286 & 772  & 189 & 182 \\
13 & 50 & 716  & 280 & 558  & 184 & 176 \\
14 & 63 & 1143 & 407 & 1087 & 246 & 240 \\
15 & 48 & 496  & 388 & 431  & 255 & 246 \\
16 & 51 & 636  & 243 & 453  & 179 & 173 \\
17 & 59 & 900  & 318 & 689  & 192 & 184 \\
18 & 51 & 653  & 243 & 429  & 189 & 181 \\
19 & 50 & 744  & 305 & 632  & 207 & 194 \\
20 & 70 & 1295 & 404 & 1142 & 243 & 228 \\
21 & 45 & 715  & 289 & 579  & 183 & 172 \\
22 & 62 & 1008 & 285 & 710  & 202 & 193 \\
23 & 47 & 707  & 240 & 398  & 174 & 164 \\
24 & 72 & 1333 & 346 & 1223 & 237 & 227 \\
25 & 50 & 681  & 275 & 297  & 187 & 177 \\
26 & 78 & 1266 & 300 & 1063 & 211 & 202 \\
27 & 54 & 887  & 289 & 696  & 196 & 191 \\
28 & 55 & 867  & 242 & 779  & 192 & 181 \\
29 & 79 & 1381 & 322 & 1112 & 214 & 206 \\
30 & 62 & 1011 & 308 & 801  & 214 & 207 \\
31 & 83 & 1546 & 514 & 1525 & 252 & 244 \\
32 & 43 & 553  & 193 & 201  & 153 & 143 \\
33 & 64 & 1112 & 364 & 876  & 212 & 206 \\
34 & 61 & 839  & 251 & 472  & 192 & 182 \\
35 & 69 & 1126 & 283 & 858  & 197 & 187 \\
36 & 54 & 773  & 302 & 657  & 199 & 190 \\
37 & 71 & 1239 & 331 & 953  & 237 & 232 \\
38 & 73 & 1124 & 294 & 847  & 211 & 200 \\
39 & 53 & 763  & 241 & 650  & 185 & 176 \\
40 & 65 & 925  & 308 & 835  & 189 & 180 \\
41 & 56 & 710  & 212 & 536  & 174 & 167 \\
42 & 74 & 1317 & 418 & 1041 & 231 & 229 \\
43 & 57 & 755  & 296 & 650  & 224 & 216 \\
44 & 53 & 601  & 231 & 384  & 165 & 156 \\
45 & 54 & 685  & 345 & 456  & 196 & 187 \\
46 & 83 & 1396 & 320 & 1218 & 224 & 220 \\
47 & 56 & 816  & 277 & 563  & 185 & 176 \\
48 & 60 & 865  & 341 & 645  & 225 & 214 \\
49 & 54 & 469  & 277 & 336  & 195 & 185 \\
\end{longtable}
\endgroup

\subsection{Results of 6 modules}\label{app:detailed_res_6}
\begingroup
\small
\setlength{\tabcolsep}{7pt}
\begin{longtable}{rrrrrrr}
\caption{Comparison of BSDP and Gurobi for six $3\times3$ modules with identical greedy initial schedules, fixed horizons, 11 threads, and a 300~s time limit.}\label{tab:six_modules_fixed_T}\\
\toprule
Instance & $T$ & Initial SoC & BSDP SoC & Gurobi SoC & BSDP LB & Gurobi LB \\
\midrule\endfirsthead
\toprule
Instance & $T$ & Initial SoC & BSDP SoC & Gurobi SoC & BSDP LB & Gurobi LB \\
\midrule\endhead
\midrule\endfoot
\bottomrule\endlastfoot
0  & 74 & 1338 & 497 & 1124 & 280 & 277 \\
1  & 80 & 1669 & 441 & 1587 & 296 & 292 \\
2  & 62 & 948  & 319 & 719  & 235 & 227 \\
3  & 95 & 2102 & 486 & 1958 & 299 & 297 \\
4  & 61 & 1114 & 487 & 941  & 265 & 259 \\
5  & 62 & 1297 & 443 & 1207 & 283 & 272 \\
6  & 73 & 1379 & 446 & 1261 & 264 & 258 \\
7  & 67 & 1480 & 408 & 1384 & 270 & 260 \\
8  & 65 & 1247 & 419 & 1074 & 257 & 247 \\
9  & 95 & 2242 & 497 & 2103 & 305 & 303 \\
10 & 64 & 1129 & 419 & 1044 & 242 & 241 \\
11 & 64 & 1115 & 491 & 851  & 269 & 256 \\
12 & 71 & 1226 & 453 & 1109 & 293 & 284 \\
13 & 70 & 1302 & 425 & 1155 & 250 & 246 \\
14 & 71 & 1291 & 378 & 1037 & 233 & 227 \\
15 & 84 & 1472 & 457 & 1246 & 269 & 263 \\
16 & 64 & 1161 & 390 & 975  & 253 & 241 \\
17 & 69 & 1075 & 447 & 877  & 271 & 265 \\
18 & 60 & 851  & 297 & 680  & 218 & 217 \\
19 & 68 & 1313 & 462 & 1126 & 277 & 267 \\
20 & 82 & 1530 & 455 & 1443 & 273 & 271 \\
21 & 68 & 1166 & 401 & 965  & 223 & 222 \\
22 & 70 & 1181 & 396 & 1023 & 257 & 252 \\
23 & 75 & 1330 & 370 & 1043 & 257 & 248 \\
24 & 69 & 1475 & 429 & 1321 & 263 & 257 \\
25 & 70 & 1040 & 385 & 834  & 233 & 226 \\
26 & 86 & 1605 & 470 & 1489 & 311 & 310 \\
27 & 91 & 1809 & 494 & 1582 & 307 & 301 \\
28 & 93 & 1636 & 835 & 1550 & 319 & 313 \\
29 & 76 & 1500 & 512 & 1217 & 280 & 277 \\
30 & 87 & 1682 & 463 & 1637 & 277 & 271 \\
31 & 69 & 1502 & 354 & 1204 & 234 & 228 \\
32 & 64 & 1116 & 450 & 880  & 234 & 228 \\
33 & 76 & 1512 & 391 & 1242 & 246 & 244 \\
34 & 84 & 1559 & 400 & 1267 & 249 & 247 \\
35 & 64 & 1156 & 447 & 936  & 253 & 244 \\
36 & 52 & 1023 & 328 & 793  & 218 & 210 \\
37 & 74 & 1524 & 422 & 1398 & 266 & 265 \\
38 & 55 & 909  & 357 & 778  & 258 & 247 \\
39 & 85 & 1938 & 622 & 1895 & 312 & 309 \\
40 & 90 & 1861 & 557 & 1819 & 316 & 314 \\
41 & 70 & 1398 & 474 & 1195 & 271 & 264 \\
42 & 80 & 1704 & 507 & 1624 & 308 & 303 \\
43 & 70 & 1123 & 307 & 953  & 237 & 227 \\
44 & 76 & 1565 & 534 & 1499 & 300 & 296 \\
45 & 65 & 1474 & 531 & 1383 & 310 & 314 \\
46 & 80 & 1714 & 417 & 1612 & 279 & 279 \\
47 & 90 & 1913 & 564 & 1880 & 286 & 286 \\
48 & 85 & 1719 & 461 & 1661 & 280 & 275 \\
49 & 65 & 1080 & 388 & 878  & 225 & 220 \\
\end{longtable}
\endgroup

\subsection{Results of 7 modules}\label{app:detailed_res_7}
\begingroup
\small
\setlength{\tabcolsep}{7pt}
\begin{longtable}{rrrrrrr}
\caption{Comparison of BSDP and Gurobi for seven $3\times3$ modules with identical greedy initial schedules, fixed horizons, 11 threads, and a 300~s time limit.}\label{tab:seven_modules_fixed_T}\\
\toprule
Instance & $T$ & Initial SoC & BSDP SoC & Gurobi SoC & BSDP LB & Gurobi LB \\
\midrule\endfirsthead
\toprule
Instance & $T$ & Initial SoC & BSDP SoC & Gurobi SoC & BSDP LB & Gurobi LB \\
\midrule\endhead
\midrule\endfoot
\bottomrule\endlastfoot
0  & 71  & 1459 & 489 & 1108 & 299 & 293 \\
1  & 89  & 2444 & 800 & 2395 & 372 & 377 \\
2  & 96  & 2287 & 571 & 2218 & 332 & 338 \\
3  & 92  & 1967 & 594 & 1917 & 346 & 345 \\
4  & 98  & 2440 & 647 & 2383 & 370 & 373 \\
5  & 80  & 1484 & 585 & 1448 & 325 & 322 \\
6  & 74  & 1436 & 881 & 1343 & 332 & 336 \\
7  & 89  & 2065 & 601 & 2021 & 323 & 325 \\
8  & 74  & 1616 & 500 & 1566 & 308 & 306 \\
9  & 99  & 2186 & 561 & 2156 & 319 & 327 \\
10 & 93  & 2067 & 598 & 1991 & 324 & 327 \\
11 & 98  & 2211 & 591 & 1931 & 327 & 332 \\
12 & 104 & 2543 & 705 & 2506 & 383 & 392 \\
13 & 93  & 2223 & 647 & 2196 & 368 & 369 \\
14 & 93  & 2170 & 552 & 2005 & 312 & 314 \\
15 & 89  & 2289 & 663 & 2212 & 352 & 358 \\
16 & 101 & 2513 & 597 & 2437 & 367 & 378 \\
17 & 71  & 1529 & 582 & 1413 & 299 & 297 \\
18 & 109 & 3022 & 857 & 3022 & 390 & 393 \\
19 & 69  & 1472 & 535 & 1336 & 290 & 287 \\
20 & 90  & 2053 & 628 & 1713 & 342 & 339 \\
21 & 104 & 2623 & 748 & 2593 & 402 & 405 \\
22 & 92  & 2122 & 577 & 2050 & 323 & 325 \\
23 & 86  & 1960 & 551 & 1867 & 322 & 326 \\
24 & 105 & 2668 & 643 & 2646 & 367 & 375 \\
25 & 81  & 1692 & 496 & 1496 & 323 & 324 \\
26 & 80  & 1782 & 471 & 1640 & 295 & 295 \\
27 & 90  & 2388 & 607 & 2327 & 357 & 361 \\
28 & 97  & 2113 & 611 & 2020 & 366 & 366 \\
29 & 99  & 2270 & 630 & 2181 & 384 & 389 \\
30 & 106 & 2642 & 661 & 2558 & 364 & 372 \\
31 & 78  & 1971 & 568 & 1772 & 326 & 323 \\
32 & 100 & 2219 & 478 & 1904 & 294 & 300 \\
33 & 91  & 2046 & 636 & 1843 & 312 & 315 \\
34 & 89  & 2009 & 628 & 1904 & 331 & 335 \\
35 & 94  & 2686 & 809 & 2686 & 379 & 387 \\
36 & 87  & 1893 & 590 & 1841 & 335 & 337 \\
37 & 109 & 2667 & 783 & 2667 & 398 & 412 \\
38 & 98  & 1983 & 563 & 1963 & 309 & 311 \\
39 & 83  & 1613 & 528 & 1529 & 299 & 303 \\
40 & 98  & 2191 & 570 & 2084 & 314 & 321 \\
41 & 71  & 1720 & 523 & 1509 & 293 & 298 \\
42 & 70  & 1580 & 521 & 1509 & 328 & 329 \\
43 & 93  & 2469 & 779 & 2446 & 367 & 382 \\
44 & 82  & 2115 & 770 & 1945 & 358 & 368 \\
45 & 73  & 1407 & 475 & 1403 & 321 & 325 \\
46 & 92  & 1884 & 566 & 1571 & 309 & 320 \\
47 & 93  & 2686 & 675 & 2581 & 358 & 364 \\
48 & 68  & 1537 & 431 & 1405 & 250 & 248 \\
49 & 84  & 1715 & 520 & 1665 & 323 & 326 \\
\end{longtable}
\endgroup

\end{document}

%% file: preamble.tex
\usepackage{times}
\usepackage{geometry}
\usepackage{xcolor}
\usepackage{array}
\usepackage{bbding}
\usepackage{multirow}
\usepackage[fleqn]{amsmath}
\usepackage{amssymb}
\usepackage{amsthm}
\usepackage{graphicx}
\usepackage{mathrsfs}
\usepackage{natbib}
\usepackage{lscape}
\usepackage[hyphens]{url}
\usepackage[hidelinks]{hyperref}
\usepackage{comment}
\usepackage{bm}
\usepackage[title]{appendix}
\usepackage{longtable}
\usepackage{mathtools}
\usepackage{chngcntr}
\usepackage{authblk}
\usepackage{babel}
\usepackage{dsfont}
\usepackage{subcaption}
\usepackage{fancyhdr}
\usepackage{cleveref}
\usepackage{abstract}
\usepackage[short]{optidef}

\usepackage{booktabs}

\usepackage{algorithm}
\usepackage[noend]{algpseudocode}

\algrenewcommand\algorithmicrequire{\textbf{Input:}}
\algrenewcommand\algorithmicensure{\textbf{Output:}}

\makeatletter
\allowdisplaybreaks
\date{}

\fancypagestyle{plain}{%
	\fancyhf{} 
	
    \lhead{}
	\chead{\footnotesize \textcolor{gray}{\textit{\shortauthors -- \runningtitle}}}
	\rhead{\footnotesize \textcolor{gray}{\thepage}}
}
\makeatletter
\def\blfootnote{\xdef\@thefnmark{}\@footnotetext}
\makeatother

\makeatletter
\def\titlepageext{
	\begin{center}	
	{\parindent0pt
		\rule{0.9\textwidth}{1pt}
		\begin{minipage}[t]{0.25\textwidth}
			\small {\it Keywords:}\\
			\keyword
		\end{minipage}%
		\hspace{3mm}
		\begin{minipage}[t]{0.6\textwidth}
			\small \abstract
		\end{minipage}%
		
		\rule{0.9\textwidth}{2pt}
	}
	\end{center}

	\blfootnote{* Corresponding author. E-mail address: \href{mailto:\corresemail}{\corresemail}.}
}
\makeatother

\usepackage[mathlines, pagewise]{lineno}
\usepackage{etoolbox} 

\newcommand*\linenomathpatchAMS[1]{%
	\expandafter\pretocmd\csname #1\endcsname {\linenomathAMS}{}{}%
	\expandafter\pretocmd\csname #1*\endcsname{\linenomathAMS}{}{}%
	\expandafter\apptocmd\csname end#1\endcsname {\endlinenomath}{}{}%
	\expandafter\apptocmd\csname end#1*\endcsname{\endlinenomath}{}{}%
}

\expandafter\ifx\linenomath\linenomathWithnumbers
\let\linenomathAMS\linenomathWithnumbers
\patchcmd\linenomathAMS{\advance\postdisplaypenalty\linenopenalty}{}{}{}
\else
\let\linenomathAMS\linenomathNonumbers
\fi

\linenomathpatchAMS{gather}
\linenomathpatchAMS{multline}
\linenomathpatchAMS{align}
\linenomathpatchAMS{alignat}
\linenomathpatchAMS{flalign}

\nolinenumbers

\newtheorem{theorem}{Theorem}

\newtheorem{definition}{Definition}
\newtheorem{lemma}{Lemma}

\usepackage{tikz}
\usetikzlibrary{arrows.meta}

\definecolor{natureblue}{RGB}{0,114,178}
\definecolor{natureorange}{RGB}{230,159,0}
\definecolor{natureteal}{RGB}{0,128,115}
\definecolor{naturevermillion}{RGB}{213,94,0}

\newlength{\cgspImageWidth}